\def\draftversion{false}

\RequirePackage{ifthen}
\ifthenelse{\equal{\draftversion}{true}}{
  \documentclass[aps,prl,10pt,galley,amsmath,amssymb,longbibliography]{revtex4-2}
}{
  \documentclass[aps,prl,10pt,twocolumn,amsmath,amssymb,longbibliography,nofootinbib]{revtex4-2}
}

\usepackage{graphicx}
\usepackage[usenames,dvipsnames]{color} 
\usepackage{bm} 
\usepackage{soul} 
\usepackage{booktabs}
\usepackage{mathrsfs}
\usepackage{float}
\usepackage{dcolumn}
\usepackage{siunitx}
\newcolumntype{C}{>{\centering\arraybackslash}p{3cm}}

\ifthenelse{\equal{\draftversion}{true}}{
  \marginparwidth 2.7in
  \marginparsep 0.5in
  \newcounter{comm} 
  \def\commnext{\stepcounter{comm}}
  \def\commtext{{\bf\color{blue}[\arabic{comm}]}}
  \def\commmar{{\bf\color{blue}[\arabic{comm}]}}
  \def\dvm#1{\commnext\marginpar{\small DV\commmar: #1}\commtext}
  \def\dsm#1{\commnext\marginpar{\small DS\commmar: #1}\commtext}
  \def\jcm#1{\commnext\marginpar{\small JC\commmar: #1}\commtext}
  \def\mlab#1{\marginpar{\small\bf #1}}
  
}{
  \def\dvm#1{}
  \def\dsm#1{}
  \def\jcm#1{}
  \def\mlab#1{}
  
}

\newcommand{\beq}{\begin{equation}}
\newcommand{\eeq}{\end{equation}}
\newcommand{\bea}{\begin{eqnarray}}
\newcommand{\eea}{\end{eqnarray}}

\def\z2{$\mathbb{Z}_2$}

\def\phd{\phantom{(D) }}

\DeclareMathAlphabet\mathbfcal{OMS}{cmsy}{b}{n}   

\begin{document}


\title{Inducing topological flat bands in bilayer graphene with electric and magnetic superlattices}

\author{Daniel Seleznev}
\affiliation{Department of Physics and Astronomy, Center for Materials Theory, Rutgers University, Piscataway, New Jersey 08854, USA}
\email{dms632@physics.rutgers.edu}

\author{Jennifer Cano}
\affiliation{Department of Physics and Astronomy, Stony Brook University,
Stony Brook, New York 11794, USA}
\affiliation{Center for Computational Quantum Physics, Flatiron Institute, New York, New York 10010, USA}

\author{David Vanderbilt}
\affiliation{Department of Physics and Astronomy, Center for Materials Theory, Rutgers University, Piscataway, New Jersey 08854, USA}
\begin{abstract}

It was recently argued that Bernal stacked bilayer graphene (BLG) exposed to a 2D superlattice (SL) potential exhibits a variety of intriguing behaviors [Ghorashi \textit{et al}., Phys. Rev. Lett. \textbf{130}, 196201 (2023)]. Chief among them is the appearance of flat Chern bands that are favorable to the appearance of fractional Chern insulator states. Here, we explore the application of spatially periodic out-of-plane orbital magnetic fields to the model of Ghorashi \textit{et al}.~to find additional means of inducing flat Chern bands. We focus on fields that vary on length scales much larger than the atomic spacing in BLG, generating what we refer to as magnetic SLs. The magnetic SLs we investigate either introduce no net magnetic flux to the SL unit cell, or a single quantum of flux. We find that magnetic SLs acting on their own can induce topological flat bands, but richer behavior, such as the appearance of flat and generic bands with high Chern numbers, can be observed when the magnetic SLs act in conjunction with commensurate electric SLs. Finally, we propose a method of generating unit-flux-quantum magnetic SLs along with concomitant electric SLs. The magnetic SL is generated by periodic arrays of flux vortices originating from type II superconductors, while the electric SL arises due to a magnetic SL-induced charge density on the surface of a magnetoelectric material. Tuning the vortex lattice and the magnetoelectric coupling permits control of both SLs, and we study their effects on the band structure of BLG.

\end{abstract}

\maketitle


\section{Introduction}
\label{Intro}

The past few years have witnessed much excitement surrounding the field of twisted moiré materials. Beginning with the discovery of Mott insulating \cite{cao-nat18b} and superconducting \cite{cao-nat18a,yankowitz-sciadv19} behavior in twisted bilayer graphene (TBG), twisted heterostructures -- including those not based on graphene alone -- have since revealed themselves as highly tunable platforms capable of generating a rich variety of interaction-driven phenomena \cite{cao-nat18b,chen-natphys19,regan-nat20,li-nat21b,cao-nat18a,yankowitz-sciadv19,chen-nat19,li-nat21a,sharpe-sci2019,lu-nat19,serlin-sci20,chen-nat20,polshyn-nat20,tschirhart-sci21,xie-nat21,xu-prx23,cai-nat23,park-nat23,zeng-nat23,lu-nat24,liu-nat20,burg-prl19,he-natphys20,shen-natphys20,cao-nat20,park-nat21,xu-natphys21,chen-natphys20,hao-sci21,xu-nat2020,tang-nat20,wang-natmat20,xu-natnano21,song-sci21,can-natphys21,zhao-sci23,li-nat21,guo-arxiv24,xia-arxiv24,ji-arxiv24,redekop-arxiv24}. Underlying the appearance of correlated electron physics in these systems are flat bands induced by the emergence of spatially modulated interlayer couplings acting on the moiré length scale to quench the kinetic energy \cite{bistritzer-pnas11}.

Impeding further progress in the field, however, is the high sensitivity of twisted moiré materials to disorder stemming from twist angle inhomogeneities, lattice relaxation, domain formation, and substrate effects. Collectively, these factors hinder sample reproducibility \cite{lau-nat22}, thus making alternative methods of reproducing moiré flat-band physics highly sought after. 

Recently, the authors of Refs.~\cite{ghorashi-prl23,ghorashi-prb23,zeng-prb24} demonstrated that gated Bernal stacked bilayer graphene (BLG) exposed to a spatially modulated superlattice (SL) potential provides such an alternative platform, whereby the potential emulates the effect of the moiré interlayer coupling. In particular, Ref.~\cite{ghorashi-prl23} indicated that under experimentally feasible conditions, two distinct flat band regimes may occur. The first ``stack of flat bands" scenario features many topologically trivial almost perfectly flat bands spanning a range of energies relevant to the low-energy Hamiltonian describing the system. The second regime features topologically nontrivial flat bands with nonzero Chern numbers $C$, including those with $|C|>1$. 

The latter regime is especially interesting as it provides the opportunity to explore the interplay between topology and electronic correlations. In twisted heterostructures, this interplay can lead to the appearance of correlation-induced effects and states such as orbital Chern ferromagnetism and the quantum anomalous Hall effect \cite{sharpe-sci2019,lu-nat19,serlin-sci20,chen-nat20,polshyn-nat20,tschirhart-sci21}, as well as the fractional quantum anomalous Hall effect at fractional fillings \cite{xie-nat21,xu-prx23,cai-nat23,park-nat23,zeng-nat23,lu-nat24,wang-prl24,ji-arxiv24,redekop-arxiv24}. Bands with $|C|>1$ have also been proposed to lead to fractional Chern insulating states without Landau level analogues and to other exotic phenomena \cite{liu-prl12,yang-prb12,wang-prb12,sterdyniak-prb13,wu-prl13,barkeshli-prx12,barkeshli-prb13,moller-prl15,wu-prb15,behrmann-prl16,andrews-prb18,wang-prl22,liu-book24}. Furthermore, it has been argued that topology and interactions may be vital to the emergence of superconductivity in these materials \cite{khalaf-sciadv21,chatterjee-prb22,torma-natrevphys22}. In the case of the SL-exposed BLG (SL-BLG) model of Ref. \cite{ghorashi-prl23}, some of the topologically nontrivial flat bands were shown to potentially host fractional Chern insulating states at fractional fillings. Correlated insulating phases have also been recently claimed to have been observed in this setup \cite{sun-arxiv23}. 

Motivated by the possibility of correlated phases in SL-BLG stemming from topological flat bands, the present paper investigates extensions of the model of Ref.~\cite{ghorashi-prl23} to find additional means of inducing flat Chern bands. We focus on the effects of externally applied out-of-plane magnetic fields, due to their tendency to flatten bands and induce band topology. As the model of SL-BLG in Ref.~\cite{ghorashi-prl23} is meant to emulate TBG, we are also motivated to incorporate magnetic fields due to their observed ability to induce interesting correlated behavior in TBG \cite{dean-nat13,pierce-natphys21,saito-natphys21,das-prl22,herzog-prl22,finney-pnas22}. In particular, we study spatially periodic magnetic fields whose variation is on length scales much larger than the intralayer atomic spacing in BLG, generating what we refer to as magnetic SLs (MSLs). Furthermore, we consider two types of MSLs: those that introduce no net magnetic flux to the SL unit cell, and those that introduce a quantum of flux. Due to the size of the SL unit cell, a flux quantum may be realized with magnetic fields on the order of a few Tesla. We find that MSLs on their own can induce topological flat bands, but richer behavior, which can include the appearance of multiple flat and generic (non-flat) bands with high Chern numbers, can be observed when the MSLs act in conjunction with commensurate electric SLs (ESLs).

This paper is organized as follows. In Sec.~\ref{CanoModelReview}, we review the model of SL-BLG introduced in Ref.~\cite{ghorashi-prl23}. In Sec.~\ref{Methods} we discuss the computational methods we employ to calculate the band structure of SL-BLG exposed to periodic magnetic fields that either do or do not introduce flux to the SL unit cell. Using the methods outlined in the section prior, Sec.~\ref{Results_Triangle} explores the qualitative impacts of MSLs on SL-BLG using a model of a triangular MSL. 
In Sec.~\ref{Results_SC}, we propose a method of generating MSLs featuring a flux quantum along with concomitant ESLs. The MSL is generated by a periodic array of flux vortices originating from a type II superconductor, while the ESL appears due to an MSL-induced charge density on the surface of a magnetoelectric material. Tuning the vortex lattice and the magnetoelectric coupling permits control of both SLs, and we study their effects on the band structure of BLG. Finally, we summarize the findings of this paper in Sec.~\ref{Summary}. 

\section{Bilayer Graphene Exposed to an Electric Superlattice}
\label{CanoModelReview}
\begin{figure}[t]
\centering
\includegraphics[width= 9.0cm]{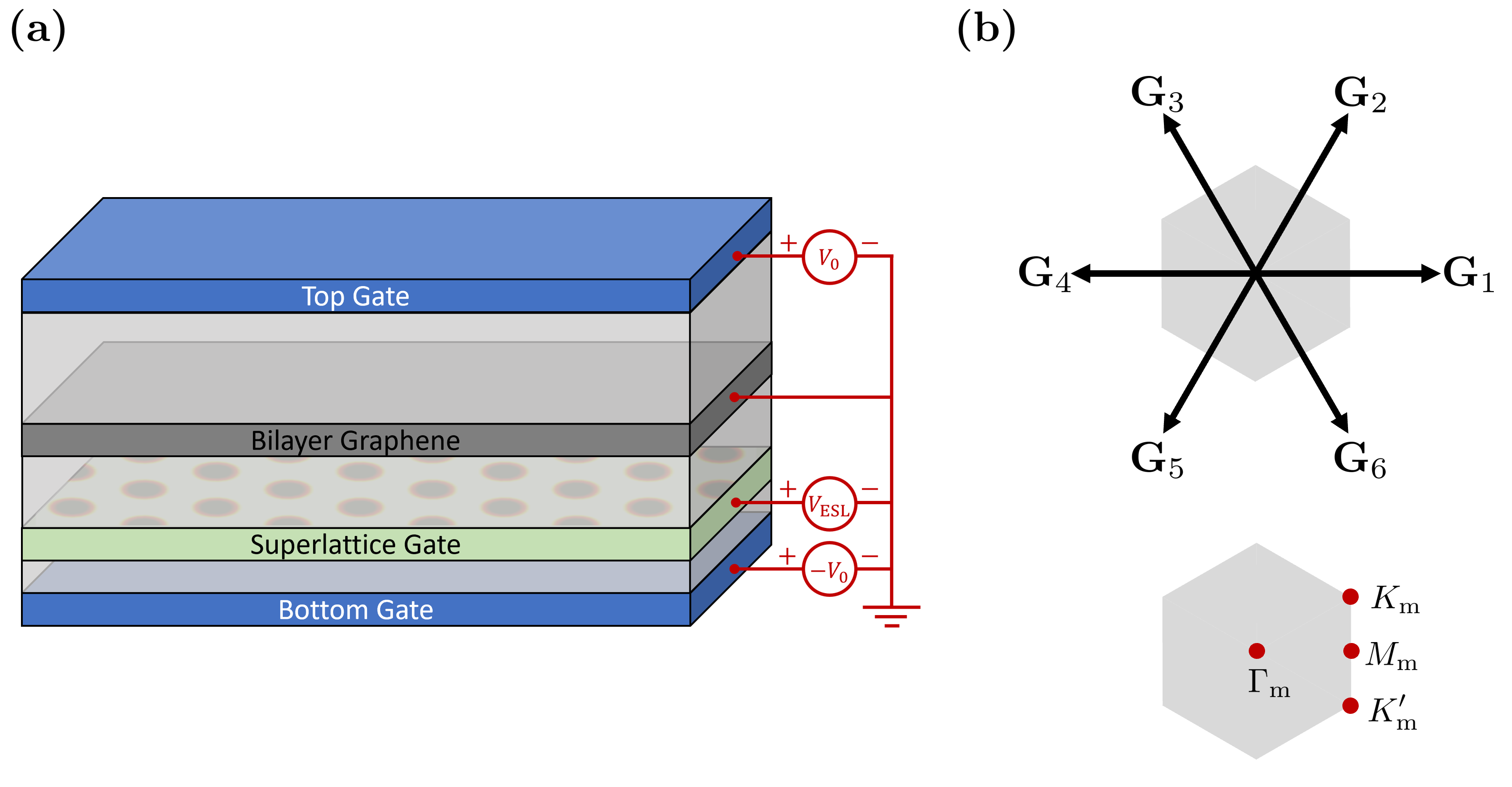}
\caption{(a) Schematic depiction of the experimental setup modeled by the SL-BLG Hamiltonian of Ghorashi \textit{et al}. \cite{ghorashi-prl23}. The system is tuned by a displacement field $V_0$ and the ESL potential $V_{\text{ESL}}$. (b) Scattering vectors of the triangular ESL of Eq.~(\ref{eq:HESL}), as well as the corresponding mBZ and its high-symmetry points.}
\label{fig:fig1}
\end{figure}

In this section we briefly review the model of SL-BLG introduced by Ghorashi \textit{et al}.~in Ref.~\cite{ghorashi-prl23}. The suggested experimental setup is depicted schematically in Fig.~\ref{fig:fig1}(a), in which the BLG sample is subject to top and bottom gates of equal and opposite voltage, as well as to a spatially varying ESL gate. The latter may be realized in practice by a patterned dielectric \cite{forsythe-natnano18}, and has been employed in studies of monolayer graphene \cite{forsythe-natnano18,li-natnano21,barcons-natcomm22}. The BLG exposed to these gates was then modeled by a continuum Hamiltonian
\begin{equation}
    H=H_{\text{BLG}}+H_{V_0}+H_{\text{ESL}}. \label{eq:H_SL}
\end{equation}
Because the spin-orbit coupling is extremely weak in BLG, this Hamiltonian may be considered spin degenerate, and we will thus ignore spin labels. The first term on the right-hand side of Eq.~(\ref{eq:H_SL}) describes the low-energy Hamiltonian of intrinsic BLG near the $K$ and $K'$ valleys of the BLG Brillouin zone (BZ) and is given by the four-band Hamiltonian
\begin{equation}
    H_{\text{BLG}}=\hbar v\tau^0(-i\mu\partial_x\sigma^x-i\partial_y\sigma^y)+\frac{t}{2}(\tau^x\sigma^x-\tau^y\sigma^y).\label{eq:HBLG}
\end{equation} Here $\tau$ and $\sigma$ denote Pauli matrices describing the layer and sublattice spaces, respectively, and $\mu=\pm1$ indicates the $K$ or $K'$ valley.

We take the Fermi velocity to be $v=10^6$ m/s and adopt $t=400$\,meV as the interlayer hopping between the vertically aligned $A$ and $B$ sublattice sites of the two layers. We shall be interested in energies on the order of tens of meV, i.e., $E \ll t$.  In this limit there are two remote high-energy bands near $E=\pm t$, as well as a pair of bands in the low-energy sector that touch quadratically with dispersions $E(\textbf{k})=\pm\gamma|\textbf{k}|^2$ (in the limit of large $t$), where $\gamma=\hbar^2v^2/t$ and $\textbf{k}$ is defined relative to the $K$ or $K'$ point \cite{mccann-prb06,mccann-prl06,neto-rmp09,mccann-ropp13}.

The displacement field term is given by \begin{equation}
    H_{V_0}=V_0\tau^z\sigma^0,\label{eq:HV0}
\end{equation} and applies equal and opposite potential energies on the top and bottom layers. This term opens a gap of size $2|V_0|$ between the low-energy bands, which now disperse as 
\begin{equation}
E(\textbf{k})=\pm\sqrt{ V_0^2+\gamma^2|\textbf{k}|^4 } ,\label{eq:eldisp}  
\end{equation}
and as a result are also slightly flattened near $\textbf{k}=0$
\cite{ohta-sci06,mccann-prb06,mccann-prl06,mccann-ropp13,min-prb07,castro-prl07}. In the presence of a non-zero $V_0$, BLG becomes a valley Chern insulator exhibiting opposite signs of Berry curvature in opposite valleys \cite{martin-prl08,neto-rmp09,zhang-prl11,mccann-ropp13}.

The potential term $H_{\text{ESL}}$ that is primarily featured in the model of SL-BLG is triangular and symmetric under 2D inversion (i.e., $C_{2z}$) and takes the form \begin{equation}
    H_{\text{ESL}}=\frac{V_{\text{ESL}}}{2}[(\tau^0+\tau^z)+\alpha(\tau^0-\tau^z)]\sigma^0\sum^6_{n=1}\exp(i\textbf{G}_n\cdot\textbf{r}), \label{eq:HESL}
\end{equation} where $\textbf{r}=(x,y)$ denotes in-plane position and the $\textbf{G}_n$ are the scattering wave vectors of the ESL potential. Specifically, $\textbf{G}_n=Q(\cos\phi_n,\sin\phi_n)$ where $Q=4\pi/\sqrt{3}L$, $\phi_n=\pi(n-1)/3$, and $L$ is the SL lattice constant. The $\textbf{G}_n$ vectors introduce a ``mini Brillouin zone" (mBZ) as shown in Fig.~\ref{fig:fig1}(b), with its center $\Gamma_\text{m}$ coinciding with the $K$ or $K'$ point of the original BLG BZ. The parameter $\alpha$ is the ratio of SL potentials felt by the top and bottom graphene layers, and takes into account that the top graphene layer is located further away from the SL gate than the bottom layer and experiences a different screening environment. In the rest of this paper, we treat $\alpha$ as a phenomenological parameter and set it to $0.3$ \cite{rokni-scirep17,ghorashi-prl23}, but we note that it may be computed self-consistently \cite{zeng-prb24}. 

The ESL potential opens gaps at the mBZ boundaries, and in doing so may generate isolated valley Chern bands. Varying $V_0$, $V_{\text{ESL}}$, and $L$ then permits control of the band widths as well as their Chern numbers. It is important to note however, that while increasing $L$ will lead to greater band flattening, SLs with large $L$ are also more susceptible to disorder \cite{forsythe-natnano18}. Additionally, greater $L$ values tend to diminish interband spacings, increasing the sensitivity of band topology to small parameter changes. These considerations must be taken into account when constructing the SL in order to generate isolated topological bands.

\section{Bilayer Graphene Exposed to Magnetic Superlattices}
\label{Methods}
We now turn to a discussion of the calculation of the continuum electronic band structure of SL-BLG exposed to out-of-plane magnetic fields $\textbf{B}(\textbf{r})$ that vary on length scales much larger than the atomic spacings in BLG. The fields we consider either introduce no net flux to the SL unit cell, or else a single quantum of flux. For both cases, the minimal coupling prescription introduces the magnetic field $\textbf{B}(\textbf{r})$ to the SL-BLG Hamiltonian via the vector potential $\textbf{A}(\textbf{r})$, where $\textbf{B}(\textbf{r})=\nabla\times\textbf{A}(\textbf{r})$. The vector potential in turn produces an effective periodic scattering potential in both cases. In the presence of a flux quantum, however, it additionally introduces Landau levels (LLs) to the problem. As the magnetic fields we consider are strictly out-of-plane, we will treat them as scalar fields $B(\textbf{r})$. The electromagnetic gauge can then be chosen so that the vector potential is in-plane: $\textbf{A}(\textbf{r})=[A_x(\textbf{r}),A_y(\textbf{r})]$.

The electron spin $g$-factor in BLG is close to $2$ \cite{mccann-prl06,eich-prx18,kurzmann-prl19,lee-prl20,kurzmann-natcomm2021} and the magnetic fields we consider are only a few Tesla in strength. (In the case of unit magnetic flux per supercell, we consider superlattice sizes with area about 10 times that of magic-angle TBG, where the corresponding field would be a few tens of Tesla.) As a result, the spin splitting is negligible and we will continue to neglect the spin degree of freedom in the remainder of the paper.

\subsection{Zero net flux magnetic superlattices}
\label{Methods_zeroflux}
When the magnetic field is spatially periodic and introduces no net flux in the periodic domain (unit cell), it can be expressed as a Fourier series without a zeroth harmonic, \begin{equation}
B(\textbf{r})=\sum_{\textbf{G}\neq0}B(\textbf{G})\exp(i\textbf{G}\cdot\textbf{r}), \label{eq:ZeroFluxB}
\end{equation} where $B(\textbf{G})$ denotes a Fourier coefficient. It is then possible to select an electromagnetic gauge in which the vector potential $\textbf{A}(\textbf{r})$ shares the same periodicity as $B(\textbf{r})$:
\begin{equation}
    \textbf{A}(\textbf{r})=\sum_{\textbf{G}\neq0}\textbf{A}(\textbf{G})\exp(i\textbf{G}\cdot\textbf{r}). \label{eq:ZeroFluxA}
\end{equation} Note that since we are working with in-plane vector potentials, the Fourier coefficients $\textbf{A}(\textbf{G})$ are in-plane as well. As the magnetic field is given by the curl of the vector potential, it is then easy to see that \begin{equation}
    \textbf{A}(\textbf{G})=\frac{B(\textbf{G})\,\hat{\textbf{z}}\times\textbf{G}}{i|\textbf{G}|^2}. \label{eq:ZeroFluxA_G}
\end{equation} 

The magnetic field is incorporated into the Hamiltonian of Eq.~(\ref{eq:H_SL}) via minimal coupling. In particular, the kinetic energy in Eq.~(\ref{eq:HBLG}) is modified to become \begin{equation}
   \hbar v\tau^0\left[\textbf{k}-\frac{e}{\hbar}\textbf{A}(\textbf{r})\right]\cdot\boldsymbol{\sigma}. \label{eq:MinCoupKE}
\end{equation} In the above line, $e<0$ is the electron charge, $\textbf{k}=(k_x,k_y)=(-i\partial_x,-i\partial_y)$, $\boldsymbol{\sigma}=(\sigma^x,\sigma^y)$, and we focus on the $K$ valley ($\mu=1$). Taking advantage of the fact that the kinetic energy is linear in momentum, we may split Eq.~(\ref{eq:MinCoupKE}) into two parts, \begin{equation}
    \hbar v\tau^0\textbf{k}\cdot\boldsymbol{\sigma}-ev\tau^0\textbf{A}(\textbf{r})\cdot\boldsymbol{\sigma},
\end{equation}
resulting in the full SL-BLG Hamiltonian of Eq.~(\ref{eq:H_SL}) acquiring an additional term \begin{equation}
    H_{\text{MSL}}=-ev\tau^0\textbf{A}(\textbf{r})\cdot\boldsymbol{\sigma} .
    \label{eq:HMSL}
\end{equation} In the case of zero net magnetic flux, we may therefore view the periodic magnetic field as generating an effective scattering potential via $\textbf{A}(\textbf{r})$ that is superimposed on the one produced by the ESL. Unlike the potential of Eq.~(\ref{eq:HESL}), however, we note that it now acts to mix the two sublattices.

If we further assume that $H_{\text{MSL}}$ is commensurate with $H_{\text{ESL}}$, then the magnetic Hamiltonian remains periodic, just as in the original nonmagnetic case, and the energy eigenstates take the form of Bloch waves. All Bloch waves are expressible as superpositions of plane waves, and so we diagonalize the Hamiltonian using the latter. Numerically, convergence of the resulting band structure is ensured by selecting a sufficiently high momentum cutoff for the plane waves. The Chern numbers of isolated bands are then computed by tiling the mBZ with plaquettes of sufficiently small size and summing the discretized Berry phases around the boundary of each plaquette \cite{vanderbilt-book2018}.

\subsection{Magnetic superlattices with a quantum of flux}
\label{Methods_oneflux}
When a net magnetic flux permeates the SL unit cell, Eq.~(\ref{eq:ZeroFluxB}) is modified by the addition of a zeroth harmonic term, so that \begin{equation}
B(\textbf{r})=B_{\Phi}+\sum_{\textbf{G}\neq0}B(\textbf{G})\exp(i\textbf{G}\cdot\textbf{r}), \label{eq:FluxB}
\end{equation} where $B_{\Phi}$ is a constant. By superposition, the vector potential corresponding to this magnetic field can then be expressed as \begin{equation}
    \textbf{A}(\textbf{r})=\textbf{A}_{\Phi}(\textbf{r})+\sum_{\textbf{G}\neq0}\textbf{A}(\textbf{G})\exp(i\textbf{G}\cdot\textbf{r}).\label{eq:FluxA}
\end{equation} $\textbf{A}_{\Phi}(\textbf{r})$ is taken to be a linear function of position, as in the Landau or symmetric gauge, in order to generate a uniform $B_{\Phi}$. The second term, which we will denote as $\textbf{A}_{0}(\textbf{r})$, is the same as that of Eq.~(\ref{eq:ZeroFluxA}), with the Fourier coefficients $\textbf{A}(\textbf{G})$ given by the formula of Eq.~(\ref{eq:ZeroFluxA_G}).

Under the minimal coupling scheme, the kinetic energy in Eq.~(\ref{eq:HBLG}) reads \begin{equation}
   \hbar v\tau^0\left[\textbf{k}-\frac{e}{\hbar}\textbf{A}_{\Phi}(\textbf{r})\right]\cdot\boldsymbol{\sigma}+H_{\text{MSL}}, \label{eq:Phi_Ham} 
\end{equation} where $H_{\text{MSL}}=-ev\tau^0\textbf{A}_{0}(\textbf{r})\cdot\boldsymbol{\sigma}$. We thus see that the periodic piece $\textbf{A}_0(\textbf{r})$ of the full vector potential again generates an effective scattering potential, and we will restrict our considerations to MSLs that are commensurate with the ESL. The resulting minimally-coupled Hamiltonian therefore describes the low-energy electrons of BLG as being subject to an effective scattering potential $H_{\text{SL}}=H_{\text{ESL}}+H_{\text{MSL}}$, a displacement field, and a constant out-of-plane magnetic field $B_{\Phi}$. The presence of the latter, however, ultimately results in the Hamiltonian losing its periodicity. 

Nonetheless, under certain conditions it is still possible to generate a band structure for this Hamiltonian; we shall now briefly outline the reasons. For more thorough expositions on the subject of band structures in the presence of orbital magnetic fields, we refer the reader to Refs.~\cite{brown-pr64,zak-pr64a,zak-pr64b,zak-pr64c,zak-pr65,hofstadter-prb76,thouless-prl82,xiao-rmp10,herzog-prl20,ozawa-prb21,wang-prl21,herzog-prb22,wang-prb22,wang-prb23,fang-prb23,singh-natcomm24}. 

Under the action of a translation operator $T_{\textbf{R}}=e^{i\textbf{p}\cdot\textbf{R}/\hbar}$, where $\textbf{R}$ is any nonzero lattice vector, the magnetic Hamiltonian (which now incorporates $B_{\Phi}$ and all scattering terms) is generally not left invariant, since $\textbf{A}_{\Phi}(\textbf{r})\neq\textbf{A}_{\Phi}(\textbf{r}+\textbf{R})$ in general. As $\textbf{A}_{\Phi}(\textbf{r})$ is a linear function of position, however, $\textbf{A}_{\Phi}(\textbf{r}+\textbf{R})$ must generate the same magnetic field $B_{\Phi}$, and so must be related to $\textbf{A}_{\Phi}(\textbf{r})$ via a gauge transformation,\begin{equation}
    \textbf{A}_{\Phi}(\textbf{r}+\textbf{R})=\textbf{A}_{\Phi}(\textbf{r})+\nabla\phi_{\textbf{R}}(\textbf{r}),
\end{equation} with $\nabla\phi_{\textbf{R}}(\textbf{r})=\textbf{A}_{\Phi}(\textbf{r}+\textbf{R})-\textbf{A}_{\Phi}(\textbf{r})$. $\nabla\phi_{\textbf{R}}(\textbf{r})$ must therefore be independent of $\textbf{r}$, resulting in $\phi_{\textbf{R}}(\textbf{r})$ being linear in $\textbf{r}$. By the same token, $\nabla\phi_{\textbf{R}}(\textbf{r})$ is linear in $\textbf{R}$. The specific form of $\phi_{\textbf{R}}(\textbf{r})$ is determined by the choice of gauge for $\textbf{A}_{\Phi}(\textbf{r})$.

The magnetic Hamiltonian is then found to be invariant under the action of the magnetic translation operator \begin{equation}
    \widetilde{T}_{\textbf{R}}=e^{i\phi_{\textbf{R}}(\textbf{r})}T_{\textbf{R}}.
\end{equation} Letting $\widetilde{T}_1$ and $\widetilde{T}_2$ denote the magnetic translation operators corresponding to translations by the unit cell vectors $\textbf{a}_1$ and $\textbf{a}_2$, respectively, it can be shown that generally these operators do not commute, with \begin{equation}
\widetilde{T}_1\widetilde{T}_2=e^{2\pi i\Phi/\Phi_0}\widetilde{T}_2\widetilde{T}_1,\end{equation} where  $\Phi=B_{\Phi}|\textbf{a}_1\times\textbf{a}_2|$ is the flux through the nonmagnetic unit cell and $\Phi_0=h/|e|$ is the flux quantum. When an integer number of flux quanta pierce the unit cell, the magnetic translation operators do commute, and it is possible to generate a band structure. If instead $\Phi/\Phi_0=p/q$ where $p$ and $q$ are coprime, then the unit cell must be enlarged $q$ times along one of the lattice vectors in order to generate a new magnetic cell with $p$ flux quanta. Each band of the $\textbf{A}_{\Phi}(\textbf{r})=0$ Hamiltonian is then split into $q$ subbands upon inclusion of $B_{\Phi}$. Finally, when $\Phi/\Phi_0$ is an irrational number, no band structure may be obtained and other methods must be used to solve the Hamiltonian \cite{hofstadter-prb76}.

Although it is tempting to diagonalize the Hamiltonian in the basis of plane waves, as was done in the case of zero magnetic flux, this can no longer be done when an integer number of flux quanta penetrate the magnetic unit cell. The eigenstates of the magnetic translation operators obey \begin{equation}
\widetilde{T}_{\textbf{R}}\psi_{\textbf{k}}(\textbf{r})=e^{i\textbf{k}\cdot\textbf{R}}\psi_{\textbf{k}}(\textbf{r}),    
\end{equation} and can always be written in a Bloch-like form as \begin{equation}
 \psi_{\textbf{k}}(\textbf{r})=e^{i\textbf{k}\cdot\textbf{r}}u_{\textbf{k}}(\textbf{r}).   
\end{equation} However, in light of $\psi_{\textbf{k}}(\textbf{r})$ being an eigenstate of $\widetilde{T}_{\textbf{R}}$, rather than $T_{\textbf{R}}$, $u_{\textbf{k}}(\textbf{r})$ must obey twisted boundary conditions: \begin{equation}
    u_{\textbf{k}}(\textbf{r}+\textbf{R})=e^{-i\phi_{\textbf{R}}(\textbf{r})}u_{\textbf{k}}(\textbf{r}).
\end{equation} Due to this fact, the magnetic translation eigenstates cannot be expanded using a plane wave basis. 

In the absence of any periodic scattering potential and when $\textbf{A}_{\Phi}(\textbf{r})$ solely generates the magnetic field, the energy eigenstates belong to degenerate LLs. A band structure representation of the energy levels may be obtained by selecting a unit cell with $\Phi=\Phi_0$ and realizing that specific linear combinations of energy eigenstates within a single LL produce energy eigenstates in Bloch form. When a periodic potential is reintroduced to the problem, resulting in a magnetic cell with $p$ flux quanta, each original LL is split into $p$ magnetic subbands. Although states from a single LL no longer form Bloch energy eigenstates, they can be used to construct a basis of magnetic translation eigenstates $|\textbf{k},n\rangle$, where $\textbf{k}$ denotes the wavevector and $n$ indexes the LLs, in which the magnetic Hamiltonian can be diagonalized \cite{brown-pr64,zak-pr64c,zak-pr65}. Numerical convergence of the resulting band structure is ensured by a sufficiently high LL cutoff. 

The effective kinetic energy in Eq.~(\ref{eq:Phi_Ham}) (first term) can be rewritten in terms of $\boldsymbol{\pi}=\textbf{p}-e\textbf{A}_{\Phi}(\textbf{r})$, where the components of $\boldsymbol{\pi}$ obey $[\pi_x,\pi_y]=i\hbar eB_{\Phi}$. The introduction of the ladder operators \begin{equation}
     a=\frac{\pi_x+i\pi_y}{\sqrt{2\hbar eB_{\Phi}}}, \text{ }a^\dagger=\frac{\pi_x-i\pi_y}{\sqrt{2\hbar eB_{\Phi}}},\text{ }[a,a^\dagger]=1,
\end{equation} then allows for straightforward, gauge-independent evaluation of the matrix elements of the kinetic energy in the $|\textbf{k},n\rangle$ basis; the matrix elements solely depend on the LL $n$, and not $\textbf{k}$. The same cannot be said of the matrix elements of the scattering potential, which further depend on the choice of gauge for the magnetic vector potential.

Reference~\cite{herzog-prb22} (see also Supplementary Notes 11 and 12 of Ref.~\cite{singh-natcomm24} for further discussion) recently demonstrated that the issue of the gauge dependence of the scattering potential matrix elements can be avoided. By expressing the magnetic translation operators via gauge-invariant LL guiding center momenta, and by generating the magnetic translation eigenstates $|\textbf{k},n\rangle$ using these operators, the authors derived expressions for the scattering matrix elements that are gauge invariant.

Following the results of this work, for a given scattering potential expressible as a Fourier series with reciprocal lattice vectors $\textbf{G}$, we compute a matrix element $\langle\textbf{k},m|e^{i\textbf{G}\cdot\textbf{r}}|\textbf{k},n\rangle$ as  
\begin{equation} e^{-i\pi (G_1G_2+G_1+G_2)-2\pi i(G_1k_2-G_2k_1)}\mathcal{H}^{\textbf{G}}_{mn}.\label{eq:ScatAmpMain}
\end{equation} 
$G_1$ and $G_2$ are the internal coordinates of $\textbf{G}$, while $k_1$ and $k_2$ are those of $\textbf{k}$. $\mathcal{H}^{\textbf{G}}_{mn}$ denotes the purely LL-dependent part of the scattering amplitude, and is expressed via associated Laguerre polynomials. 

We note that the above scattering amplitude is a slight modification of the one explicitly presented in Ref.~\cite{herzog-prb22}, implementing a suggestion of J.~Herzog-Arbeitman \cite{herzog-privcomm23}. The modification follows the observation that the original formalism is found to feature an implicit selection of the BZ origin $\Gamma$. However, the selected origin may not necessarily be the $\textbf{k}$-point of maximal symmetry (i.e., the little group of $\Gamma$ may not equal the point group of the system), and the point of highest symmetry may instead be located elsewhere in the BZ. The scattering amplitude as written in Eq.~(\ref{eq:ScatAmpMain}) circumvents this issue to ensure that $\Gamma$ is the point of highest symmetry. For further details we refer the reader to the Appendix.

The Chern numbers of isolated bands can no longer be computed using the simple plaquette approach outlined previously. A Bloch energy eigenstate at wavevector $\textbf{k}$ can be expressed as a linear combination
$|\psi_{m\textbf{k}}\rangle=\sum_n \xi_{m\textbf{k}}(n)\,|\textbf{k},n\rangle $ of the $|\textbf{k},n\rangle$ basis states, which themselves represent topologically nontrivial LLs. The Berry curvature at $\textbf{k}$ therefore arises from two contributions: the Berry curvature of the underlying LLs, and the Berry curvature of the coefficient vectors $\xi_{m\textbf{k}}$ that give the change in character of the Bloch state in $\textbf{k}$-space \cite{wang-prb06,wang-prb07}. The former contribution is simple, as the Berry curvature of any LL is uniform, and yields a Chern number $C=-1$ \cite{ozawa-prb21,wang-prl21,herzog-prb22}, while the latter contribution can be computed using the plaquette method \cite{wang-prb06,wang-prb07}. Thus the Chern number of a band is given by the sum of the LL and character Chern numbers. We employ this method in our calculations.

We note that an alternative approach to computing the Chern numbers involves obtaining the winding numbers of the eigenvalues of appropriately constructed Wilson loops, as shown by Ref.~\cite{herzog-prb22}. We confirmed that the Chern numbers computed in this way match those found using the plaquette-based method outlined above.

\section{Triangular magnetic superlattice}
\label{Results_Triangle}
In this section we apply the methods described in Sec.~\ref{Methods} to obtain a qualitative understanding of the effects of MSLs on the band structure of SL-BLG. To this end, we focus on the relatively simple case of a triangular MSL generated by \begin{equation}
B(\textbf{r})=\sum^6_{n=1}B_0\cos(\textbf{G}_n\cdot\textbf{r}), \label{eq:TriField}
\end{equation} where the reciprocal vectors $\textbf{G}_n$ are the same as those depicted in Fig.~\ref{fig:fig1}(b), and where the field is commensurate with the ESL potential of Eq.~(\ref{eq:HESL}). We consider both the zero and single flux quantum cases, where the latter is obtained by adding the constant $B_{\Phi}=(2\Phi_0/\sqrt{3}L^2)$ to $B(\textbf{r})$. For each flux case, we first study the effect of the magnetic field acting by itself without the ESL, and subsequently study its action in conjunction with the ESL. We find that MSLs acting alone are capable of generating topological flat bands with $|C|=1$ and $2$, but the inclusion of the ESL can further enrich the topology of the band structures by generating an increased number of not only flat, but also generic (non-flat) bands with even higher Chern numbers. In the band structures to be shown, topological bands are highlighted in red. However, we do not highlight those that feature gaps to neighboring bands that are smaller than $0.5$\,meV, as the topology of these bands is particularly sensitive to changes in model parameters.

We will indicate the topological flat bands in tables corresponding to the band structures. The featured data are Chern numbers $C$, band widths $W$, energies at the BZ origin $E_{\Gamma_m}$, and energy gaps to the bands immediately below and above, $E_{g-}$ and $E_{g+}$, respectively. A listed gap value is taken to be the smaller of the indirect or direct gaps; however, in the case when an indirect gap is less than $0.5$\,meV, we list the smallest direct gap and indicate this fact with a ``(D)" next to the gap value. We consider a band to be flat when $W<2$\,meV.

\begin{figure}[t]
\centering
\includegraphics[width= \columnwidth]{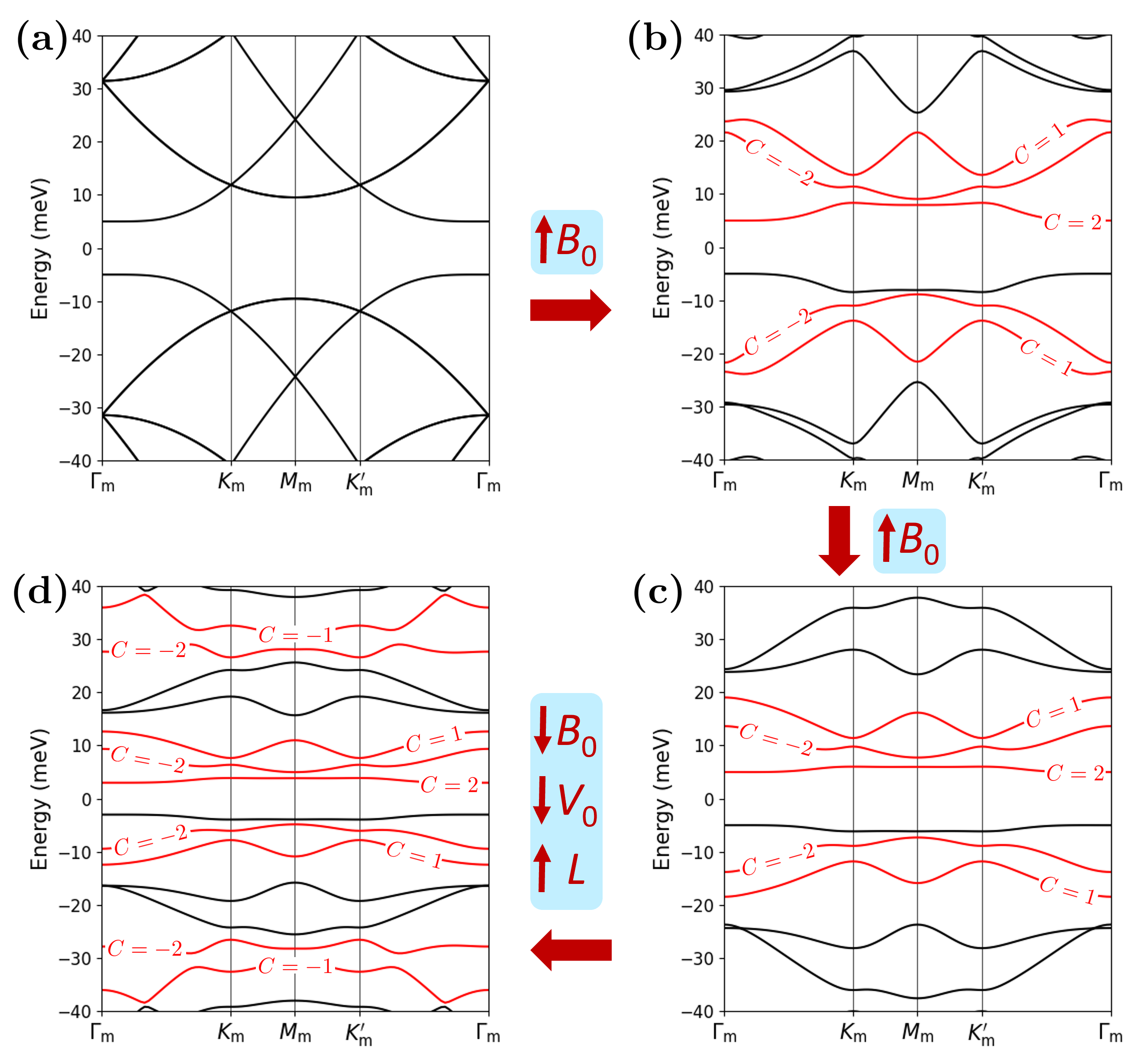}
\caption{Evolution of low-energy bands in BLG induced by a triangular MSL, here with zero net flux ($B_\Phi=0$) and in the absence of the triangular ESL ($V_\textrm{ESL}=0$). The high-symmetry points are defined in Fig.~\ref{fig:fig1}(b). (a) Folded bands characterize an empty-lattice starting configuration ($B_0=0$\,T, $V_0=5$\,meV, $L=40$\,nm). (b) As $B_0$ is tuned to $2.5$\,T, topological bands emerge. (c) Further increasing $B_0$ to $5$\,T flattens the bands and produces a flat $C=2$ band close to the Fermi energy. (d) Decreasing the magnetic field and gate strength while increasing the SL lattice constant ($B_0=3$\,T, $V_0=3$\,meV, $L=50$\,nm) introduces more topological bands and further flattens preexisting bands.}
\label{fig:fig2}
\end{figure}

We note that periodic zero-flux fields, including the one we will employ, may in practice be generated by the stray fields of periodic arrays of ferromagnetic nanorods, and have been utilized and studied in the context of spintronic applications \cite{nielsch-apl01,nielsch-jmmm02,bruno-prl04,taillefumier-prb08}.
\subsection{Zero net flux}
\label{Results_noflux}

\begin{figure*}[t]
\centering
\includegraphics[width= 17.9 cm]{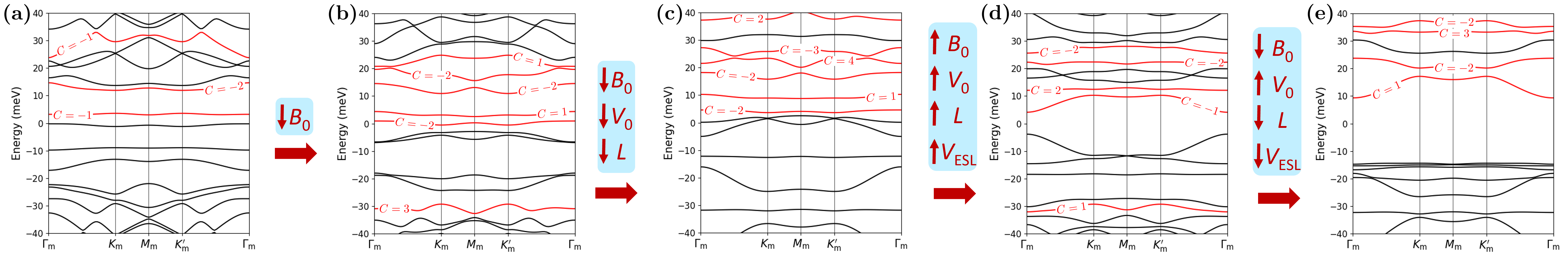}
\caption{Band structures of BLG plotted in the presence of both the triangular ESL and zero-net-flux triangular MSL. (a)~$B_0=0$ T, $V_{\text{ESL}}=10$\,meV, $V_0=-5$\,meV, $L=50$\,nm. (b)~$B_0=-2$\,T, $V_{\text{ESL}}=10$\,meV, $V_0=-5$\,meV, $L=50$\,nm. (c)~$B_0=-5$\,T, $V_{\text{ESL}}=10$\,meV, $V_0=-10$\,meV, $L=44$\,nm. (d)~$B_0=3$\,T, $V_{\text{ESL}}=15$\,meV, $V_0=-5$\,meV, $L=50$\,nm. (e)~$B_0=-5$\,T, $V_{\text{ESL}}=10$\,meV, $V_0=10$\,meV, $L=45$\,nm.}
\label{fig:fig3}
\end{figure*}

Throughout this work, we present results for the $K$ valley. Corresponding results for the $K'$ valley can be obtained with an appropriate modification of the parameters. As a reminder, we work in the low-energy regime relative to the scale of interlayer hopping $t$, so that all of the bands shown emerge from the two low-energy bands of the continuum model of Eq.~(\ref{eq:eldisp}).

Fig.~\ref{fig:fig2} depicts the evolution of the band structure as $B_0$, $V_0$, and $L$ are varied, while the ESL is absent. Starting with the MSL absent as well ($B_0=0$\,T, $V_0=5$\,meV, $L=40$\,nm), turning the MSL on (increasing $B_0$) opens gaps at the high-symmetry points of the band structure. Isolated Chern bands appear as $B_0$ is first increased to $2.5$\,T, and are subsequently flattened as $B_0$ is further increased to $5$\,T. A flat Chern band with $C=2$ and a band width $W$ on the order of 1\,meV then appears close to the Fermi energy, as detailed in Table~\ref{tab:NoFlux_NoESL}. The same band may be further flattened by increasing $L$ to $50$\,nm, while simultaneously decreasing $B_0$ and $V_0$ to $3$\,T and $3$\,meV. The increase in lattice constant also introduces more Chern bands into the energy window, though the bands are not necessarily flat. We therefore see that the MSL may introduce flat topological bands, including those with $|C|>1$, without the ESL. Increasing the field strength reduces band widths, as does increasing the SL size; the latter, however, also increases the density of states and reduces the gaps between isolated bands, potentially increasing the sensitivity of the band topology to changes in model parameters.

\begin{table}[t]
\caption{\label{tab:NoFlux_NoESL}
Details of flat bands appearing in Fig.~\ref{fig:fig2}. Energy at $\Gamma_m$, 
Chern number, band width, gap below, and gap above,
respectively (energies in meV).}
\begin{ruledtabular}
\begin{tabular}{lcrccc}
& $E_{\Gamma_m}$ & $C$ & $W$ & $E_{g-}$ & $E_{g+}$ \\
\colrule
Panel (c) 
& 5.00 & 2 & 1.02 & 9.99 & 1.75 \\
Panel (d)
& 3.00 & $2$ & 0.88 & 6.00 & 1.14 \\
\end{tabular}
\end{ruledtabular}
\end{table}

Next we reintroduce $H_{\text{ESL}}$ to the Hamiltonian. Beginning with the magnetic field switched off in Fig.~\ref{fig:fig3}(a) (see figure caption for details of the parameters), a $C=-1$ flat band is observed near the Fermi energy, as well as a generic $C=-2$ band \cite{ghorashi-prl23}. When the magnetic field is turned on ($B_0=-2$\,T) in Fig.~\ref{fig:fig3}(b), the number of topological bands belonging to the same energy window increases, with $C=-2$ and $C=1$ flat bands being found particularly close to the Fermi energy. A band with $C=3$ can also be observed at lower energies. Further strengthening $B_0$ while modifying the top and bottom gate voltages and lattice constant, shown in Fig.~\ref{fig:fig3}(c),  further flattens the $C=-2$ and $C=1$ bands near the Fermi energy, while also introducing generic bands with high Chern numbers $C=4$ and $C=-3$. As the magnetic field is diminished in strength and changed in sign while $V_{\text{ESL}}$, $V_0$, and $L$ are increased in Fig.~\ref{fig:fig3}(d), two flat bands with $C=2$ and $-2$ are found at higher energies. Finally, in Fig.~\ref{fig:fig3}(e), another parameter set is found that generates a flat band with $C=3$ at higher energies, along with other topological bands. Details of the flat bands may be found in Table~\ref{tab:NoFlux_ESL}.

The detailed nature of these results should not obscure our main point, which is that the addition of the magnetic degree of freedom associated with the strength $B_0$ of the MSL gives additional opportunities for tuning the system, raising the likelihood of achieving topological flat bands close to the Fermi energy.

\subsection{Single flux quantum}
\label{Results_oneflux}

We now consider the case that the background magnetic field $B_\Phi$ of Eq.~(\ref{eq:FluxB}) is present, and has been chosen such that the unit cell contains a single flux quantum.

\begin{table}[b]
\caption{\label{tab:NoFlux_ESL}
Details of flat bands appearing in Fig.~\ref{fig:fig3}.
Energy at $\Gamma_m$, Chern number, band width, gap below, and gap above,
respectively (energies in meV).}
\begin{ruledtabular}
\begin{tabular}{ldrccc}
& \multicolumn{1}{r}{$E_{\Gamma_m}$} & $C$ & $W$ & $E_{g-}$ & $E_{g+}$ \\
\colrule
Panel (a) 
& 3.20 & $-1$ & 0.55 & 3.26& 8.34\\
Panel (b) 
& 0.94 & $-2$ & 1.46 & 2.37 & 1.62 \\
&4.38 & $1$ & 1.80 & 1.62 & 6.57 \\
Panel (c)
& 4.62 & $-2$ & 0.96 & 1.15 & 4.17 \\
& 10.28 & $1$ & 1.48 & 4.17 & 5.55 \\
Panel (d)
& 12.09 & $2$ & 0.90 & 1.82 & 2.00 \\
& 22.29 & $-2$ & 0.92 & 1.52 & 3.23 \\
Panel (e)
& 33.51 & $-3$ & 0.87 & 2.44 & 1.33 \\
\end{tabular}
\end{ruledtabular}
\end{table}

\begin{figure}[b]
\centering
\includegraphics[width= \columnwidth]{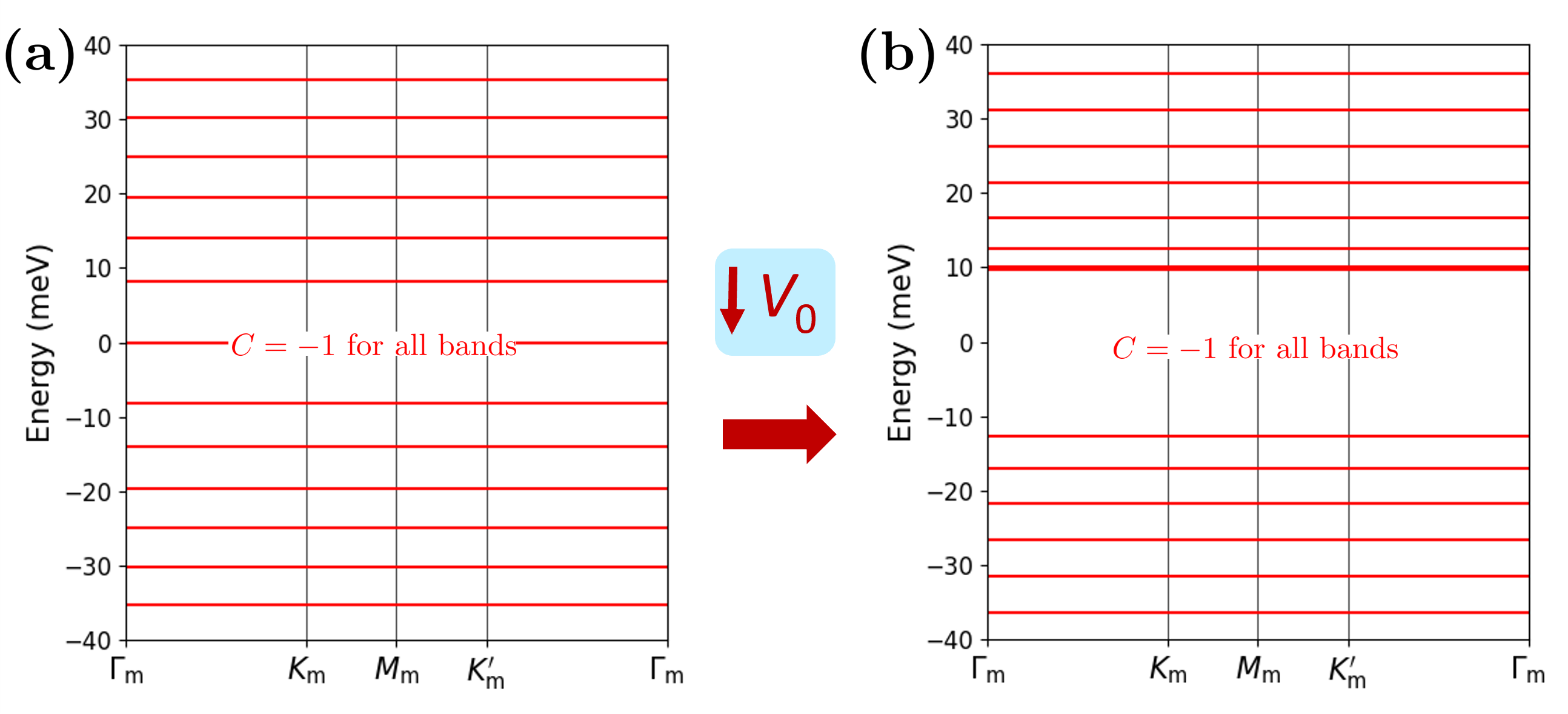}
\caption{Band structures of BLG plotted in the presence of the field $B_{\Phi}$ and without any scattering potentials. All bands are perfectly flat with $C=-1$, and directly correspond to LLs. (a) Band structure for $V_0=0$. The spectrum is particle-hole symmetric and the bands at $E=0$ are doubly degenerate. (b) Band structure for $V_0=-10$. Particle-hole symmetry is broken, and the formerly doubly degenerate bands at $E=0$ acquire a splitting.}
\label{fig:fig4}
\end{figure}

We first plot the magnetic band structures of low-energy continuum BLG in the absence of any periodic scattering potentials; these are just the ordinary LLs of low-energy BLG. Figure~\ref{fig:fig4}(a) depicts the band structure in the empty lattice approximation for $L=50$ nm and $V_0=0$\,meV. All of the bands are perfectly flat and feature Chern numbers of $-1$ \cite{thouless-prl82,ozawa-prb21}. The bands at $E=0$ are doubly degenerate; as we are considering the states of BLG originating from the $K$ valley, these LL states are predominantly formed from orbitals on the $A$ sublattice of the bottom layer. For states in the $K'$ valley, the $E=0$ states are instead formed from orbitals at the $B$ sublattice sites of the top layer \cite{mccann-prl06}. In Fig.~\ref{fig:fig4}(b) the displacement field is turned on with $V_0=-10$\,meV. All of the bands retain their original Chern numbers and remain flat, but are shifted in energy, resulting in broken particle-hole symmetry in the valley. The previously doubly degenerate bands at $E=0$ are also shifted upwards and feature a small energy splitting \cite{mccann-prl06}. In the opposite $K'$ valley, the energy shifts occur in the opposite direction.

\begin{figure*}[t]
\centering
\includegraphics[width= 13cm]{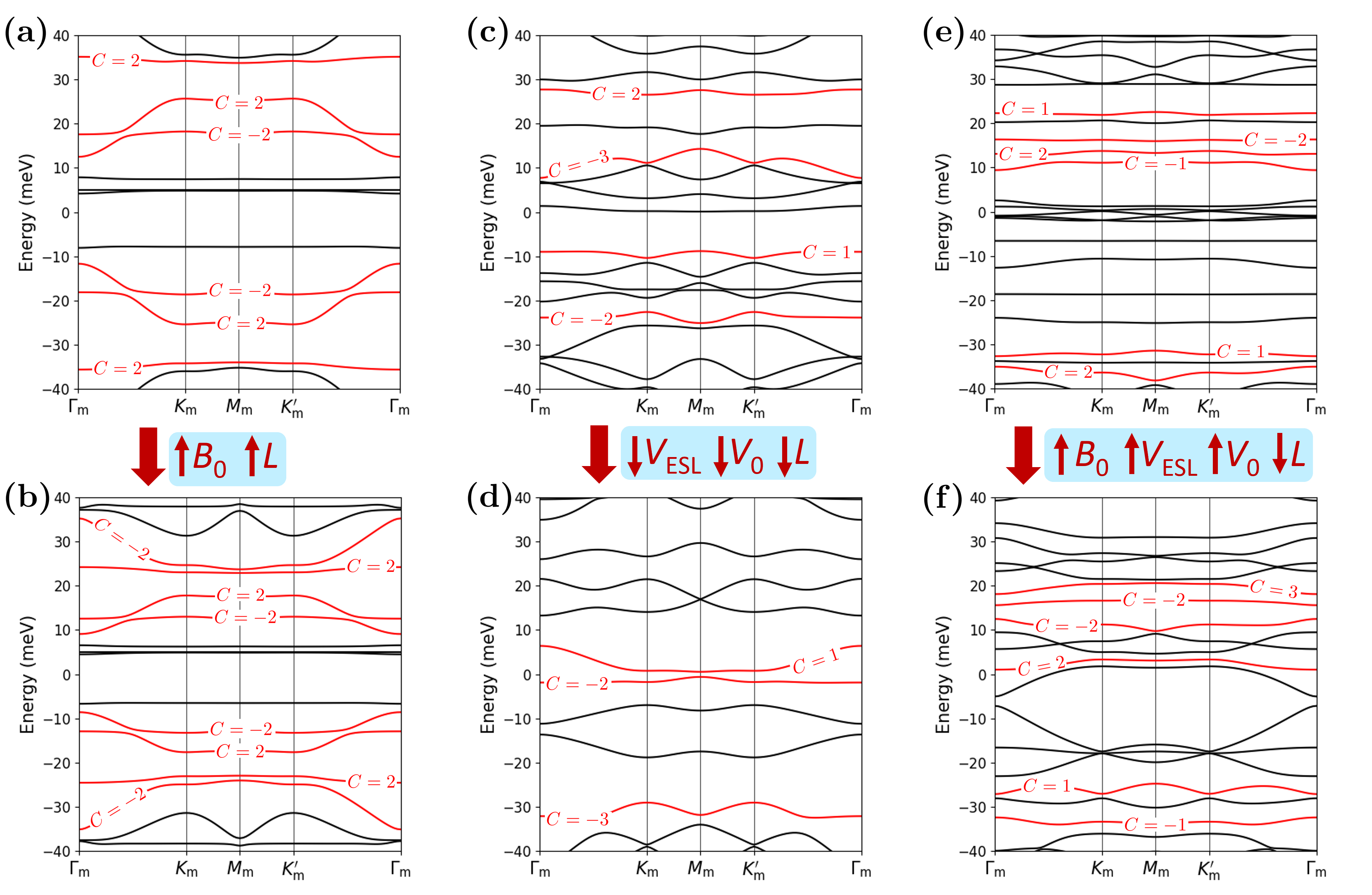}
\caption{Band structures of BLG plotted in the presence of the $B_{\Phi}$ background field and one or both of the magnetic or electric triangular SLs. The band structures in (a,b) do not feature the ESL ($V_{\text{ESL}}=0$), those in (c,d) do not feature the MSL ($B_0=0$), while (e,f) incorporate both SLs. The specific parameters employed are: (a)~$B_0=-5$\,T, $V_{\text{ESL}}=0$\,meV, $V_0=-5$\,meV, $L=40$\,nm. (b)~$B_0=-3$\,T, $V_{\text{ESL}}=0$\,meV, $V_0=-5$\,meV, $L=50$\,nm. (c)~$B_0=0$\,T, $V_{\text{ESL}}=20$\,meV, $V_0=10$\,meV, $L=50$\,nm. (d)~$B_0=0$\,T, $V_{\text{ESL}}=15$\,meV, $V_0=-3$\,meV, $L=40$\,nm. (e)~$B_0=-3$\,T, $V_{\text{ESL}}=10$\,meV, $V_0=-5$\,meV, $L=60$\,nm. (f)~$B_0=5$\,T, $V_{\text{ESL}}=20$\,meV, $V_0=10$\,meV, $L=51$\,nm.}
\label{fig:fig5}
\end{figure*}

Next we include the effective scattering potentials stemming from the ESL and the inhomogeneous component of the total magnetic field and depict the resulting band structures in Fig.~\ref{fig:fig5} (see figure caption for details of the parameters). Further information on the identified flat bands appears in Table~\ref{tab:Flux_ESL}. We omit flat bands with $|C|=1$ from the tables, as perfectly flat bands with those Chern numbers can be obtained without employing MSLs. The first column in Fig.~\ref{fig:fig5} depicts band structures with $V_{\text{ESL}}=0$, the middle column those with $B_0=0$, while the last column combines both the ESL and MSL. Starting with the parameters $B_0=-5$ T, $V_0=-5$\,meV, and $L=40$\,nm, the band structure in Fig.~\ref{fig:fig5}(a) exhibits flat bands with $C=2$ near the top and bottom of the energy window. Other bands with $|C|=2$ are present as well. As $B_0$ is weakened to $-3$\,T and the lattice constant is increased to $50$\,nm [Fig.~\ref{fig:fig5}(b)], all band widths are decreased.

\begin{table}[!b]
\caption{\label{tab:Flux_ESL}
Details of flat bands appearing in Fig.~\ref{fig:fig5}.
Energy at $\Gamma_m$, Chern number, band width, gap below, and gap above, respectively (energies in meV). Flat bands with $|C|=1$ are omitted.}
\begin{ruledtabular}
\begin{tabular}{ldrc@{\hskip -1em}c@{\hskip -2em}c}
& \multicolumn{1}{c}{$E_{\Gamma_m}$} & $C$ & $W$ & $E_{g-}$ & $E_{g+}$ \\
\colrule
Panel (a) 
& -35.59 & $2$ & 1.60 & \phd 1.14 (D) & 8.62\\
& 35.12 & $2$ & 1.38 & 8.09& \phd 1.22 (D)\\
Panel (b) 
& 24.22 & $2$ & 1.31 & 5.12& \phd 0.54 (D)\\
Panel (c)
& 27.72 & $2$ & 1.17 & 6.86 & 1.96 \\
Panel (d)
& -1.87 & $-2$ & 1.28 & 5.06 & 1.23 \\
Panel (e)
& 13.10 & $2$ & 0.85 & 1.36 & 2.24 \\
& 16.36 & $-2$ & 0.37 & 2.24 & 3.70 \\
Panel (f)
& 15.61 & $-2$ & 1.19 & 3.18 & 1.31 \\
\end{tabular}
\end{ruledtabular}
\end{table}

Now focusing on the $B_0=0$ regime, Fig.~\ref{fig:fig5}(c) reveals the appearance of topological flat bands with $C=2$, as well as generic topological bands with Chern numbers $-2$ and $-3$. As the ESL potential and the displacement field are weakened while the SL constant is lowered in going to Fig.~\ref{fig:fig5}(d), a single flat topological band with $C=-2$ appears close to the Fermi energy, while a $C=-3$ band is present at lower energies

When the MSL and ESL are combined in Fig.~\ref{fig:fig5}(e), we observe a flat $C=2$ band, as well as a single particularly flat $C=-2$ band with a bandwidth on the order of $0.4$\,meV. Subsequently in Fig.~\ref{fig:fig5}(f), $B_0$, $V_{\text{ESL}}$, and $V_0$ are increased while $L$ is decreased, resulting in an observed $C=-2$ flat band, as indicated in Table~\ref{tab:Flux_ESL}.

We again emphasize that the main point of these results is to demonstrate that the size $L$ of the MSL, in addition to its strength $B_0$, provide additional degrees of freedom for tuning the system to obtain topological flat bands close to the Fermi energy.

\section{Flux quantum magnetic superlattices and concomitant electric superlattices}
\label{Results_SC}
\begin{figure}[t]
\centering
\includegraphics[width= \columnwidth]{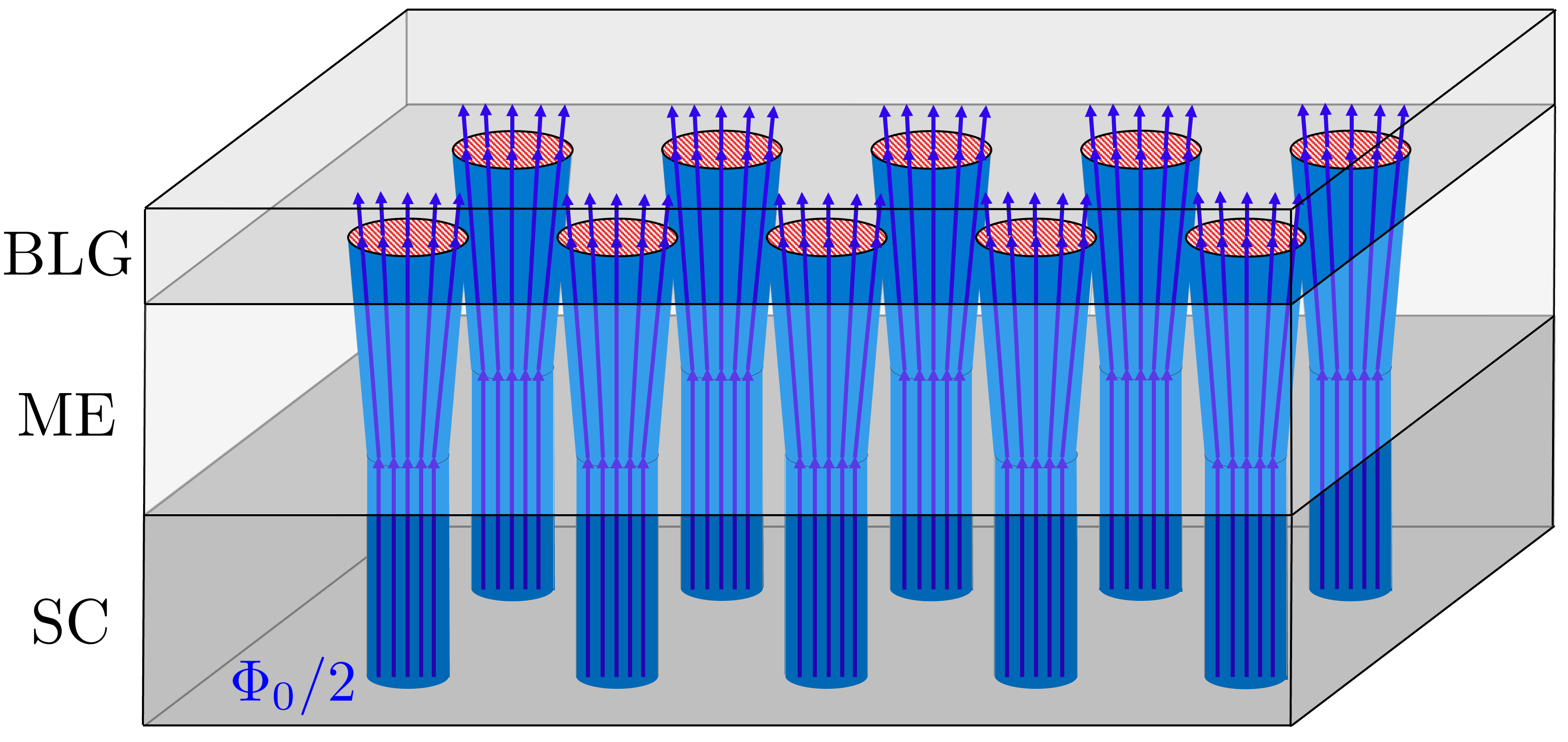}
\caption{Schematic depiction of the setup proposed to generate flux quantum MSLs with concomitant ESLs. The MSLs are generated by the magnetic fields (blue lines) emanating from the vortices (blue cylinders) of a mixed state type II superconductor (SC), with each vortex introducing half of a flux quantum $\Phi_0/2$. Prior to penetrating the BLG sample, the magnetic fields pass through a magnetoelectric (ME) material characterized by a ME coupling tensor $\boldsymbol{\alpha}$, inducing a polarization in the bulk and a periodic charge density on the surface. The charge density (red areas) on the surface closest to the BLG generates the ESL potential.}
\label{fig:fig6}
\end{figure}

We presently propose a method of generating single-flux-quantum MSLs and concomitant ESLs, for which we envision an experimental setup as illustrated in Fig.~\ref{fig:fig6}. The MSL is generated by the magnetic fluxes emanating from the vortices of a type II superconductor in the mixed state. Since we have adopted the notation $\Phi_0=h/|e|$ for the fundamental flux quantum, each superconducting vortex contributes a flux amount of $h/2|e|=\Phi_0/2$. Thus, a band structure may be generated when the SL unit cell contains an even number of vortices. For simplicity, we model the in-plane profiles of the magnetic fields stemming from individual vortices by isotropic Gaussians. Letting $\rho$ be the width of each Gaussian, the total magnetic field piercing the 2D MSL is then given by

\begin{equation}
    B(\textbf{r})=\frac{\Phi_0}{2}\sum_{\textbf{R},\boldsymbol{\tau}}\frac{\exp[-|\textbf{r}-\textbf{R}-\boldsymbol{\tau}|^2/(2\rho^2)]}{2\pi\rho^2},
\end{equation}
where $\textbf{R}$ denotes the origins of unit cells and $\boldsymbol{\tau}$ indicates the basis (locations of vortex centers relative to the origin of a unit cell). In the limit of large $\rho$, $B(\textbf{r})$ approaches the homogeneous value of $\Phi_0/A$, where $A$ is the cell area, and we recover the ordinary LL band structure.

In the scenario envisioned in Fig.~\ref{fig:fig6}, the concomitant ESL stems from the periodic surface charge density $\sigma(\textbf{r})$ induced by placing a magnetoelectric material above the superconductor. The vortex magnetic fields penetrating the magnetoelectric induce columns of electric polarization, and the latter in turn produce the surface charge density. The surface charge density is then equal to the $z$-component of the polarization,
\begin{equation}
    \sigma_\textrm{surf}(\textbf{r})=P_z(\textbf{r})=\frac{\alpha_{zz}}{\mu_0}B(\textbf{r}),\label{eq:ME_charge}
\end{equation} where $\alpha_{zz}$ is the relevant component of the magnetoelectric coupling tensor. The resulting potential is then obtained from the 2D integral \begin{equation}
    H_{\text{ESL}}=\frac{[(\tau^0+\tau^z)+\alpha(\tau^0-\tau^z)]\sigma^0}{2}\int\frac{e\sigma_\textrm{surf}(\textbf{r}')d^2r'}{4\pi\epsilon_0|\textbf{r}-\textbf{r}'|}.
\end{equation}
Employing the convolution theorem and Eq.~(\ref{eq:ME_charge}), the ESL potential can then be expressed as \begin{equation}
    H_{\text{ESL}}=[(\tau^0+\tau^z)+\alpha(\tau^0-\tau^z)]\sigma^0\sum_{\textbf{G}\neq0}\frac{e\alpha_{zz}c^2}{4|\textbf{G}|}B(\textbf{G})e^{i\textbf{G}\cdot\textbf{r}},
\end{equation} where $c$ is the speed of light
and
\begin{equation}
    B(\textbf{G})=\frac{\Phi_0}{2A}\Big[\sum_{\boldsymbol{\tau}}e^{-i\textbf{G}\cdot\boldsymbol{\tau}} \Big] e^{-|\textbf{G}|^2\rho^2/2}.
\end{equation}
The $\textbf{G}=0$ term is omitted from $H_{\text{ESL}}$ since the magnetoelectric material is net neutral, with the bottom and top surface charges canceling at large distances. The expression in the brackets corresponds to the structure factor.

\begin{figure}[t]
\centering
\includegraphics[width= \columnwidth]{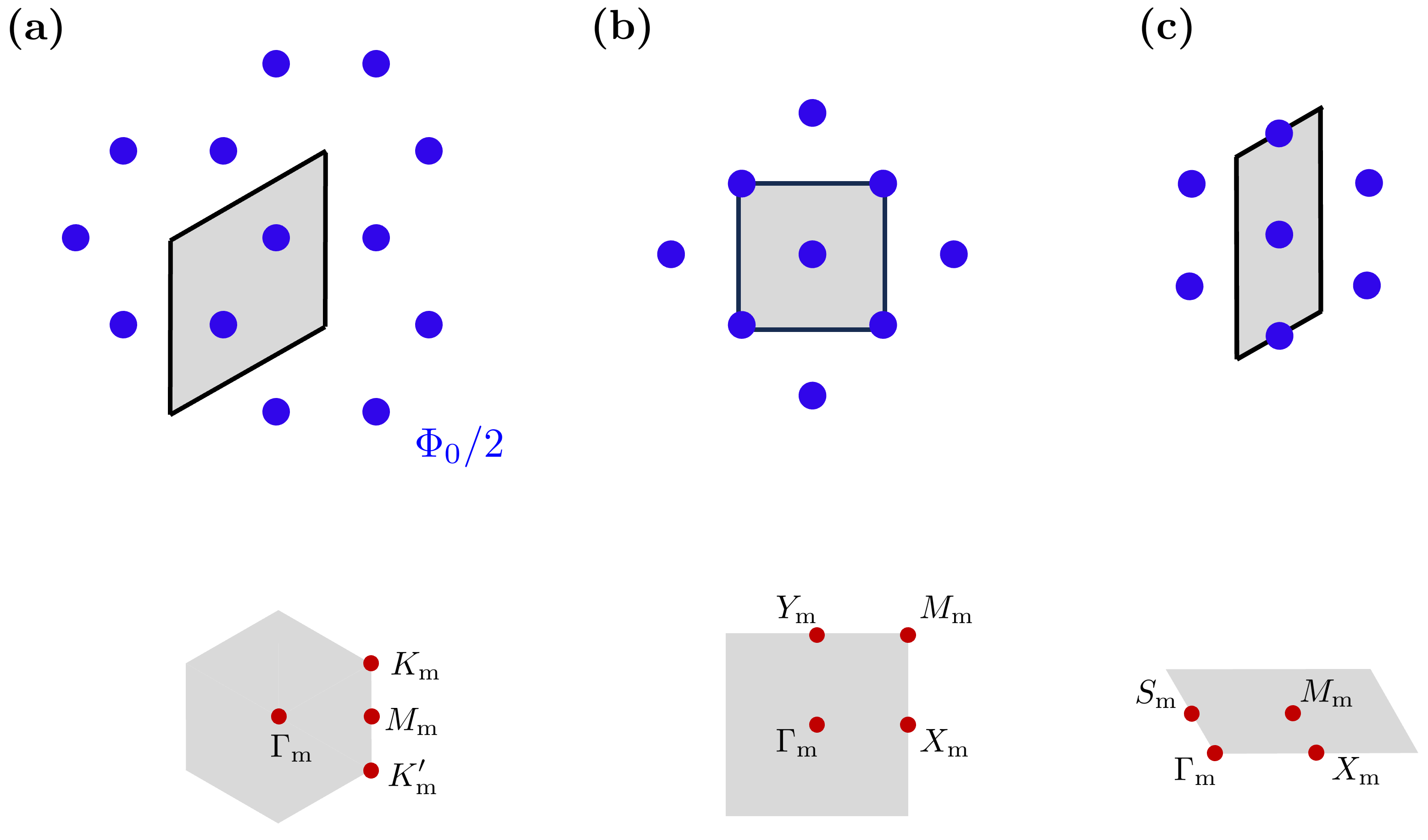}
\caption{Magnetic unit cells (top row) and corresponding mBZ domains (bottom row) for the (a) honeycomb, (b) square, and (c) triangular vortex lattices. Each blue dot represents a superconducting flux quantum of $\Phi_0/2$. (a,b) Employ the Wigner-Seitz reciprocal lattice cell, with high symmetry points are indicated. (c) Shows the primitive reciprocal unit cell and labels points defining the straight-line paths in the mBZ along which the band structures will be plotted.}
\label{fig:fig7}
\end{figure}

In the following, we will consider three types of vortex lattices: honeycomb, square, and triangular, all of which are depicted in Fig.~\ref{fig:fig7}. Although triangular vortex lattices are most commonly found in practice, square lattices have been identified in a number of materials \cite{yaron-nat96,dewilde-prl97,eskildsen-prl97,yethiraj-prl97,brown-prl04} and proposals for honeycomb lattices have been put forth \cite{reichhardt-prb01,han-prl04,meng-prb14}. We consider the vortex spacing to be controllable using temperature or magnetic field strength, and the magnetoelectric constants we will employ belong to the typical range of $0.1$ to $100$ ps/m. We again emphasize that the detailed nature of the following results is not meant to obscure the main point of this being a highly tunable setup that enables the generation of topological flat bands close to the Fermi energy.

\begin{figure*}[t]
\centering
\includegraphics[width= 17.9 cm]{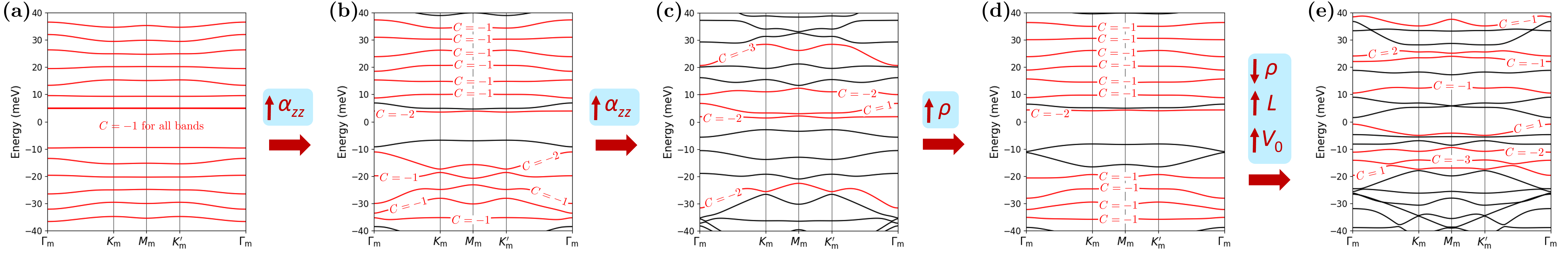}
\caption{Band structures of BLG exposed to MSLs and ESLs generated by honeycomb vortex lattices. Starting from a configuration with $\alpha_{zz}=0$\,ps/m (a), the topological characters of the bands are identical to those of LLs. As the ME constant is increased through (b) and (c), trivial bands and topological bands with $|C|>1$ appear. The width $\rho$ is subsequently increased in (d), resulting the bands reverting back to LL character. Modifying the other parameters (e) can then result in the appearance of more topological flat bands. The specific parameters employed are: (a) $L=50$\,nm, $\rho=10$\,nm, $V_0=-5$\,meV, $\alpha_{zz}=0$\,ps/m. (b) $L=50$\,nm, $\rho=10$\,nm, $V_0=-5$\,meV, $\alpha_{zz}=30$\,ps/m. (c) $L=50$\,nm, $\rho=10$\,nm, $V_0=-5$\,meV, $\alpha_{zz}=80$\,ps/m. (d) $L=50$\,nm, $\rho=15$\,nm, $V_0=-5$\,meV, $\alpha_{zz}=80$\,ps/m. (e) $L=60$\,nm, $\rho=10$\,nm, $V_0=5$\,meV, $\alpha_{zz}=80$\,ps/m.}
\label{fig:fig8}
\end{figure*}

\subsection{Honeycomb Vortex Lattice}
\label{HexagonalVortLat}
We consider a honeycomb lattice as depicted in Fig.~\ref{fig:fig7}(a), with the internal coordinates of the two vortex centers within the unit cell being $(1/3,1/3)$ and $(2/3,2/3)$. For this arrangement of vortices, the magnetic cell is identical to the geometric cell.

We track the evolution of the resulting BLG band structure in Fig.~\ref{fig:fig8}, which also contains the details of the specific parameters employed. More information on the identified $|C|>1$ flat bands may be found in Table~\ref{tab:Honeycomb_SC}. We initially consider a parameter set with $\alpha_{zz}=0$ (no ESL) as shown in Fig.~\ref{fig:fig8}(a). The resulting bands are all found to be topological, but feature the same Chern numbers as the original BLG LLs. Upon increasing $\alpha_{zz}$ to $30$\,ps/m [Fig.~\ref{fig:fig8}(b)], several band touchings occur, with band gaps closing and reopening, such that a number of bands become trivial, with a single particularly flat topological band with $C=-2$ emerging near the Fermi energy, as detailed in Table~\ref{tab:Honeycomb_SC}. There is also a single generic band with $C=-2$, while all other bands are found to exhibit $C=-1$. Upon further increasing the ME coupling to $80$\,ps/m (Fig.~\ref{fig:fig8}(c)), additional band touchings occur, resulting in more trivial bands, as well as generic bands with higher Chern numbers, including one with $C=-3$. The flat $C=-2$ band close to the Fermi energy remains, but its band width is slightly increased. 

\begin{table}[b]
\caption{\label{tab:Honeycomb_SC}
Details of flat bands appearing in Fig.~\ref{fig:fig8}.
Energy at $\Gamma_m$, Chern number, band width, gap below, and gap above, respectively (energies in meV). Flat bands with $|C|=1$ are omitted.}
\begin{ruledtabular}
\begin{tabular}{ldrc@{\hskip -1em}d@{\hskip 0.5em}c}
& \multicolumn{1}{c}{$E_{\Gamma_m}$} & $C$ & $W$ & \multicolumn{1}{c}{$E_{g-}$} & $E_{g+}$ \\
\colrule
Panel (b) 
& 3.91 & $-2$ & 0.38 & 10.32 & 0.74\\
Panel (c)
& 1.90 & $-2$ & 0.72 & 4.30 & 0.87 \\
Panel (d)
& 4.31 & $-2$ & 0.31 & 12.19 & 0.64 \\
Panel (e)
& -14.15 & $-3$ & 1.40 & 0.65 & 2.89 \\
& -11.18 & $-2$ & 1.70 & 2.89 & 0.83 \\
& 24.09 & $2$ & 1.89 & 1.21 \text{ (D)} & 2.20 \\
\end{tabular}
\end{ruledtabular}
\end{table}

Increasing the width $\rho$ of the vortex fields recovers the LL character of many of the bands, as evidenced by Fig.~\ref{fig:fig8}(d), although the flat $C=-2$ close to the Fermi energy remains. Subsequently reducing $\rho$ while increasing the cell size $L$ and displacement field $V_0$ [Fig.~\ref{fig:fig8}(e)] introduces more topological flat bands, including those with higher Chern numbers: $C=-3$, $-2$, and $2$. Detailed information on these bands is contained in Table~\ref{tab:Honeycomb_SC}.

\subsection{Square Vortex Lattice}
\label{SquareVortLat}
We now turn to the consideration of square flux lattices of the type depicted in Fig.~\ref{fig:fig7}(b), with the internal coordinates of the two vortex centers within the unit cell being $(0,0)$ and $(1/2,1/2)$. Unlike the honeycomb lattice case, the present magnetic cell is twice the size of the geometric cell. The parameter $L$ is now taken to be the lattice constant of the geometric cell, not that of the magnetic cell as was the case in the previous sections.

\begin{figure*}[t]
\centering
\includegraphics[width= 13cm]{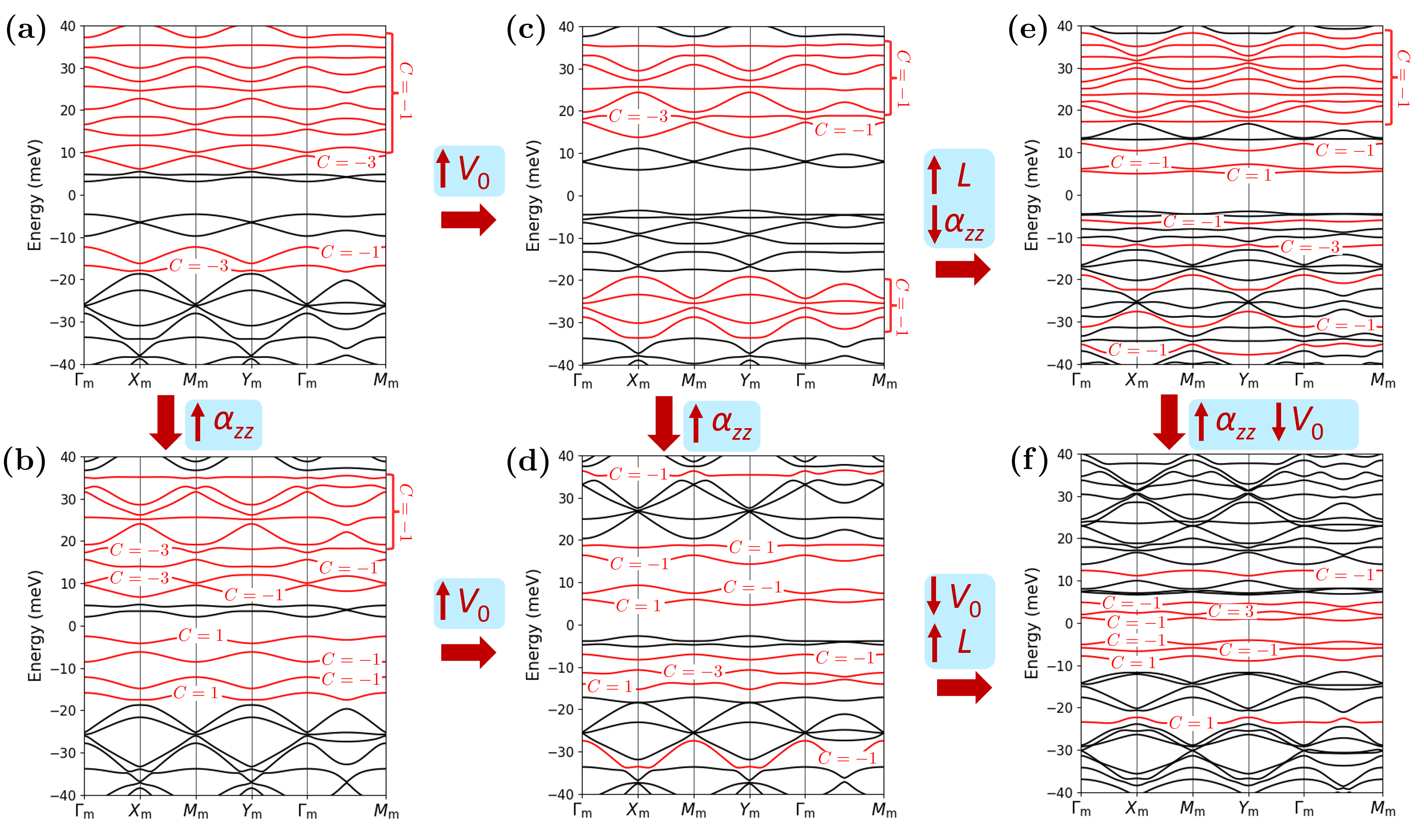}
\caption{Band structures of BLG exposed to MSLs and ESLs generated by square vortex lattices. All band structures shown employ $\rho=10$\,nm. (a,b) both feature $L=40$\,nm and $V_0=-5$\,meV, with $\alpha_{zz}$ increased from $50$ to $80$\,ps/m in going from (a) to (b). $V_0$ is subsequently increased to $5$\,meV in (c,d), with $\alpha_{zz}$ again varying between $50$ and $80$\,ps/m in going from (c) to (d). In (e,f) $L$ is increased to $50$\,nm, with (e) additionally featuring $V_0=5$\,meV and $\alpha_{zz}=30$\,ps/m, while (f) has $V_0=-5$\,meV and $\alpha_{zz}=80$\,ps/m. In the course of varying the parameters, some of the bands retain the Chern numbers of the original LLs, but generic, as well as flat topological bands with $|C|>1$ may be found near the Fermi energy, or at higher energies.}
\label{fig:fig9}
\end{figure*}

The evolution of the resulting BLG band structure as a function of the model parameters is depicted in Fig.~\ref{fig:fig9} (see figure caption for parameter values employed), with further flat $|C|>1$ band details indicated in Table~\ref{tab:Square_SC}. In all plots shown, $\rho=10$\,nm. Beginning with Fig.~\ref{fig:fig9}(a), we initially consider the case of $L=40$\,nm, $V_0=-5$\,meV, and $\alpha_{zz}=50$\,ps/m. While the majority of the topological bands exhibit $C=-1$, two bands are found to feature $C=-3$, with one of them featuring a band width on the order of $1.5$\,meV. Increasing $\alpha_{zz}$ to $80$\,ps/m in Fig.~\ref{fig:fig9}(b) results in several band touchings, with an increased number of bands exhibiting $C\neq-1$, and with a flat $C=-3$ band occurring once again.

Figures~\ref{fig:fig9}(c) and (d) employ the same parameters as Figs.~\ref{fig:fig9}(a) and (b), respectively, but with $V_0=5$\,meV. We again find that as $\alpha_{zz}$ is increased from $50$ to $80$\,ps/m in going from Fig.~\ref{fig:fig9}(c) to (d), a smaller number of bands exhibits the Chern numbers of LLs. In both Figs.~\ref{fig:fig9}(c) and (d), a $C=-3$ flat band appears.

In going to Figs.~\ref{fig:fig9}(e) and (f), $L$ is increased to $50$\,nm. The rest of the parameters in Fig.~\ref{fig:fig9}(e) take the values of $V_0=5$\,meV, $\alpha_{zz}=30$\,ps/m, while in Fig.~\ref{fig:fig9}(f) they are $V_0=-5$\,meV and $\alpha_{zz}=80$\,ps/m. Independently of the fact that $V_0$ changes sign in going between the two band structures, we find that that increasing the ME coupling again results in a smaller number of bands sharing the same Chern numbers as bands representing LLs. There are also topological flat bands with $|C|=3$ as before, but due to the increase in unit cell size the spacings between the bands are diminished.

A notable feature of the results presented above is that the Chern numbers are all odd integers.  Since we start from $C=-1$ in the trivial limit, Fig.~\ref{fig:fig8}(a), this means that Chern numbers transfer between bands in units of two when bands touch as parameters are varied. This can be understood from the fact that the unit cell has to be doubled in order to accommodate a single quantum of flux, and will also be observed in the results presented in the next subsection.

\begin{table}[b]
\caption{\label{tab:Square_SC}
Details of flat bands appearing in Fig.~\ref{fig:fig9}. Energy at $\Gamma_m$,
Chern number, band width, gap below, and gap above, respectively (energies in meV). Flat bands with $|C|=1$ are omitted.}
\begin{ruledtabular}
\begin{tabular}{ldrc@{\hskip -1em}d@{\hskip -1.5em}d@{\hskip 0.9em}}
& \multicolumn{1}{c}{$E_{\Gamma_m}$} & $C$ & $W$ & \multicolumn{1}{r}{$E_{g-}$} & \multicolumn{1}{c}{$E_{g+}$} \\
\colrule
Panel (a) 
& -16.75 & $-3$ & 1.50 & 0.79 \text{ (D)} & 0.63\text{ }\\
Panel (b)
& 17.28 & $-3$ & 0.99 & 1.88 & 0.78 \\
Panel (c)
& 19.70 & $-3$ & 0.63 & 1.07 & 0.90 \\
Panel (d)
& -11.40 & $-3$ & 1.51 & 0.75 & 2.47 \\
Panel (e)
& -11.90 & $-3$ & 0.49 & 0.71 & 0.88 \\
Panel (f)
& 1.97 & $3$ & 1.26 & 0.80 & 0.59 \\
\end{tabular}
\end{ruledtabular}
\end{table}

\subsection{Triangular Vortex Lattice}
\label{TriangularVortLat}
The triangular vortex lattice and our chosen magnetic unit cell are depicted in Fig.~\ref{fig:fig7}(c), with vortex centers located at $(1/2,0)$ and $(1/2,1/2)$ in internal coordinates. By $L$ we  again denote the geometric rather than magnetic cell lattice constant, and note that as in the square vortex lattice case earlier, the magnetic unit cell is twice the size of the geometric cell. 

\begin{figure}[t]
\centering
\includegraphics[width= \columnwidth]{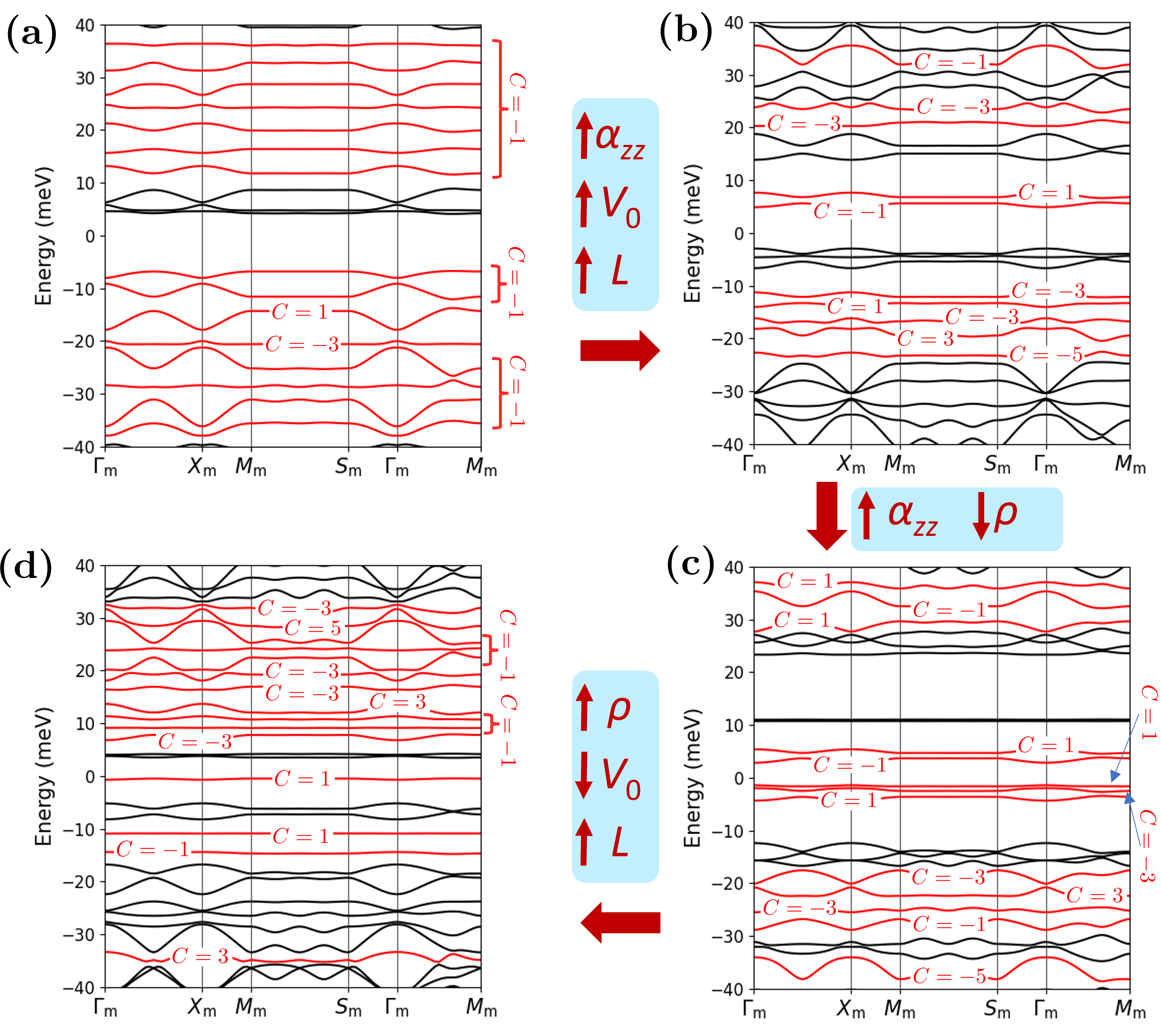}
\caption{Band structures of BLG exposed to MSLs and ESLs generated by triangular vortex lattices. (a) In the initial configuration with $L=40$\,nm, $\rho=10$\,nm, $V_0=-5$\,meV, and $\alpha_{zz}=50$\,ps/m, the majority of the bands share the topology of the LLs. (b) The  increase of $L$, $V_0$ and $\alpha_{zz}$ to values of 45\,nm, 5\,meV and 80\,ps/m, respectively, is marked by the appearance of multiple topological flat bands, including one with $C=-5$. Decreasing $\rho$ to 8\,nm, while increasing $\alpha_{zz}$ to 100\,ps/m in (c) also results in the appearance of a number of topological flat bands, as well as generic bands with $|C|>1$. Finally (d), increasing $L$ and $\rho$ to 50 and 10\,nm respectively, while decreasing $V_0$ and $\alpha_{zz}$ to $-5$\,meV and 80\,ps/m, also yields nontrivial flat bands.}
\label{fig:fig10}
\end{figure}

Band structures as a function of the model parameters are shown in Fig.~\ref{fig:fig10} (see figure caption for parameter details), with Table~\ref{tab:Triangle_SC} giving some details of selected bands shown in the figure. Starting with $L=40$\,nm, $\rho=10$\,nm, $V_0=-5$\,meV, and $\alpha_{zz}=50$\,ps/m [Fig. \ref{fig:fig10}(a)], we find that most of the bands retain LL topology, with Chern numbers $C=-1$. A single band is found to have $C=1$, and there is a flat band with $C=-3$, as detailed in Table~\ref{tab:Triangle_SC}.

Subsequently increasing $L$, $V_0$, and $\alpha_{zz}$ (while keeping $\rho$ fixed), results in an increase in the number of bands with $|C|>1$, with many of them being flat. Remarkably, a flat band with $C=-5$ can be found at lower energies, and several bands with $C=-3$ and band widths on the order of 1\,meV are also present.

We subsequently decrease $\rho$ to $8$ nm and increase $\alpha_{zz}$ to 100\,ps/m while keeping $L$ and $V_0$ fixed [Fig.~\ref{fig:fig10}(c)]. This time there are not as many flat bands with higher Chern numbers, but we do find generic bands with higher Chern numbers, including ones with $C=-5$ and $C=-3$. At lower energies we also encounter two adjacent flat bands with $C=-3$ and $C=3$.

In Fig.~\ref{fig:fig10}(d), $L=50$\,nm, $\rho=10$\,nm, $V_0=-5$\,meV, and $\alpha_{zz}=80$\,ps/m. We find a number of topological flat bands, including a variety of $|C|=3$ bands of varying band widths and at various energies. A generic $C=-5$ band also appears at higher energies. 

\begin{table}[t]
\caption{\label{tab:Triangle_SC}
Details of flat bands appearing in Fig.~\ref{fig:fig10}. Energy at $\Gamma_m$, Chern number, band width, gap below, and gap above, respectively (energies in meV). Flat bands with $|C|=1$ are omitted.}
\begin{ruledtabular}
\begin{tabular}{ldrc@{\hskip -1em}d@{\hskip -1.5em}d@{\hskip 0.9em}}
& \multicolumn{1}{c}{$E_{\Gamma_m}$} & $C$ & $W$ & \multicolumn{1}{r}{$E_{g-}$} & \multicolumn{1}{c}{$E_{g+}$} \\
\colrule
Panel (a) 
& -20.02 & $-3$ & 0.53 & 0.66 & 2.33 \\
Panel (b) 
& -22.69 & $-5$ & 0.91 & 1.05 & 2.07 \\
& -16.18 & $-3$ & 0.81 & 0.99 & 2.20 \\
& -11.27 & $-3$ & 0.92 & 1.11 & 4.62 \\
& 20.30 & $-3$ & 1.09 & 1.56 & 1.45 \\
& 23.84 & $-3$ & 1.76 & 1.45 & 0.64 \\
Panel (c) 
& -25.52 & $-3$ & 0.89 & 1.38 & 2.14\\
& -20.77 & $3$ & 1.52 & 2.14 & 0.98\\
& -1.99 & $-3$ & 0.71 & 0.75 & 0.52 \text{ (D)}\\
Panel (d) 
& -33.39 & $3$ & 1.92 & 0.54 & 0.65 \text{ (D)} \\
& 6.79 & $-3$ & 1.10 & 2.58 & 1.19 \\
& 13.68 & $3$ & 1.98 & 0.88 \text{ (D)} & 2.68 \\
& 16.41 & $-3$ & 0.92 & 2.68 & 0.92 \\
& 18.09 & $-3$ & 1.36 & 0.92 & 0.56 \\
& 32.47 & $-3$ & 1.47 & 0.89 \text{ (D)} & 0.71 \\
\end{tabular}
\end{ruledtabular}
\end{table}

\section{Summary}
\label{Summary}
In this paper we have studied the application of MSLs generated by out-of-plane orbital magnetic fields as an additional means of inducing topological flat bands in BLG. We considered MSLs that either introduce no net flux or a quantum of flux to the SL unit cell. In the latter case, we employed the recently developed gauge-invariant method of Ref.~\cite{herzog-prb22} to perform band structure calculations. We also explored the consequences of MSLs acting in conjunction with ESLs on the band structure of BLG.

To obtain a qualitative understanding of the action of MSLs on the BLG band structure, we initially explored the application of a simple triangular MSL in the first-harmonic approximation. When acting by itself, with or without a quantum of flux, the MSL was found to induce topological flat bands of Chern number $|C|>1$ when the field strength, displacement field voltage and SL lattice constant were varied. Increasing the strength of the field or increasing the SL size was found to consistently reduce band widths. Generic topological bands were also obtained, including those of higher Chern number. Upon introduction of the commensurate triangular ESL primarily studied by Ghorashi \textit{et al}.~\cite{ghorashi-prl23}, the topology of the band structures may be enriched, with more topological flat and generic bands appearing. In the flux quantum case particularly, topological flat bands could be more readily induced.

We subsequently proposed a platform for the realization of flux quantum MSLs with concomitant ESL potentials. Our proposed setup involved generating the MSL using superconductor vortex lattices, and exposing a magnetoelectric material to the magnetic fields of the vortex lattice; the subsequently induced commensurate surface charge density generated the ESL potential. We explored this setup and its application to BLG with honeycomb, square, and triangular vortex lattices. In all cases, band flatness could be tuned by the SL cell size as well as the width of the in-plane profiles of the magnetic fields in the vortices, although large widths eventually resulted in the recovery of LL topology. Increasing the magnetoelectric constant promoted band gap closings and reopenings, leading to a greater number of bands (flat or generic) with $|C|>1$. The square and triangular vortex lattices proved to be especially conducive for the appearance of bands of particularly high Chern numbers; however, the doubling of the magnetic unit cell relative to the geometric unit cell yielded smaller band gaps between neighboring bands.

\vspace{1em}\paragraph{Acknowledgments.}
We thank Jonah Herzog-Arbeitman for useful discussions regarding the gauge-independent calculations of band structures in the presence of flux quanta. We also thank Yongxin Zeng for assistance in creating Figure 1. D.S. acknowledges support from the National Science Foundation under Grant No.~DGE-1842213. D.S. and D.V. also acknowledge the National Science Foundation under Grant No.~DMR-1954856. J.C. acknowledges the Air Force Office of Scientific Research under Grant No.~FA9550-20-1-0260, the Alfred P.~Sloan Foundation through a Sloan Research Fellowship, and the Simons Foundation. 

\appendix

\section{Modification of the magnetic Bloch state scattering amplitudes}
\label{AppA}
As discussed in Sec.~\ref{Methods_oneflux}, we employ a gauge-invariant formalism developed in Ref.~\cite{herzog-prb22} to compute the continuum electronic band structures of SL-BLG in the presence of magnetic flux quanta $\Phi_0$. The primary advantage of this formalism is the gauge-invariant evaluation of the scattering matrix elements  $\langle\textbf{k},m|e^{i\textbf{G}\cdot\textbf{r}}|\textbf{k},n\rangle$, where $|\textbf{k},n\rangle$ is a magnetic translation eigenstate, with $\textbf{k}$ denoting the wavevector and $n$ indicating the index of the LL used to construct the state. Ref.~\cite{herzog-prb22} demonstrated that \begin{equation}\langle\textbf{k},m|e^{i\textbf{G}\cdot\textbf{r}}|\textbf{k},n\rangle=e^{-i\pi G_1G_2-2\pi i(G_1k_2-G_2k_1)}\mathcal{H}^{\textbf{G}}_{mn},
\label{eq:ScatAmp}\end{equation} where $G_1$ and $G_2$ are the internal coordinates of $\textbf{G}$ and $k_1$ and $k_2$ are those of $\textbf{k}$. $\mathcal{H}^{\textbf{G}}_{mn}$ is a LL-dependent quantity that is expressed in terms of associated Laguerre polynomials.

The gauge-invariant evaluation of the scattering amplitudes is enabled by introducing translation operators written in terms of gauge-invariant LL guiding center momenta $\textbf{Q}$, defined as \begin{equation}
    \textbf{Q}=\boldsymbol{\pi}-eB_{\Phi}\textbf{r}\times\hat{\textbf{z}}.
\end{equation} Here, $\boldsymbol{\pi}=\textbf{p}-e\textbf{A}_{\Phi}(\textbf{r})$, where $\textbf{p}$ is the canonical momentum operator, $\textbf{A}_{\Phi}(\textbf{r})$ is the vector potential corresponding to the field $B_{\Phi}$, and $\textbf{r}=(x,y,0)$. The magnetic translation eigenstates $|\textbf{k},n\rangle$ are then generated with the help of the translation operators corresponding to primitive translations $T_{\textbf{a}_i}=\exp(i\textbf{a}_i\cdot\textbf{Q}/\hbar)$, and $[T_{\textbf{a}_1},T_{\textbf{a}_2}]=0$ when an integer number of flux quanta thread the unit cell.

\begin{figure}[t]
\centering
\includegraphics[width= \columnwidth]{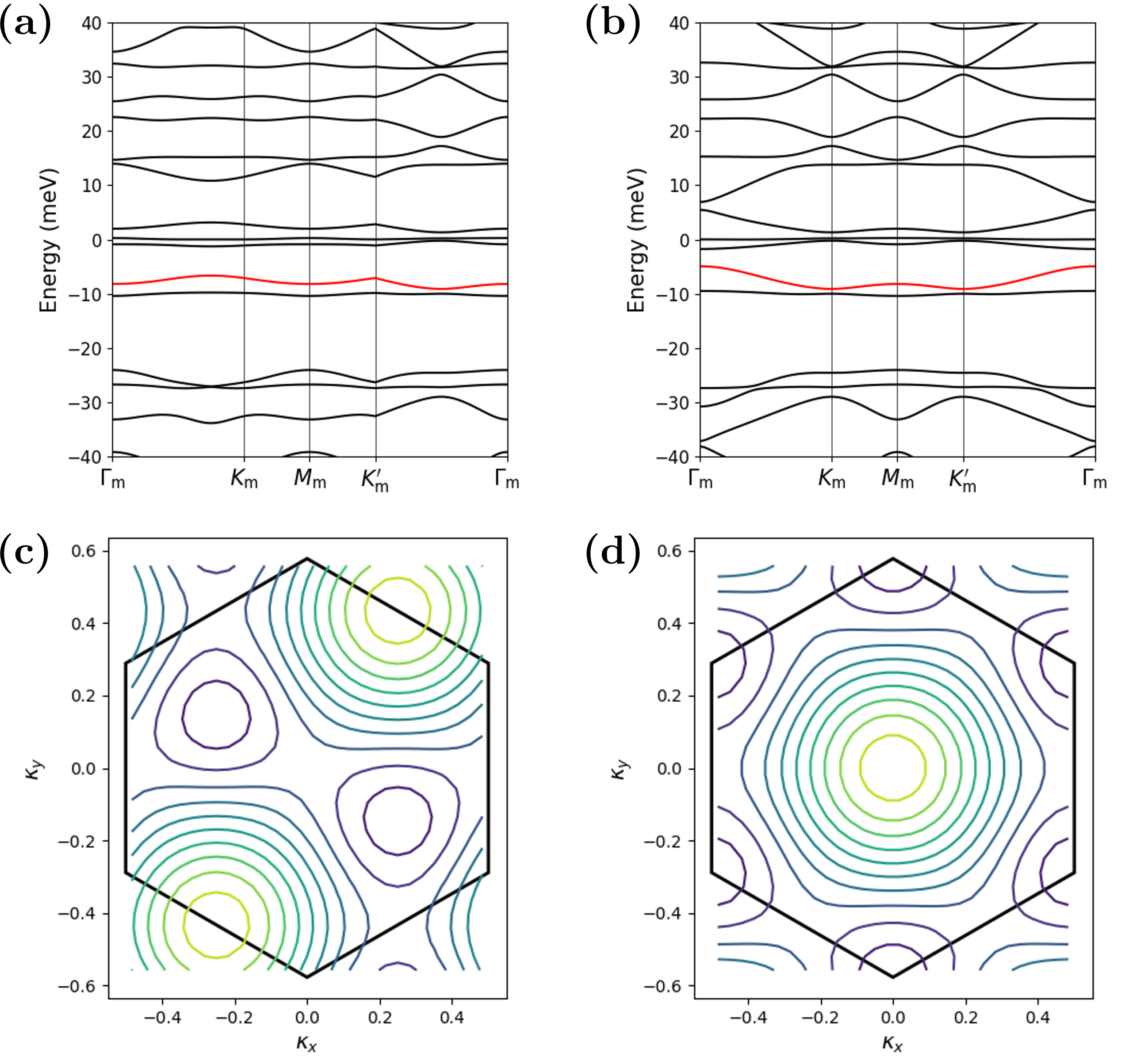}
\caption{(a,b) Band structures of the triangular SL lattice model with flux presented in Sec.~\ref{Results_oneflux} computed using (a) the original scattering amplitude of Eq.~(\ref{eq:ScatAmp}) or (b) the modified scattering amplitude in Eq.~(\ref{eq:NewScatAmp}). The model parameters are $B_0=-2$\,T, $V_{\text{ESL}}=10$\,meV, $V_0=-5$\,meV, $L=50$ nm. (c,d) Contour plots for the band highlighted in red in panels (a,b) respectively. Here $\kappa_i=k_i/Q$, where $Q$ is the reciprocal lattice constant, and the hexagon outlines the Wigner-Seitz mBZ.}
\label{fig:fig11}
\end{figure}

A subtlety of this formalism is the choice of origin of the magnetic BZ. In the case of zero-flux band structures, the BZ origin $\Gamma$ is the point of highest symmetry in the band structure, and shares the point group symmetry of the system under consideration. The formulation of Herzog-Arbeitman \textit{et al}.~\cite{herzog-prb22} features an implicit selection of the BZ origin, but the selected $\Gamma$ may not be the $\textbf{k}$-point with maximal symmetry. We see an example of this in Fig.~\ref{fig:fig11}(a), in which a band structure of the triangular MSL model of Sec.~\ref{Results_Triangle} is shown, with a contour plot corresponding to the highlighted red band shown in the panel below. It is clear from the contour plot that $\Gamma_m$ is not the point of highest symmetry, and in the band structure we find that other high-symmetry points do not have the expected symmetry either. For example, while $K_m$ -- a point with 3-fold rotational symmetry -- is expected to feature zero group velocity, it is clear from the plotted band structure that is not the case.

From the contour plot in Fig.~\ref{fig:fig11}(a), however, it appears that a rigid translation of the band structure would result in $\Gamma_m$ becoming the highest symmetry point. A shift of the band structure in $\textbf{k}$-space may occur by effecting a gauge transformation whereby $\textbf{A}_{\Phi}(\textbf{r})\to\textbf{A}_{\Phi}(\textbf{r})+\textbf{A}_0$, where $\textbf{A}_0$ is a constant; under such a transformation, $\textbf{k}\to\textbf{k}-\frac{e}{\hbar}\textbf{A}_0$. The choice of $\textbf{A}_0$ is therefore implicit in the formalism of Ref.~\cite{herzog-prb22}. In order to ensure that $\textbf{A}_0$ is always chosen so that $\Gamma$ is the highest symmetry point, the scattering amplitude of Eq.~(\ref{eq:ScatAmp}) may be modified to be \cite{herzog-privcomm23} \begin{equation}
\langle\textbf{k},m|e^{i\textbf{G}\cdot\textbf{r}}|\textbf{k},n\rangle=e^{-i\pi(G_1G_2+G_1+G_2)-2\pi i(G_1k_2-G_2k_1)}\mathcal{H}^{\textbf{G}}_{mn},\label{eq:NewScatAmp}
\end{equation} where an additional phase factor of $e^{-i\pi(G_1+G_2)}$ now multiplies the original scattering amplitude. With this modified scattering amplitude we find that the band structure, plotted in Fig.~\ref{fig:fig11}(b), now obeys the symmetries of the system. We employ the modified scattering amplitude to calculate all band structures plotted in the main text.

A second subtlety in this formalism arises with the choice of spatial origin of the unit cell relative to the guiding centers. A change in the origin manifests itself in a constant shift of the guiding center momentum operator $\textbf{Q}$, and in turn modifies the translation operators $T_{\textbf{a}_i}$ by constant phase factors. These phase factors result in a shift of the $\textbf{k}$-points and therefore band structure, and thus may again result in $\Gamma$ not being the point of maximal symmetry in the BZ. 

In the case of the triangular MSL model of Sec.~\ref{Results_Triangle}, the obvious choice of origin centered on the maximum of $B(\textbf{r})$ automatically ensures the desired symmetry in reciprocal space. For the structures in Sec.~\ref{Results_SC} with two vortices per cell, however, we find that we need to take care to make a choice that places the vortices at high-symmetry positions, as other choices result in a violation of the desired $\textbf{k}$-space symmetry.

\bibliography{pap}

\begin{thebibliography}{119}%
\makeatletter
\providecommand \@ifxundefined [1]{%
 \@ifx{#1\undefined}
}%
\providecommand \@ifnum [1]{%
 \ifnum #1\expandafter \@firstoftwo
 \else \expandafter \@secondoftwo
 \fi
}%
\providecommand \@ifx [1]{%
 \ifx #1\expandafter \@firstoftwo
 \else \expandafter \@secondoftwo
 \fi
}%
\providecommand \natexlab [1]{#1}%
\providecommand \enquote  [1]{``#1''}%
\providecommand \bibnamefont  [1]{#1}%
\providecommand \bibfnamefont [1]{#1}%
\providecommand \citenamefont [1]{#1}%
\providecommand \href@noop [0]{\@secondoftwo}%
\providecommand \href [0]{\begingroup \@sanitize@url \@href}%
\providecommand \@href[1]{\@@startlink{#1}\@@href}%
\providecommand \@@href[1]{\endgroup#1\@@endlink}%
\providecommand \@sanitize@url [0]{\catcode `\\12\catcode `\$12\catcode `\&12\catcode `\#12\catcode `\^12\catcode `\_12\catcode `\%12\relax}%
\providecommand \@@startlink[1]{}%
\providecommand \@@endlink[0]{}%
\providecommand \url  [0]{\begingroup\@sanitize@url \@url }%
\providecommand \@url [1]{\endgroup\@href {#1}{\urlprefix }}%
\providecommand \urlprefix  [0]{URL }%
\providecommand \Eprint [0]{\href }%
\providecommand \doibase [0]{https://doi.org/}%
\providecommand \selectlanguage [0]{\@gobble}%
\providecommand \bibinfo  [0]{\@secondoftwo}%
\providecommand \bibfield  [0]{\@secondoftwo}%
\providecommand \translation [1]{[#1]}%
\providecommand \BibitemOpen [0]{}%
\providecommand \bibitemStop [0]{}%
\providecommand \bibitemNoStop [0]{.\EOS\space}%
\providecommand \EOS [0]{\spacefactor3000\relax}%
\providecommand \BibitemShut  [1]{\csname bibitem#1\endcsname}%
\let\auto@bib@innerbib\@empty
\bibitem [{\citenamefont {Cao}\ \emph {et~al.}(2018{\natexlab{a}})\citenamefont {Cao}, \citenamefont {Fatemi}, \citenamefont {Demir}, \citenamefont {Fang}, \citenamefont {Tomarken}, \citenamefont {Luo}, \citenamefont {Sanchez-Yamagishi}, \citenamefont {Watanabe}, \citenamefont {Taniguchi}, \citenamefont {Kaxiras}, \citenamefont {Ashoori},\ and\ \citenamefont {Jarillo-Herrero}}]{cao-nat18b}%
  \BibitemOpen
  \bibfield  {author} {\bibinfo {author} {\bibfnamefont {Y.}~\bibnamefont {Cao}}, \bibinfo {author} {\bibfnamefont {V.}~\bibnamefont {Fatemi}}, \bibinfo {author} {\bibfnamefont {A.}~\bibnamefont {Demir}}, \bibinfo {author} {\bibfnamefont {S.}~\bibnamefont {Fang}}, \bibinfo {author} {\bibfnamefont {S.~L.}\ \bibnamefont {Tomarken}}, \bibinfo {author} {\bibfnamefont {J.~Y.}\ \bibnamefont {Luo}}, \bibinfo {author} {\bibfnamefont {J.~D.}\ \bibnamefont {Sanchez-Yamagishi}}, \bibinfo {author} {\bibfnamefont {K.}~\bibnamefont {Watanabe}}, \bibinfo {author} {\bibfnamefont {T.}~\bibnamefont {Taniguchi}}, \bibinfo {author} {\bibfnamefont {E.}~\bibnamefont {Kaxiras}}, \bibinfo {author} {\bibfnamefont {R.~C.}\ \bibnamefont {Ashoori}},\ and\ \bibinfo {author} {\bibfnamefont {P.}~\bibnamefont {Jarillo-Herrero}},\ }\bibfield  {title} {\bibinfo {title} {Correlated insulator behaviour at half-filling in magic-angle graphene superlattices},\ }\href {https://doi.org/10.1038/nature26154} {\bibfield  {journal} {\bibinfo  {journal}
  {Nature}\ }\textbf {\bibinfo {volume} {556}},\ \bibinfo {pages} {80–84} (\bibinfo {year} {2018}{\natexlab{a}})}\BibitemShut {NoStop}%
\bibitem [{\citenamefont {Cao}\ \emph {et~al.}(2018{\natexlab{b}})\citenamefont {Cao}, \citenamefont {Fatemi}, \citenamefont {Fang}, \citenamefont {Watanabe}, \citenamefont {Taniguchi}, \citenamefont {Kaxiras},\ and\ \citenamefont {Jarillo-Herrero}}]{cao-nat18a}%
  \BibitemOpen
  \bibfield  {author} {\bibinfo {author} {\bibfnamefont {Y.}~\bibnamefont {Cao}}, \bibinfo {author} {\bibfnamefont {V.}~\bibnamefont {Fatemi}}, \bibinfo {author} {\bibfnamefont {S.}~\bibnamefont {Fang}}, \bibinfo {author} {\bibfnamefont {K.}~\bibnamefont {Watanabe}}, \bibinfo {author} {\bibfnamefont {T.}~\bibnamefont {Taniguchi}}, \bibinfo {author} {\bibfnamefont {E.}~\bibnamefont {Kaxiras}},\ and\ \bibinfo {author} {\bibfnamefont {P.}~\bibnamefont {Jarillo-Herrero}},\ }\bibfield  {title} {\bibinfo {title} {Unconventional superconductivity in magic-angle graphene superlattices},\ }\href {https://doi.org/10.1038/nature26160} {\bibfield  {journal} {\bibinfo  {journal} {Nature}\ }\textbf {\bibinfo {volume} {556}},\ \bibinfo {pages} {43–50} (\bibinfo {year} {2018}{\natexlab{b}})}\BibitemShut {NoStop}%
\bibitem [{\citenamefont {Yankowitz}\ \emph {et~al.}(2019)\citenamefont {Yankowitz}, \citenamefont {Chen}, \citenamefont {Polshyn}, \citenamefont {Zhang}, \citenamefont {Watanabe}, \citenamefont {Taniguchi}, \citenamefont {Graf}, \citenamefont {Young},\ and\ \citenamefont {Dean}}]{yankowitz-sciadv19}%
  \BibitemOpen
  \bibfield  {author} {\bibinfo {author} {\bibfnamefont {M.}~\bibnamefont {Yankowitz}}, \bibinfo {author} {\bibfnamefont {S.}~\bibnamefont {Chen}}, \bibinfo {author} {\bibfnamefont {H.}~\bibnamefont {Polshyn}}, \bibinfo {author} {\bibfnamefont {Y.}~\bibnamefont {Zhang}}, \bibinfo {author} {\bibfnamefont {K.}~\bibnamefont {Watanabe}}, \bibinfo {author} {\bibfnamefont {T.}~\bibnamefont {Taniguchi}}, \bibinfo {author} {\bibfnamefont {D.}~\bibnamefont {Graf}}, \bibinfo {author} {\bibfnamefont {A.~F.}\ \bibnamefont {Young}},\ and\ \bibinfo {author} {\bibfnamefont {C.~R.}\ \bibnamefont {Dean}},\ }\bibfield  {title} {\bibinfo {title} {Tuning superconductivity in twisted bilayer graphene},\ }\href {https://doi.org/10.1126/science.aav1910} {\bibfield  {journal} {\bibinfo  {journal} {Science}\ }\textbf {\bibinfo {volume} {363}},\ \bibinfo {pages} {1059–1064} (\bibinfo {year} {2019})}\BibitemShut {NoStop}%
\bibitem [{\citenamefont {Chen}\ \emph {et~al.}(2019{\natexlab{a}})\citenamefont {Chen}, \citenamefont {Jiang}, \citenamefont {Wu}, \citenamefont {Lyu}, \citenamefont {Li}, \citenamefont {Chittari}, \citenamefont {Watanabe}, \citenamefont {Taniguchi}, \citenamefont {Shi}, \citenamefont {Jung}, \citenamefont {Zhang},\ and\ \citenamefont {Wang}}]{chen-natphys19}%
  \BibitemOpen
  \bibfield  {author} {\bibinfo {author} {\bibfnamefont {G.}~\bibnamefont {Chen}}, \bibinfo {author} {\bibfnamefont {L.}~\bibnamefont {Jiang}}, \bibinfo {author} {\bibfnamefont {S.}~\bibnamefont {Wu}}, \bibinfo {author} {\bibfnamefont {B.}~\bibnamefont {Lyu}}, \bibinfo {author} {\bibfnamefont {H.}~\bibnamefont {Li}}, \bibinfo {author} {\bibfnamefont {B.~L.}\ \bibnamefont {Chittari}}, \bibinfo {author} {\bibfnamefont {K.}~\bibnamefont {Watanabe}}, \bibinfo {author} {\bibfnamefont {T.}~\bibnamefont {Taniguchi}}, \bibinfo {author} {\bibfnamefont {Z.}~\bibnamefont {Shi}}, \bibinfo {author} {\bibfnamefont {J.}~\bibnamefont {Jung}}, \bibinfo {author} {\bibfnamefont {Y.}~\bibnamefont {Zhang}},\ and\ \bibinfo {author} {\bibfnamefont {F.}~\bibnamefont {Wang}},\ }\bibfield  {title} {\bibinfo {title} {Evidence of a gate-tunable {Mott} insulator in a trilayer graphene moiré superlattice},\ }\href {https://doi.org/10.1038/s41567-018-0387-2} {\bibfield  {journal} {\bibinfo  {journal} {Nature Physics}\ }\textbf {\bibinfo
  {volume} {15}},\ \bibinfo {pages} {237–241} (\bibinfo {year} {2019}{\natexlab{a}})}\BibitemShut {NoStop}%
\bibitem [{\citenamefont {Regan}\ \emph {et~al.}(2020)\citenamefont {Regan}, \citenamefont {Wang}, \citenamefont {Jin}, \citenamefont {Bakti~Utama}, \citenamefont {Gao}, \citenamefont {Wei}, \citenamefont {Zhao}, \citenamefont {Zhao}, \citenamefont {Zhang}, \citenamefont {Yumigeta}, \citenamefont {Blei}, \citenamefont {Carlstr\"{o}m}, \citenamefont {Watanabe}, \citenamefont {Taniguchi}, \citenamefont {Tongay}, \citenamefont {Crommie}, \citenamefont {Zettl},\ and\ \citenamefont {Wang}}]{regan-nat20}%
  \BibitemOpen
  \bibfield  {author} {\bibinfo {author} {\bibfnamefont {E.~C.}\ \bibnamefont {Regan}}, \bibinfo {author} {\bibfnamefont {D.}~\bibnamefont {Wang}}, \bibinfo {author} {\bibfnamefont {C.}~\bibnamefont {Jin}}, \bibinfo {author} {\bibfnamefont {M.~I.}\ \bibnamefont {Bakti~Utama}}, \bibinfo {author} {\bibfnamefont {B.}~\bibnamefont {Gao}}, \bibinfo {author} {\bibfnamefont {X.}~\bibnamefont {Wei}}, \bibinfo {author} {\bibfnamefont {S.}~\bibnamefont {Zhao}}, \bibinfo {author} {\bibfnamefont {W.}~\bibnamefont {Zhao}}, \bibinfo {author} {\bibfnamefont {Z.}~\bibnamefont {Zhang}}, \bibinfo {author} {\bibfnamefont {K.}~\bibnamefont {Yumigeta}}, \bibinfo {author} {\bibfnamefont {M.}~\bibnamefont {Blei}}, \bibinfo {author} {\bibfnamefont {J.~D.}\ \bibnamefont {Carlstr\"{o}m}}, \bibinfo {author} {\bibfnamefont {K.}~\bibnamefont {Watanabe}}, \bibinfo {author} {\bibfnamefont {T.}~\bibnamefont {Taniguchi}}, \bibinfo {author} {\bibfnamefont {S.}~\bibnamefont {Tongay}}, \bibinfo {author} {\bibfnamefont {M.}~\bibnamefont
  {Crommie}}, \bibinfo {author} {\bibfnamefont {A.}~\bibnamefont {Zettl}},\ and\ \bibinfo {author} {\bibfnamefont {F.}~\bibnamefont {Wang}},\ }\bibfield  {title} {\bibinfo {title} {Mott and generalized {Wigner} crystal states in {WSe$_2$}/{WS$_2$} moiré superlattices},\ }\href {https://doi.org/10.1038/s41586-020-2092-4} {\bibfield  {journal} {\bibinfo  {journal} {Nature}\ }\textbf {\bibinfo {volume} {579}},\ \bibinfo {pages} {359–363} (\bibinfo {year} {2020})}\BibitemShut {NoStop}%
\bibitem [{\citenamefont {Li}\ \emph {et~al.}(2021{\natexlab{a}})\citenamefont {Li}, \citenamefont {Jiang}, \citenamefont {Li}, \citenamefont {Zhang}, \citenamefont {Kang}, \citenamefont {Zhu}, \citenamefont {Watanabe}, \citenamefont {Taniguchi}, \citenamefont {Chowdhury}, \citenamefont {Fu}, \citenamefont {Shan},\ and\ \citenamefont {Mak}}]{li-nat21b}%
  \BibitemOpen
  \bibfield  {author} {\bibinfo {author} {\bibfnamefont {T.}~\bibnamefont {Li}}, \bibinfo {author} {\bibfnamefont {S.}~\bibnamefont {Jiang}}, \bibinfo {author} {\bibfnamefont {L.}~\bibnamefont {Li}}, \bibinfo {author} {\bibfnamefont {Y.}~\bibnamefont {Zhang}}, \bibinfo {author} {\bibfnamefont {K.}~\bibnamefont {Kang}}, \bibinfo {author} {\bibfnamefont {J.}~\bibnamefont {Zhu}}, \bibinfo {author} {\bibfnamefont {K.}~\bibnamefont {Watanabe}}, \bibinfo {author} {\bibfnamefont {T.}~\bibnamefont {Taniguchi}}, \bibinfo {author} {\bibfnamefont {D.}~\bibnamefont {Chowdhury}}, \bibinfo {author} {\bibfnamefont {L.}~\bibnamefont {Fu}}, \bibinfo {author} {\bibfnamefont {J.}~\bibnamefont {Shan}},\ and\ \bibinfo {author} {\bibfnamefont {K.~F.}\ \bibnamefont {Mak}},\ }\bibfield  {title} {\bibinfo {title} {Continuous {Mott} transition in semiconductor moiré superlattices},\ }\href {https://doi.org/10.1038/s41586-021-03853-0} {\bibfield  {journal} {\bibinfo  {journal} {Nature}\ }\textbf {\bibinfo {volume} {597}},\ \bibinfo
  {pages} {350–354} (\bibinfo {year} {2021}{\natexlab{a}})}\BibitemShut {NoStop}%
\bibitem [{\citenamefont {Chen}\ \emph {et~al.}(2019{\natexlab{b}})\citenamefont {Chen}, \citenamefont {Sharpe}, \citenamefont {Gallagher}, \citenamefont {Rosen}, \citenamefont {Fox}, \citenamefont {Jiang}, \citenamefont {Lyu}, \citenamefont {Li}, \citenamefont {Watanabe}, \citenamefont {Taniguchi}, \citenamefont {Jung}, \citenamefont {Shi}, \citenamefont {Goldhaber-Gordon}, \citenamefont {Zhang},\ and\ \citenamefont {Wang}}]{chen-nat19}%
  \BibitemOpen
  \bibfield  {author} {\bibinfo {author} {\bibfnamefont {G.}~\bibnamefont {Chen}}, \bibinfo {author} {\bibfnamefont {A.~L.}\ \bibnamefont {Sharpe}}, \bibinfo {author} {\bibfnamefont {P.}~\bibnamefont {Gallagher}}, \bibinfo {author} {\bibfnamefont {I.~T.}\ \bibnamefont {Rosen}}, \bibinfo {author} {\bibfnamefont {E.~J.}\ \bibnamefont {Fox}}, \bibinfo {author} {\bibfnamefont {L.}~\bibnamefont {Jiang}}, \bibinfo {author} {\bibfnamefont {B.}~\bibnamefont {Lyu}}, \bibinfo {author} {\bibfnamefont {H.}~\bibnamefont {Li}}, \bibinfo {author} {\bibfnamefont {K.}~\bibnamefont {Watanabe}}, \bibinfo {author} {\bibfnamefont {T.}~\bibnamefont {Taniguchi}}, \bibinfo {author} {\bibfnamefont {J.}~\bibnamefont {Jung}}, \bibinfo {author} {\bibfnamefont {Z.}~\bibnamefont {Shi}}, \bibinfo {author} {\bibfnamefont {D.}~\bibnamefont {Goldhaber-Gordon}}, \bibinfo {author} {\bibfnamefont {Y.}~\bibnamefont {Zhang}},\ and\ \bibinfo {author} {\bibfnamefont {F.}~\bibnamefont {Wang}},\ }\bibfield  {title} {\bibinfo {title} {Signatures of
  tunable superconductivity in a trilayer graphene moiré superlattice},\ }\href {https://doi.org/10.1038/s41586-019-1393-y} {\bibfield  {journal} {\bibinfo  {journal} {Nature}\ }\textbf {\bibinfo {volume} {572}},\ \bibinfo {pages} {215–219} (\bibinfo {year} {2019}{\natexlab{b}})}\BibitemShut {NoStop}%
\bibitem [{\citenamefont {Li}\ \emph {et~al.}(2021{\natexlab{b}})\citenamefont {Li}, \citenamefont {Li}, \citenamefont {Regan}, \citenamefont {Wang}, \citenamefont {Zhao}, \citenamefont {Kahn}, \citenamefont {Yumigeta}, \citenamefont {Blei}, \citenamefont {Taniguchi}, \citenamefont {Watanabe}, \citenamefont {Tongay}, \citenamefont {Zettl}, \citenamefont {Crommie},\ and\ \citenamefont {Wang}}]{li-nat21a}%
  \BibitemOpen
  \bibfield  {author} {\bibinfo {author} {\bibfnamefont {H.}~\bibnamefont {Li}}, \bibinfo {author} {\bibfnamefont {S.}~\bibnamefont {Li}}, \bibinfo {author} {\bibfnamefont {E.~C.}\ \bibnamefont {Regan}}, \bibinfo {author} {\bibfnamefont {D.}~\bibnamefont {Wang}}, \bibinfo {author} {\bibfnamefont {W.}~\bibnamefont {Zhao}}, \bibinfo {author} {\bibfnamefont {S.}~\bibnamefont {Kahn}}, \bibinfo {author} {\bibfnamefont {K.}~\bibnamefont {Yumigeta}}, \bibinfo {author} {\bibfnamefont {M.}~\bibnamefont {Blei}}, \bibinfo {author} {\bibfnamefont {T.}~\bibnamefont {Taniguchi}}, \bibinfo {author} {\bibfnamefont {K.}~\bibnamefont {Watanabe}}, \bibinfo {author} {\bibfnamefont {S.}~\bibnamefont {Tongay}}, \bibinfo {author} {\bibfnamefont {A.}~\bibnamefont {Zettl}}, \bibinfo {author} {\bibfnamefont {M.~F.}\ \bibnamefont {Crommie}},\ and\ \bibinfo {author} {\bibfnamefont {F.}~\bibnamefont {Wang}},\ }\bibfield  {title} {\bibinfo {title} {Imaging two-dimensional generalized {Wigner} crystals},\ }\href
  {https://doi.org/10.1038/s41586-021-03874-9} {\bibfield  {journal} {\bibinfo  {journal} {Nature}\ }\textbf {\bibinfo {volume} {597}},\ \bibinfo {pages} {650–654} (\bibinfo {year} {2021}{\natexlab{b}})}\BibitemShut {NoStop}%
\bibitem [{\citenamefont {Sharpe}\ \emph {et~al.}(2019)\citenamefont {Sharpe}, \citenamefont {Fox}, \citenamefont {Barnard}, \citenamefont {Finney}, \citenamefont {Watanabe}, \citenamefont {Taniguchi}, \citenamefont {Kastner},\ and\ \citenamefont {Goldhaber-Gordon}}]{sharpe-sci2019}%
  \BibitemOpen
  \bibfield  {author} {\bibinfo {author} {\bibfnamefont {A.~L.}\ \bibnamefont {Sharpe}}, \bibinfo {author} {\bibfnamefont {E.~J.}\ \bibnamefont {Fox}}, \bibinfo {author} {\bibfnamefont {A.~W.}\ \bibnamefont {Barnard}}, \bibinfo {author} {\bibfnamefont {J.}~\bibnamefont {Finney}}, \bibinfo {author} {\bibfnamefont {K.}~\bibnamefont {Watanabe}}, \bibinfo {author} {\bibfnamefont {T.}~\bibnamefont {Taniguchi}}, \bibinfo {author} {\bibfnamefont {M.~A.}\ \bibnamefont {Kastner}},\ and\ \bibinfo {author} {\bibfnamefont {D.}~\bibnamefont {Goldhaber-Gordon}},\ }\bibfield  {title} {\bibinfo {title} {Emergent ferromagnetism near three-quarters filling in twisted bilayer graphene},\ }\href {https://doi.org/10.1126/science.aaw3780} {\bibfield  {journal} {\bibinfo  {journal} {Science}\ }\textbf {\bibinfo {volume} {365}},\ \bibinfo {pages} {605} (\bibinfo {year} {2019})}\BibitemShut {NoStop}%
\bibitem [{\citenamefont {Lu}\ \emph {et~al.}(2019)\citenamefont {Lu}, \citenamefont {Stepanov}, \citenamefont {Yang}, \citenamefont {Xie}, \citenamefont {Aamir}, \citenamefont {Das}, \citenamefont {Urgell}, \citenamefont {Watanabe}, \citenamefont {Taniguchi}, \citenamefont {Zhang}, \citenamefont {Bachtold}, \citenamefont {MacDonald},\ and\ \citenamefont {Efetov}}]{lu-nat19}%
  \BibitemOpen
  \bibfield  {author} {\bibinfo {author} {\bibfnamefont {X.}~\bibnamefont {Lu}}, \bibinfo {author} {\bibfnamefont {P.}~\bibnamefont {Stepanov}}, \bibinfo {author} {\bibfnamefont {W.}~\bibnamefont {Yang}}, \bibinfo {author} {\bibfnamefont {M.}~\bibnamefont {Xie}}, \bibinfo {author} {\bibfnamefont {M.~A.}\ \bibnamefont {Aamir}}, \bibinfo {author} {\bibfnamefont {I.}~\bibnamefont {Das}}, \bibinfo {author} {\bibfnamefont {C.}~\bibnamefont {Urgell}}, \bibinfo {author} {\bibfnamefont {K.}~\bibnamefont {Watanabe}}, \bibinfo {author} {\bibfnamefont {T.}~\bibnamefont {Taniguchi}}, \bibinfo {author} {\bibfnamefont {G.}~\bibnamefont {Zhang}}, \bibinfo {author} {\bibfnamefont {A.}~\bibnamefont {Bachtold}}, \bibinfo {author} {\bibfnamefont {A.~H.}\ \bibnamefont {MacDonald}},\ and\ \bibinfo {author} {\bibfnamefont {D.~K.}\ \bibnamefont {Efetov}},\ }\bibfield  {title} {\bibinfo {title} {Superconductors, orbital magnets and correlated states in magic-angle bilayer graphene},\ }\href
  {https://doi.org/10.1038/s41586-019-1695-0} {\bibfield  {journal} {\bibinfo  {journal} {Nature}\ }\textbf {\bibinfo {volume} {574}},\ \bibinfo {pages} {653–657} (\bibinfo {year} {2019})}\BibitemShut {NoStop}%
\bibitem [{\citenamefont {Serlin}\ \emph {et~al.}(2020)\citenamefont {Serlin}, \citenamefont {Tschirhart}, \citenamefont {Polshyn}, \citenamefont {Zhang}, \citenamefont {Zhu}, \citenamefont {Watanabe}, \citenamefont {Taniguchi}, \citenamefont {Balents},\ and\ \citenamefont {Young}}]{serlin-sci20}%
  \BibitemOpen
  \bibfield  {author} {\bibinfo {author} {\bibfnamefont {M.}~\bibnamefont {Serlin}}, \bibinfo {author} {\bibfnamefont {C.~L.}\ \bibnamefont {Tschirhart}}, \bibinfo {author} {\bibfnamefont {H.}~\bibnamefont {Polshyn}}, \bibinfo {author} {\bibfnamefont {Y.}~\bibnamefont {Zhang}}, \bibinfo {author} {\bibfnamefont {J.}~\bibnamefont {Zhu}}, \bibinfo {author} {\bibfnamefont {K.}~\bibnamefont {Watanabe}}, \bibinfo {author} {\bibfnamefont {T.}~\bibnamefont {Taniguchi}}, \bibinfo {author} {\bibfnamefont {L.}~\bibnamefont {Balents}},\ and\ \bibinfo {author} {\bibfnamefont {A.~F.}\ \bibnamefont {Young}},\ }\bibfield  {title} {\bibinfo {title} {Intrinsic quantized anomalous {Hall} effect in a moiré heterostructure},\ }\href {https://doi.org/10.1126/science.aay5533} {\bibfield  {journal} {\bibinfo  {journal} {Science}\ }\textbf {\bibinfo {volume} {367}},\ \bibinfo {pages} {900} (\bibinfo {year} {2020})}\BibitemShut {NoStop}%
\bibitem [{\citenamefont {Chen}\ \emph {et~al.}(2020{\natexlab{a}})\citenamefont {Chen}, \citenamefont {Sharpe}, \citenamefont {Fox}, \citenamefont {Zhang}, \citenamefont {Wang}, \citenamefont {Jiang}, \citenamefont {Lyu}, \citenamefont {Li}, \citenamefont {Watanabe}, \citenamefont {Taniguchi}, \citenamefont {Shi}, \citenamefont {Senthil}, \citenamefont {Goldhaber-Gordon}, \citenamefont {Zhang},\ and\ \citenamefont {Wang}}]{chen-nat20}%
  \BibitemOpen
  \bibfield  {author} {\bibinfo {author} {\bibfnamefont {G.}~\bibnamefont {Chen}}, \bibinfo {author} {\bibfnamefont {A.~L.}\ \bibnamefont {Sharpe}}, \bibinfo {author} {\bibfnamefont {E.~J.}\ \bibnamefont {Fox}}, \bibinfo {author} {\bibfnamefont {Y.-H.}\ \bibnamefont {Zhang}}, \bibinfo {author} {\bibfnamefont {S.}~\bibnamefont {Wang}}, \bibinfo {author} {\bibfnamefont {L.}~\bibnamefont {Jiang}}, \bibinfo {author} {\bibfnamefont {B.}~\bibnamefont {Lyu}}, \bibinfo {author} {\bibfnamefont {H.}~\bibnamefont {Li}}, \bibinfo {author} {\bibfnamefont {K.}~\bibnamefont {Watanabe}}, \bibinfo {author} {\bibfnamefont {T.}~\bibnamefont {Taniguchi}}, \bibinfo {author} {\bibfnamefont {Z.}~\bibnamefont {Shi}}, \bibinfo {author} {\bibfnamefont {T.}~\bibnamefont {Senthil}}, \bibinfo {author} {\bibfnamefont {D.}~\bibnamefont {Goldhaber-Gordon}}, \bibinfo {author} {\bibfnamefont {Y.}~\bibnamefont {Zhang}},\ and\ \bibinfo {author} {\bibfnamefont {F.}~\bibnamefont {Wang}},\ }\bibfield  {title} {\bibinfo {title} {Tunable correlated
  {Chern} insulator and ferromagnetism in a moiré superlattice},\ }\href {https://doi.org/10.1038/s41586-020-2049-7} {\bibfield  {journal} {\bibinfo  {journal} {Nature}\ }\textbf {\bibinfo {volume} {579}},\ \bibinfo {pages} {56–61} (\bibinfo {year} {2020}{\natexlab{a}})}\BibitemShut {NoStop}%
\bibitem [{\citenamefont {Polshyn}\ \emph {et~al.}(2020)\citenamefont {Polshyn}, \citenamefont {Zhu}, \citenamefont {Kumar}, \citenamefont {Zhang}, \citenamefont {Yang}, \citenamefont {Tschirhart}, \citenamefont {Serlin}, \citenamefont {Watanabe}, \citenamefont {Taniguchi}, \citenamefont {MacDonald},\ and\ \citenamefont {Young}}]{polshyn-nat20}%
  \BibitemOpen
  \bibfield  {author} {\bibinfo {author} {\bibfnamefont {H.}~\bibnamefont {Polshyn}}, \bibinfo {author} {\bibfnamefont {J.}~\bibnamefont {Zhu}}, \bibinfo {author} {\bibfnamefont {M.~A.}\ \bibnamefont {Kumar}}, \bibinfo {author} {\bibfnamefont {Y.}~\bibnamefont {Zhang}}, \bibinfo {author} {\bibfnamefont {F.}~\bibnamefont {Yang}}, \bibinfo {author} {\bibfnamefont {C.~L.}\ \bibnamefont {Tschirhart}}, \bibinfo {author} {\bibfnamefont {M.}~\bibnamefont {Serlin}}, \bibinfo {author} {\bibfnamefont {K.}~\bibnamefont {Watanabe}}, \bibinfo {author} {\bibfnamefont {T.}~\bibnamefont {Taniguchi}}, \bibinfo {author} {\bibfnamefont {A.~H.}\ \bibnamefont {MacDonald}},\ and\ \bibinfo {author} {\bibfnamefont {A.~F.}\ \bibnamefont {Young}},\ }\bibfield  {title} {\bibinfo {title} {Electrical switching of magnetic order in an orbital {Chern} insulator},\ }\href {https://doi.org/10.1038/s41586-020-2963-8} {\bibfield  {journal} {\bibinfo  {journal} {Nature}\ }\textbf {\bibinfo {volume} {588}},\ \bibinfo {pages} {66–70} (\bibinfo
  {year} {2020})}\BibitemShut {NoStop}%
\bibitem [{\citenamefont {Tschirhart}\ \emph {et~al.}(2021)\citenamefont {Tschirhart}, \citenamefont {Serlin}, \citenamefont {Polshyn}, \citenamefont {Shragai}, \citenamefont {Xia}, \citenamefont {Zhu}, \citenamefont {Zhang}, \citenamefont {Watanabe}, \citenamefont {Taniguchi}, \citenamefont {Huber},\ and\ \citenamefont {Young}}]{tschirhart-sci21}%
  \BibitemOpen
  \bibfield  {author} {\bibinfo {author} {\bibfnamefont {C.~L.}\ \bibnamefont {Tschirhart}}, \bibinfo {author} {\bibfnamefont {M.}~\bibnamefont {Serlin}}, \bibinfo {author} {\bibfnamefont {H.}~\bibnamefont {Polshyn}}, \bibinfo {author} {\bibfnamefont {A.}~\bibnamefont {Shragai}}, \bibinfo {author} {\bibfnamefont {Z.}~\bibnamefont {Xia}}, \bibinfo {author} {\bibfnamefont {J.}~\bibnamefont {Zhu}}, \bibinfo {author} {\bibfnamefont {Y.}~\bibnamefont {Zhang}}, \bibinfo {author} {\bibfnamefont {K.}~\bibnamefont {Watanabe}}, \bibinfo {author} {\bibfnamefont {T.}~\bibnamefont {Taniguchi}}, \bibinfo {author} {\bibfnamefont {M.~E.}\ \bibnamefont {Huber}},\ and\ \bibinfo {author} {\bibfnamefont {A.~F.}\ \bibnamefont {Young}},\ }\bibfield  {title} {\bibinfo {title} {Imaging orbital ferromagnetism in a moiré {Chern} insulator},\ }\href {https://doi.org/10.1126/science.abd3190} {\bibfield  {journal} {\bibinfo  {journal} {Science}\ }\textbf {\bibinfo {volume} {372}},\ \bibinfo {pages} {1323} (\bibinfo {year}
  {2021})}\BibitemShut {NoStop}%
\bibitem [{\citenamefont {Xie}\ \emph {et~al.}(2021)\citenamefont {Xie}, \citenamefont {Pierce}, \citenamefont {Park}, \citenamefont {Parker}, \citenamefont {Khalaf}, \citenamefont {Ledwith}, \citenamefont {Cao}, \citenamefont {Lee}, \citenamefont {Chen}, \citenamefont {Forrester}, \citenamefont {Watanabe}, \citenamefont {Taniguchi}, \citenamefont {Vishwanath}, \citenamefont {Jarillo-Herrero},\ and\ \citenamefont {Yacoby}}]{xie-nat21}%
  \BibitemOpen
  \bibfield  {author} {\bibinfo {author} {\bibfnamefont {Y.}~\bibnamefont {Xie}}, \bibinfo {author} {\bibfnamefont {A.~T.}\ \bibnamefont {Pierce}}, \bibinfo {author} {\bibfnamefont {J.~M.}\ \bibnamefont {Park}}, \bibinfo {author} {\bibfnamefont {D.~E.}\ \bibnamefont {Parker}}, \bibinfo {author} {\bibfnamefont {E.}~\bibnamefont {Khalaf}}, \bibinfo {author} {\bibfnamefont {P.}~\bibnamefont {Ledwith}}, \bibinfo {author} {\bibfnamefont {Y.}~\bibnamefont {Cao}}, \bibinfo {author} {\bibfnamefont {S.~H.}\ \bibnamefont {Lee}}, \bibinfo {author} {\bibfnamefont {S.}~\bibnamefont {Chen}}, \bibinfo {author} {\bibfnamefont {P.~R.}\ \bibnamefont {Forrester}}, \bibinfo {author} {\bibfnamefont {K.}~\bibnamefont {Watanabe}}, \bibinfo {author} {\bibfnamefont {T.}~\bibnamefont {Taniguchi}}, \bibinfo {author} {\bibfnamefont {A.}~\bibnamefont {Vishwanath}}, \bibinfo {author} {\bibfnamefont {P.}~\bibnamefont {Jarillo-Herrero}},\ and\ \bibinfo {author} {\bibfnamefont {A.}~\bibnamefont {Yacoby}},\ }\bibfield  {title} {\bibinfo
  {title} {Fractional {Chern} insulators in magic-angle twisted bilayer graphene},\ }\href {https://doi.org/10.1038/s41586-021-04002-3} {\bibfield  {journal} {\bibinfo  {journal} {Nature}\ }\textbf {\bibinfo {volume} {600}},\ \bibinfo {pages} {439–443} (\bibinfo {year} {2021})}\BibitemShut {NoStop}%
\bibitem [{\citenamefont {Xu}\ \emph {et~al.}(2023)\citenamefont {Xu}, \citenamefont {Sun}, \citenamefont {Jia}, \citenamefont {Liu}, \citenamefont {Xu}, \citenamefont {Li}, \citenamefont {Gu}, \citenamefont {Watanabe}, \citenamefont {Taniguchi}, \citenamefont {Tong}, \citenamefont {Jia}, \citenamefont {Shi}, \citenamefont {Jiang}, \citenamefont {Zhang}, \citenamefont {Liu},\ and\ \citenamefont {Li}}]{xu-prx23}%
  \BibitemOpen
  \bibfield  {author} {\bibinfo {author} {\bibfnamefont {F.}~\bibnamefont {Xu}}, \bibinfo {author} {\bibfnamefont {Z.}~\bibnamefont {Sun}}, \bibinfo {author} {\bibfnamefont {T.}~\bibnamefont {Jia}}, \bibinfo {author} {\bibfnamefont {C.}~\bibnamefont {Liu}}, \bibinfo {author} {\bibfnamefont {C.}~\bibnamefont {Xu}}, \bibinfo {author} {\bibfnamefont {C.}~\bibnamefont {Li}}, \bibinfo {author} {\bibfnamefont {Y.}~\bibnamefont {Gu}}, \bibinfo {author} {\bibfnamefont {K.}~\bibnamefont {Watanabe}}, \bibinfo {author} {\bibfnamefont {T.}~\bibnamefont {Taniguchi}}, \bibinfo {author} {\bibfnamefont {B.}~\bibnamefont {Tong}}, \bibinfo {author} {\bibfnamefont {J.}~\bibnamefont {Jia}}, \bibinfo {author} {\bibfnamefont {Z.}~\bibnamefont {Shi}}, \bibinfo {author} {\bibfnamefont {S.}~\bibnamefont {Jiang}}, \bibinfo {author} {\bibfnamefont {Y.}~\bibnamefont {Zhang}}, \bibinfo {author} {\bibfnamefont {X.}~\bibnamefont {Liu}},\ and\ \bibinfo {author} {\bibfnamefont {T.}~\bibnamefont {Li}},\ }\bibfield  {title} {\bibinfo {title}
  {Observation of integer and fractional quantum anomalous {Hall} effects in twisted bilayer {MoTe$_2$}},\ }\href {https://doi.org/10.1103/PhysRevX.13.031037} {\bibfield  {journal} {\bibinfo  {journal} {Phys. Rev. X}\ }\textbf {\bibinfo {volume} {13}},\ \bibinfo {pages} {031037} (\bibinfo {year} {2023})}\BibitemShut {NoStop}%
\bibitem [{\citenamefont {Cai}\ \emph {et~al.}(2023)\citenamefont {Cai}, \citenamefont {Anderson}, \citenamefont {Wang}, \citenamefont {Zhang}, \citenamefont {Liu}, \citenamefont {Holtzmann}, \citenamefont {Zhang}, \citenamefont {Fan}, \citenamefont {Taniguchi}, \citenamefont {Watanabe}, \citenamefont {Ran}, \citenamefont {Cao}, \citenamefont {Fu}, \citenamefont {Xiao}, \citenamefont {Yao},\ and\ \citenamefont {Xu}}]{cai-nat23}%
  \BibitemOpen
  \bibfield  {author} {\bibinfo {author} {\bibfnamefont {J.}~\bibnamefont {Cai}}, \bibinfo {author} {\bibfnamefont {E.}~\bibnamefont {Anderson}}, \bibinfo {author} {\bibfnamefont {C.}~\bibnamefont {Wang}}, \bibinfo {author} {\bibfnamefont {X.}~\bibnamefont {Zhang}}, \bibinfo {author} {\bibfnamefont {X.}~\bibnamefont {Liu}}, \bibinfo {author} {\bibfnamefont {W.}~\bibnamefont {Holtzmann}}, \bibinfo {author} {\bibfnamefont {Y.}~\bibnamefont {Zhang}}, \bibinfo {author} {\bibfnamefont {F.}~\bibnamefont {Fan}}, \bibinfo {author} {\bibfnamefont {T.}~\bibnamefont {Taniguchi}}, \bibinfo {author} {\bibfnamefont {K.}~\bibnamefont {Watanabe}}, \bibinfo {author} {\bibfnamefont {Y.}~\bibnamefont {Ran}}, \bibinfo {author} {\bibfnamefont {T.}~\bibnamefont {Cao}}, \bibinfo {author} {\bibfnamefont {L.}~\bibnamefont {Fu}}, \bibinfo {author} {\bibfnamefont {D.}~\bibnamefont {Xiao}}, \bibinfo {author} {\bibfnamefont {W.}~\bibnamefont {Yao}},\ and\ \bibinfo {author} {\bibfnamefont {X.}~\bibnamefont {Xu}},\ }\bibfield  {title}
  {\bibinfo {title} {Signatures of fractional quantum anomalous {Hall} states in twisted {MoTe$_2$}},\ }\href {https://doi.org/10.1038/s41586-023-06289-w} {\bibfield  {journal} {\bibinfo  {journal} {Nature}\ }\textbf {\bibinfo {volume} {622}},\ \bibinfo {pages} {63–68} (\bibinfo {year} {2023})}\BibitemShut {NoStop}%
\bibitem [{\citenamefont {Park}\ \emph {et~al.}(2023)\citenamefont {Park}, \citenamefont {Cai}, \citenamefont {Anderson}, \citenamefont {Zhang}, \citenamefont {Zhu}, \citenamefont {Liu}, \citenamefont {Wang}, \citenamefont {Holtzmann}, \citenamefont {Hu}, \citenamefont {Liu}, \citenamefont {Taniguchi}, \citenamefont {Watanabe}, \citenamefont {Chu}, \citenamefont {Cao}, \citenamefont {Fu}, \citenamefont {Yao}, \citenamefont {Chang}, \citenamefont {Cobden}, \citenamefont {Xiao},\ and\ \citenamefont {Xu}}]{park-nat23}%
  \BibitemOpen
  \bibfield  {author} {\bibinfo {author} {\bibfnamefont {H.}~\bibnamefont {Park}}, \bibinfo {author} {\bibfnamefont {J.}~\bibnamefont {Cai}}, \bibinfo {author} {\bibfnamefont {E.}~\bibnamefont {Anderson}}, \bibinfo {author} {\bibfnamefont {Y.}~\bibnamefont {Zhang}}, \bibinfo {author} {\bibfnamefont {J.}~\bibnamefont {Zhu}}, \bibinfo {author} {\bibfnamefont {X.}~\bibnamefont {Liu}}, \bibinfo {author} {\bibfnamefont {C.}~\bibnamefont {Wang}}, \bibinfo {author} {\bibfnamefont {W.}~\bibnamefont {Holtzmann}}, \bibinfo {author} {\bibfnamefont {C.}~\bibnamefont {Hu}}, \bibinfo {author} {\bibfnamefont {Z.}~\bibnamefont {Liu}}, \bibinfo {author} {\bibfnamefont {T.}~\bibnamefont {Taniguchi}}, \bibinfo {author} {\bibfnamefont {K.}~\bibnamefont {Watanabe}}, \bibinfo {author} {\bibfnamefont {J.-H.}\ \bibnamefont {Chu}}, \bibinfo {author} {\bibfnamefont {T.}~\bibnamefont {Cao}}, \bibinfo {author} {\bibfnamefont {L.}~\bibnamefont {Fu}}, \bibinfo {author} {\bibfnamefont {W.}~\bibnamefont {Yao}}, \bibinfo {author}
  {\bibfnamefont {C.-Z.}\ \bibnamefont {Chang}}, \bibinfo {author} {\bibfnamefont {D.}~\bibnamefont {Cobden}}, \bibinfo {author} {\bibfnamefont {D.}~\bibnamefont {Xiao}},\ and\ \bibinfo {author} {\bibfnamefont {X.}~\bibnamefont {Xu}},\ }\bibfield  {title} {\bibinfo {title} {Observation of fractionally quantized anomalous {Hall} effect},\ }\href {https://doi.org/10.1038/s41586-023-06536-0} {\bibfield  {journal} {\bibinfo  {journal} {Nature}\ }\textbf {\bibinfo {volume} {622}},\ \bibinfo {pages} {74–79} (\bibinfo {year} {2023})}\BibitemShut {NoStop}%
\bibitem [{\citenamefont {Zeng}\ \emph {et~al.}(2023)\citenamefont {Zeng}, \citenamefont {Xia}, \citenamefont {Kang}, \citenamefont {Zhu}, \citenamefont {Kn\"{u}ppel}, \citenamefont {Vaswani}, \citenamefont {Watanabe}, \citenamefont {Taniguchi}, \citenamefont {Mak},\ and\ \citenamefont {Shan}}]{zeng-nat23}%
  \BibitemOpen
  \bibfield  {author} {\bibinfo {author} {\bibfnamefont {Y.}~\bibnamefont {Zeng}}, \bibinfo {author} {\bibfnamefont {Z.}~\bibnamefont {Xia}}, \bibinfo {author} {\bibfnamefont {K.}~\bibnamefont {Kang}}, \bibinfo {author} {\bibfnamefont {J.}~\bibnamefont {Zhu}}, \bibinfo {author} {\bibfnamefont {P.}~\bibnamefont {Kn\"{u}ppel}}, \bibinfo {author} {\bibfnamefont {C.}~\bibnamefont {Vaswani}}, \bibinfo {author} {\bibfnamefont {K.}~\bibnamefont {Watanabe}}, \bibinfo {author} {\bibfnamefont {T.}~\bibnamefont {Taniguchi}}, \bibinfo {author} {\bibfnamefont {K.~F.}\ \bibnamefont {Mak}},\ and\ \bibinfo {author} {\bibfnamefont {J.}~\bibnamefont {Shan}},\ }\bibfield  {title} {\bibinfo {title} {Thermodynamic evidence of fractional {Chern} insulator in moiré {MoTe$_2$}},\ }\href {https://doi.org/10.1038/s41586-023-06452-3} {\bibfield  {journal} {\bibinfo  {journal} {Nature}\ }\textbf {\bibinfo {volume} {622}},\ \bibinfo {pages} {69–73} (\bibinfo {year} {2023})}\BibitemShut {NoStop}%
\bibitem [{\citenamefont {Lu}\ \emph {et~al.}(2024)\citenamefont {Lu}, \citenamefont {Han}, \citenamefont {Yao}, \citenamefont {Reddy}, \citenamefont {Yang}, \citenamefont {Seo}, \citenamefont {Watanabe}, \citenamefont {Taniguchi}, \citenamefont {Fu},\ and\ \citenamefont {Ju}}]{lu-nat24}%
  \BibitemOpen
  \bibfield  {author} {\bibinfo {author} {\bibfnamefont {Z.}~\bibnamefont {Lu}}, \bibinfo {author} {\bibfnamefont {T.}~\bibnamefont {Han}}, \bibinfo {author} {\bibfnamefont {Y.}~\bibnamefont {Yao}}, \bibinfo {author} {\bibfnamefont {A.~P.}\ \bibnamefont {Reddy}}, \bibinfo {author} {\bibfnamefont {J.}~\bibnamefont {Yang}}, \bibinfo {author} {\bibfnamefont {J.}~\bibnamefont {Seo}}, \bibinfo {author} {\bibfnamefont {K.}~\bibnamefont {Watanabe}}, \bibinfo {author} {\bibfnamefont {T.}~\bibnamefont {Taniguchi}}, \bibinfo {author} {\bibfnamefont {L.}~\bibnamefont {Fu}},\ and\ \bibinfo {author} {\bibfnamefont {L.}~\bibnamefont {Ju}},\ }\bibfield  {title} {\bibinfo {title} {Fractional quantum anomalous {Hall} effect in multilayer graphene},\ }\href {https://doi.org/10.1038/s41586-023-07010-7} {\bibfield  {journal} {\bibinfo  {journal} {Nature}\ }\textbf {\bibinfo {volume} {626}},\ \bibinfo {pages} {759–764} (\bibinfo {year} {2024})}\BibitemShut {NoStop}%
\bibitem [{\citenamefont {Liu}\ \emph {et~al.}(2020)\citenamefont {Liu}, \citenamefont {Hao}, \citenamefont {Khalaf}, \citenamefont {Lee}, \citenamefont {Ronen}, \citenamefont {Yoo}, \citenamefont {Haei~Najafabadi}, \citenamefont {Watanabe}, \citenamefont {Taniguchi}, \citenamefont {Vishwanath},\ and\ \citenamefont {Kim}}]{liu-nat20}%
  \BibitemOpen
  \bibfield  {author} {\bibinfo {author} {\bibfnamefont {X.}~\bibnamefont {Liu}}, \bibinfo {author} {\bibfnamefont {Z.}~\bibnamefont {Hao}}, \bibinfo {author} {\bibfnamefont {E.}~\bibnamefont {Khalaf}}, \bibinfo {author} {\bibfnamefont {J.~Y.}\ \bibnamefont {Lee}}, \bibinfo {author} {\bibfnamefont {Y.}~\bibnamefont {Ronen}}, \bibinfo {author} {\bibfnamefont {H.}~\bibnamefont {Yoo}}, \bibinfo {author} {\bibfnamefont {D.}~\bibnamefont {Haei~Najafabadi}}, \bibinfo {author} {\bibfnamefont {K.}~\bibnamefont {Watanabe}}, \bibinfo {author} {\bibfnamefont {T.}~\bibnamefont {Taniguchi}}, \bibinfo {author} {\bibfnamefont {A.}~\bibnamefont {Vishwanath}},\ and\ \bibinfo {author} {\bibfnamefont {P.}~\bibnamefont {Kim}},\ }\bibfield  {title} {\bibinfo {title} {Tunable spin-polarized correlated states in twisted double bilayer graphene},\ }\href {https://doi.org/10.1038/s41586-020-2458-7} {\bibfield  {journal} {\bibinfo  {journal} {Nature}\ }\textbf {\bibinfo {volume} {583}},\ \bibinfo {pages} {221–225} (\bibinfo {year}
  {2020})}\BibitemShut {NoStop}%
\bibitem [{\citenamefont {Burg}\ \emph {et~al.}(2019)\citenamefont {Burg}, \citenamefont {Zhu}, \citenamefont {Taniguchi}, \citenamefont {Watanabe}, \citenamefont {MacDonald},\ and\ \citenamefont {Tutuc}}]{burg-prl19}%
  \BibitemOpen
  \bibfield  {author} {\bibinfo {author} {\bibfnamefont {G.~W.}\ \bibnamefont {Burg}}, \bibinfo {author} {\bibfnamefont {J.}~\bibnamefont {Zhu}}, \bibinfo {author} {\bibfnamefont {T.}~\bibnamefont {Taniguchi}}, \bibinfo {author} {\bibfnamefont {K.}~\bibnamefont {Watanabe}}, \bibinfo {author} {\bibfnamefont {A.~H.}\ \bibnamefont {MacDonald}},\ and\ \bibinfo {author} {\bibfnamefont {E.}~\bibnamefont {Tutuc}},\ }\bibfield  {title} {\bibinfo {title} {Correlated insulating states in twisted double bilayer graphene},\ }\href {https://doi.org/10.1103/PhysRevLett.123.197702} {\bibfield  {journal} {\bibinfo  {journal} {Phys. Rev. Lett.}\ }\textbf {\bibinfo {volume} {123}},\ \bibinfo {pages} {197702} (\bibinfo {year} {2019})}\BibitemShut {NoStop}%
\bibitem [{\citenamefont {He}\ \emph {et~al.}(2020)\citenamefont {He}, \citenamefont {Li}, \citenamefont {Cai}, \citenamefont {Liu}, \citenamefont {Watanabe}, \citenamefont {Taniguchi}, \citenamefont {Xu},\ and\ \citenamefont {Yankowitz}}]{he-natphys20}%
  \BibitemOpen
  \bibfield  {author} {\bibinfo {author} {\bibfnamefont {M.}~\bibnamefont {He}}, \bibinfo {author} {\bibfnamefont {Y.}~\bibnamefont {Li}}, \bibinfo {author} {\bibfnamefont {J.}~\bibnamefont {Cai}}, \bibinfo {author} {\bibfnamefont {Y.}~\bibnamefont {Liu}}, \bibinfo {author} {\bibfnamefont {K.}~\bibnamefont {Watanabe}}, \bibinfo {author} {\bibfnamefont {T.}~\bibnamefont {Taniguchi}}, \bibinfo {author} {\bibfnamefont {X.}~\bibnamefont {Xu}},\ and\ \bibinfo {author} {\bibfnamefont {M.}~\bibnamefont {Yankowitz}},\ }\bibfield  {title} {\bibinfo {title} {Symmetry breaking in twisted double bilayer graphene},\ }\href {https://doi.org/10.1038/s41567-020-1030-6} {\bibfield  {journal} {\bibinfo  {journal} {Nature Physics}\ }\textbf {\bibinfo {volume} {17}},\ \bibinfo {pages} {26–30} (\bibinfo {year} {2020})}\BibitemShut {NoStop}%
\bibitem [{\citenamefont {Shen}\ \emph {et~al.}(2020)\citenamefont {Shen}, \citenamefont {Chu}, \citenamefont {Wu}, \citenamefont {Li}, \citenamefont {Wang}, \citenamefont {Zhao}, \citenamefont {Tang}, \citenamefont {Liu}, \citenamefont {Tian}, \citenamefont {Watanabe}, \citenamefont {Taniguchi}, \citenamefont {Yang}, \citenamefont {Meng}, \citenamefont {Shi}, \citenamefont {Yazyev},\ and\ \citenamefont {Zhang}}]{shen-natphys20}%
  \BibitemOpen
  \bibfield  {author} {\bibinfo {author} {\bibfnamefont {C.}~\bibnamefont {Shen}}, \bibinfo {author} {\bibfnamefont {Y.}~\bibnamefont {Chu}}, \bibinfo {author} {\bibfnamefont {Q.}~\bibnamefont {Wu}}, \bibinfo {author} {\bibfnamefont {N.}~\bibnamefont {Li}}, \bibinfo {author} {\bibfnamefont {S.}~\bibnamefont {Wang}}, \bibinfo {author} {\bibfnamefont {Y.}~\bibnamefont {Zhao}}, \bibinfo {author} {\bibfnamefont {J.}~\bibnamefont {Tang}}, \bibinfo {author} {\bibfnamefont {J.}~\bibnamefont {Liu}}, \bibinfo {author} {\bibfnamefont {J.}~\bibnamefont {Tian}}, \bibinfo {author} {\bibfnamefont {K.}~\bibnamefont {Watanabe}}, \bibinfo {author} {\bibfnamefont {T.}~\bibnamefont {Taniguchi}}, \bibinfo {author} {\bibfnamefont {R.}~\bibnamefont {Yang}}, \bibinfo {author} {\bibfnamefont {Z.~Y.}\ \bibnamefont {Meng}}, \bibinfo {author} {\bibfnamefont {D.}~\bibnamefont {Shi}}, \bibinfo {author} {\bibfnamefont {O.~V.}\ \bibnamefont {Yazyev}},\ and\ \bibinfo {author} {\bibfnamefont {G.}~\bibnamefont {Zhang}},\ }\bibfield  {title}
  {\bibinfo {title} {Correlated states in twisted double bilayer graphene},\ }\href {https://doi.org/10.1038/s41567-020-0825-9} {\bibfield  {journal} {\bibinfo  {journal} {Nature Physics}\ }\textbf {\bibinfo {volume} {16}},\ \bibinfo {pages} {520–525} (\bibinfo {year} {2020})}\BibitemShut {NoStop}%
\bibitem [{\citenamefont {Cao}\ \emph {et~al.}(2020)\citenamefont {Cao}, \citenamefont {Rodan-Legrain}, \citenamefont {Rubies-Bigorda}, \citenamefont {Park}, \citenamefont {Watanabe}, \citenamefont {Taniguchi},\ and\ \citenamefont {Jarillo-Herrero}}]{cao-nat20}%
  \BibitemOpen
  \bibfield  {author} {\bibinfo {author} {\bibfnamefont {Y.}~\bibnamefont {Cao}}, \bibinfo {author} {\bibfnamefont {D.}~\bibnamefont {Rodan-Legrain}}, \bibinfo {author} {\bibfnamefont {O.}~\bibnamefont {Rubies-Bigorda}}, \bibinfo {author} {\bibfnamefont {J.~M.}\ \bibnamefont {Park}}, \bibinfo {author} {\bibfnamefont {K.}~\bibnamefont {Watanabe}}, \bibinfo {author} {\bibfnamefont {T.}~\bibnamefont {Taniguchi}},\ and\ \bibinfo {author} {\bibfnamefont {P.}~\bibnamefont {Jarillo-Herrero}},\ }\bibfield  {title} {\bibinfo {title} {Tunable correlated states and spin-polarized phases in twisted bilayer–bilayer graphene},\ }\href {https://doi.org/10.1038/s41586-020-2260-6} {\bibfield  {journal} {\bibinfo  {journal} {Nature}\ }\textbf {\bibinfo {volume} {583}},\ \bibinfo {pages} {215–220} (\bibinfo {year} {2020})}\BibitemShut {NoStop}%
\bibitem [{\citenamefont {Park}\ \emph {et~al.}(2021)\citenamefont {Park}, \citenamefont {Cao}, \citenamefont {Watanabe}, \citenamefont {Taniguchi},\ and\ \citenamefont {Jarillo-Herrero}}]{park-nat21}%
  \BibitemOpen
  \bibfield  {author} {\bibinfo {author} {\bibfnamefont {J.~M.}\ \bibnamefont {Park}}, \bibinfo {author} {\bibfnamefont {Y.}~\bibnamefont {Cao}}, \bibinfo {author} {\bibfnamefont {K.}~\bibnamefont {Watanabe}}, \bibinfo {author} {\bibfnamefont {T.}~\bibnamefont {Taniguchi}},\ and\ \bibinfo {author} {\bibfnamefont {P.}~\bibnamefont {Jarillo-Herrero}},\ }\bibfield  {title} {\bibinfo {title} {Tunable strongly coupled superconductivity in magic-angle twisted trilayer graphene},\ }\href {https://doi.org/10.1038/s41586-021-03192-0} {\bibfield  {journal} {\bibinfo  {journal} {Nature}\ }\textbf {\bibinfo {volume} {590}},\ \bibinfo {pages} {249–255} (\bibinfo {year} {2021})}\BibitemShut {NoStop}%
\bibitem [{\citenamefont {Xu}\ \emph {et~al.}(2021{\natexlab{a}})\citenamefont {Xu}, \citenamefont {Al~Ezzi}, \citenamefont {Balakrishnan}, \citenamefont {Garcia-Ruiz}, \citenamefont {Tsim}, \citenamefont {Mullan}, \citenamefont {Barrier}, \citenamefont {Xin}, \citenamefont {Piot}, \citenamefont {Taniguchi}, \citenamefont {Watanabe}, \citenamefont {Carvalho}, \citenamefont {Mishchenko}, \citenamefont {Geim}, \citenamefont {Fal’ko}, \citenamefont {Adam}, \citenamefont {Neto}, \citenamefont {Novoselov},\ and\ \citenamefont {Shi}}]{xu-natphys21}%
  \BibitemOpen
  \bibfield  {author} {\bibinfo {author} {\bibfnamefont {S.}~\bibnamefont {Xu}}, \bibinfo {author} {\bibfnamefont {M.~M.}\ \bibnamefont {Al~Ezzi}}, \bibinfo {author} {\bibfnamefont {N.}~\bibnamefont {Balakrishnan}}, \bibinfo {author} {\bibfnamefont {A.}~\bibnamefont {Garcia-Ruiz}}, \bibinfo {author} {\bibfnamefont {B.}~\bibnamefont {Tsim}}, \bibinfo {author} {\bibfnamefont {C.}~\bibnamefont {Mullan}}, \bibinfo {author} {\bibfnamefont {J.}~\bibnamefont {Barrier}}, \bibinfo {author} {\bibfnamefont {N.}~\bibnamefont {Xin}}, \bibinfo {author} {\bibfnamefont {B.~A.}\ \bibnamefont {Piot}}, \bibinfo {author} {\bibfnamefont {T.}~\bibnamefont {Taniguchi}}, \bibinfo {author} {\bibfnamefont {K.}~\bibnamefont {Watanabe}}, \bibinfo {author} {\bibfnamefont {A.}~\bibnamefont {Carvalho}}, \bibinfo {author} {\bibfnamefont {A.}~\bibnamefont {Mishchenko}}, \bibinfo {author} {\bibfnamefont {A.~K.}\ \bibnamefont {Geim}}, \bibinfo {author} {\bibfnamefont {V.~I.}\ \bibnamefont {Fal’ko}}, \bibinfo {author} {\bibfnamefont
  {S.}~\bibnamefont {Adam}}, \bibinfo {author} {\bibfnamefont {A.~H.~C.}\ \bibnamefont {Neto}}, \bibinfo {author} {\bibfnamefont {K.~S.}\ \bibnamefont {Novoselov}},\ and\ \bibinfo {author} {\bibfnamefont {Y.}~\bibnamefont {Shi}},\ }\bibfield  {title} {\bibinfo {title} {Tunable van {Hove} singularities and correlated states in twisted monolayer–bilayer graphene},\ }\href {https://doi.org/10.1038/s41567-021-01172-9} {\bibfield  {journal} {\bibinfo  {journal} {Nature Physics}\ }\textbf {\bibinfo {volume} {17}},\ \bibinfo {pages} {619–626} (\bibinfo {year} {2021}{\natexlab{a}})}\BibitemShut {NoStop}%
\bibitem [{\citenamefont {Chen}\ \emph {et~al.}(2020{\natexlab{b}})\citenamefont {Chen}, \citenamefont {He}, \citenamefont {Zhang}, \citenamefont {Hsieh}, \citenamefont {Fei}, \citenamefont {Watanabe}, \citenamefont {Taniguchi}, \citenamefont {Cobden}, \citenamefont {Xu}, \citenamefont {Dean},\ and\ \citenamefont {Yankowitz}}]{chen-natphys20}%
  \BibitemOpen
  \bibfield  {author} {\bibinfo {author} {\bibfnamefont {S.}~\bibnamefont {Chen}}, \bibinfo {author} {\bibfnamefont {M.}~\bibnamefont {He}}, \bibinfo {author} {\bibfnamefont {Y.-H.}\ \bibnamefont {Zhang}}, \bibinfo {author} {\bibfnamefont {V.}~\bibnamefont {Hsieh}}, \bibinfo {author} {\bibfnamefont {Z.}~\bibnamefont {Fei}}, \bibinfo {author} {\bibfnamefont {K.}~\bibnamefont {Watanabe}}, \bibinfo {author} {\bibfnamefont {T.}~\bibnamefont {Taniguchi}}, \bibinfo {author} {\bibfnamefont {D.~H.}\ \bibnamefont {Cobden}}, \bibinfo {author} {\bibfnamefont {X.}~\bibnamefont {Xu}}, \bibinfo {author} {\bibfnamefont {C.~R.}\ \bibnamefont {Dean}},\ and\ \bibinfo {author} {\bibfnamefont {M.}~\bibnamefont {Yankowitz}},\ }\bibfield  {title} {\bibinfo {title} {Electrically tunable correlated and topological states in twisted monolayer–bilayer graphene},\ }\href {https://doi.org/10.1038/s41567-020-01062-6} {\bibfield  {journal} {\bibinfo  {journal} {Nature Physics}\ }\textbf {\bibinfo {volume} {17}},\ \bibinfo {pages}
  {374–380} (\bibinfo {year} {2020}{\natexlab{b}})}\BibitemShut {NoStop}%
\bibitem [{\citenamefont {Hao}\ \emph {et~al.}(2021)\citenamefont {Hao}, \citenamefont {Zimmerman}, \citenamefont {Ledwith}, \citenamefont {Khalaf}, \citenamefont {Najafabadi}, \citenamefont {Watanabe}, \citenamefont {Taniguchi}, \citenamefont {Vishwanath},\ and\ \citenamefont {Kim}}]{hao-sci21}%
  \BibitemOpen
  \bibfield  {author} {\bibinfo {author} {\bibfnamefont {Z.}~\bibnamefont {Hao}}, \bibinfo {author} {\bibfnamefont {A.~M.}\ \bibnamefont {Zimmerman}}, \bibinfo {author} {\bibfnamefont {P.}~\bibnamefont {Ledwith}}, \bibinfo {author} {\bibfnamefont {E.}~\bibnamefont {Khalaf}}, \bibinfo {author} {\bibfnamefont {D.~H.}\ \bibnamefont {Najafabadi}}, \bibinfo {author} {\bibfnamefont {K.}~\bibnamefont {Watanabe}}, \bibinfo {author} {\bibfnamefont {T.}~\bibnamefont {Taniguchi}}, \bibinfo {author} {\bibfnamefont {A.}~\bibnamefont {Vishwanath}},\ and\ \bibinfo {author} {\bibfnamefont {P.}~\bibnamefont {Kim}},\ }\bibfield  {title} {\bibinfo {title} {Electric field–tunable superconductivity in alternating-twist magic-angle trilayer graphene},\ }\href {https://doi.org/10.1126/science.abg0399} {\bibfield  {journal} {\bibinfo  {journal} {Science}\ }\textbf {\bibinfo {volume} {371}},\ \bibinfo {pages} {1133} (\bibinfo {year} {2021})}\BibitemShut {NoStop}%
\bibitem [{\citenamefont {Xu}\ \emph {et~al.}(2020)\citenamefont {Xu}, \citenamefont {Liu}, \citenamefont {Rhodes}, \citenamefont {Watanabe}, \citenamefont {Taniguchi}, \citenamefont {Hone}, \citenamefont {Elser}, \citenamefont {Mak},\ and\ \citenamefont {Shan}}]{xu-nat2020}%
  \BibitemOpen
  \bibfield  {author} {\bibinfo {author} {\bibfnamefont {Y.}~\bibnamefont {Xu}}, \bibinfo {author} {\bibfnamefont {S.}~\bibnamefont {Liu}}, \bibinfo {author} {\bibfnamefont {D.~A.}\ \bibnamefont {Rhodes}}, \bibinfo {author} {\bibfnamefont {K.}~\bibnamefont {Watanabe}}, \bibinfo {author} {\bibfnamefont {T.}~\bibnamefont {Taniguchi}}, \bibinfo {author} {\bibfnamefont {J.}~\bibnamefont {Hone}}, \bibinfo {author} {\bibfnamefont {V.}~\bibnamefont {Elser}}, \bibinfo {author} {\bibfnamefont {K.~F.}\ \bibnamefont {Mak}},\ and\ \bibinfo {author} {\bibfnamefont {J.}~\bibnamefont {Shan}},\ }\bibfield  {title} {\bibinfo {title} {Correlated insulating states at fractional fillings of moiré superlattices},\ }\href {https://doi.org/10.1038/s41586-020-2868-6} {\bibfield  {journal} {\bibinfo  {journal} {Nature}\ }\textbf {\bibinfo {volume} {587}},\ \bibinfo {pages} {214–218} (\bibinfo {year} {2020})}\BibitemShut {NoStop}%
\bibitem [{\citenamefont {Tang}\ \emph {et~al.}(2020)\citenamefont {Tang}, \citenamefont {Li}, \citenamefont {Li}, \citenamefont {Xu}, \citenamefont {Liu}, \citenamefont {Barmak}, \citenamefont {Watanabe}, \citenamefont {Taniguchi}, \citenamefont {MacDonald}, \citenamefont {Shan},\ and\ \citenamefont {Mak}}]{tang-nat20}%
  \BibitemOpen
  \bibfield  {author} {\bibinfo {author} {\bibfnamefont {Y.}~\bibnamefont {Tang}}, \bibinfo {author} {\bibfnamefont {L.}~\bibnamefont {Li}}, \bibinfo {author} {\bibfnamefont {T.}~\bibnamefont {Li}}, \bibinfo {author} {\bibfnamefont {Y.}~\bibnamefont {Xu}}, \bibinfo {author} {\bibfnamefont {S.}~\bibnamefont {Liu}}, \bibinfo {author} {\bibfnamefont {K.}~\bibnamefont {Barmak}}, \bibinfo {author} {\bibfnamefont {K.}~\bibnamefont {Watanabe}}, \bibinfo {author} {\bibfnamefont {T.}~\bibnamefont {Taniguchi}}, \bibinfo {author} {\bibfnamefont {A.~H.}\ \bibnamefont {MacDonald}}, \bibinfo {author} {\bibfnamefont {J.}~\bibnamefont {Shan}},\ and\ \bibinfo {author} {\bibfnamefont {K.~F.}\ \bibnamefont {Mak}},\ }\bibfield  {title} {\bibinfo {title} {Simulation of {Hubbard} model physics in {WSe$_2$}/{WS$_2$} moiré superlattices},\ }\href {https://doi.org/10.1038/s41586-020-2085-3} {\bibfield  {journal} {\bibinfo  {journal} {Nature}\ }\textbf {\bibinfo {volume} {579}},\ \bibinfo {pages} {353–358} (\bibinfo {year}
  {2020})}\BibitemShut {NoStop}%
\bibitem [{\citenamefont {Wang}\ \emph {et~al.}(2020)\citenamefont {Wang}, \citenamefont {Shih}, \citenamefont {Ghiotto}, \citenamefont {Xian}, \citenamefont {Rhodes}, \citenamefont {Tan}, \citenamefont {Claassen}, \citenamefont {Kennes}, \citenamefont {Bai}, \citenamefont {Kim}, \citenamefont {Watanabe}, \citenamefont {Taniguchi}, \citenamefont {Zhu}, \citenamefont {Hone}, \citenamefont {Rubio}, \citenamefont {Pasupathy},\ and\ \citenamefont {Dean}}]{wang-natmat20}%
  \BibitemOpen
  \bibfield  {author} {\bibinfo {author} {\bibfnamefont {L.}~\bibnamefont {Wang}}, \bibinfo {author} {\bibfnamefont {E.-M.}\ \bibnamefont {Shih}}, \bibinfo {author} {\bibfnamefont {A.}~\bibnamefont {Ghiotto}}, \bibinfo {author} {\bibfnamefont {L.}~\bibnamefont {Xian}}, \bibinfo {author} {\bibfnamefont {D.~A.}\ \bibnamefont {Rhodes}}, \bibinfo {author} {\bibfnamefont {C.}~\bibnamefont {Tan}}, \bibinfo {author} {\bibfnamefont {M.}~\bibnamefont {Claassen}}, \bibinfo {author} {\bibfnamefont {D.~M.}\ \bibnamefont {Kennes}}, \bibinfo {author} {\bibfnamefont {Y.}~\bibnamefont {Bai}}, \bibinfo {author} {\bibfnamefont {B.}~\bibnamefont {Kim}}, \bibinfo {author} {\bibfnamefont {K.}~\bibnamefont {Watanabe}}, \bibinfo {author} {\bibfnamefont {T.}~\bibnamefont {Taniguchi}}, \bibinfo {author} {\bibfnamefont {X.}~\bibnamefont {Zhu}}, \bibinfo {author} {\bibfnamefont {J.}~\bibnamefont {Hone}}, \bibinfo {author} {\bibfnamefont {A.}~\bibnamefont {Rubio}}, \bibinfo {author} {\bibfnamefont {A.~N.}\ \bibnamefont {Pasupathy}},\
  and\ \bibinfo {author} {\bibfnamefont {C.~R.}\ \bibnamefont {Dean}},\ }\bibfield  {title} {\bibinfo {title} {Correlated electronic phases in twisted bilayer transition metal dichalcogenides},\ }\href {https://doi.org/10.1038/s41563-020-0708-6} {\bibfield  {journal} {\bibinfo  {journal} {Nature Materials}\ }\textbf {\bibinfo {volume} {19}},\ \bibinfo {pages} {861–866} (\bibinfo {year} {2020})}\BibitemShut {NoStop}%
\bibitem [{\citenamefont {Xu}\ \emph {et~al.}(2021{\natexlab{b}})\citenamefont {Xu}, \citenamefont {Ray}, \citenamefont {Shao}, \citenamefont {Jiang}, \citenamefont {Lee}, \citenamefont {Weber}, \citenamefont {Goldberger}, \citenamefont {Watanabe}, \citenamefont {Taniguchi}, \citenamefont {Muller}, \citenamefont {Mak},\ and\ \citenamefont {Shan}}]{xu-natnano21}%
  \BibitemOpen
  \bibfield  {author} {\bibinfo {author} {\bibfnamefont {Y.}~\bibnamefont {Xu}}, \bibinfo {author} {\bibfnamefont {A.}~\bibnamefont {Ray}}, \bibinfo {author} {\bibfnamefont {Y.-T.}\ \bibnamefont {Shao}}, \bibinfo {author} {\bibfnamefont {S.}~\bibnamefont {Jiang}}, \bibinfo {author} {\bibfnamefont {K.}~\bibnamefont {Lee}}, \bibinfo {author} {\bibfnamefont {D.}~\bibnamefont {Weber}}, \bibinfo {author} {\bibfnamefont {J.~E.}\ \bibnamefont {Goldberger}}, \bibinfo {author} {\bibfnamefont {K.}~\bibnamefont {Watanabe}}, \bibinfo {author} {\bibfnamefont {T.}~\bibnamefont {Taniguchi}}, \bibinfo {author} {\bibfnamefont {D.~A.}\ \bibnamefont {Muller}}, \bibinfo {author} {\bibfnamefont {K.~F.}\ \bibnamefont {Mak}},\ and\ \bibinfo {author} {\bibfnamefont {J.}~\bibnamefont {Shan}},\ }\bibfield  {title} {\bibinfo {title} {Coexisting ferromagnetic–antiferromagnetic state in twisted bilayer {CrI$_3$}},\ }\href {https://doi.org/10.1038/s41565-021-01014-y} {\bibfield  {journal} {\bibinfo  {journal} {Nature Nanotechnology}\
  }\textbf {\bibinfo {volume} {17}},\ \bibinfo {pages} {143–147} (\bibinfo {year} {2021}{\natexlab{b}})}\BibitemShut {NoStop}%
\bibitem [{\citenamefont {Song}\ \emph {et~al.}(2021)\citenamefont {Song}, \citenamefont {Sun}, \citenamefont {Anderson}, \citenamefont {Wang}, \citenamefont {Qian}, \citenamefont {Taniguchi}, \citenamefont {Watanabe}, \citenamefont {McGuire}, \citenamefont {Stöhr}, \citenamefont {Xiao}, \citenamefont {Cao}, \citenamefont {Wrachtrup},\ and\ \citenamefont {Xu}}]{song-sci21}%
  \BibitemOpen
  \bibfield  {author} {\bibinfo {author} {\bibfnamefont {T.}~\bibnamefont {Song}}, \bibinfo {author} {\bibfnamefont {Q.-C.}\ \bibnamefont {Sun}}, \bibinfo {author} {\bibfnamefont {E.}~\bibnamefont {Anderson}}, \bibinfo {author} {\bibfnamefont {C.}~\bibnamefont {Wang}}, \bibinfo {author} {\bibfnamefont {J.}~\bibnamefont {Qian}}, \bibinfo {author} {\bibfnamefont {T.}~\bibnamefont {Taniguchi}}, \bibinfo {author} {\bibfnamefont {K.}~\bibnamefont {Watanabe}}, \bibinfo {author} {\bibfnamefont {M.~A.}\ \bibnamefont {McGuire}}, \bibinfo {author} {\bibfnamefont {R.}~\bibnamefont {Stöhr}}, \bibinfo {author} {\bibfnamefont {D.}~\bibnamefont {Xiao}}, \bibinfo {author} {\bibfnamefont {T.}~\bibnamefont {Cao}}, \bibinfo {author} {\bibfnamefont {J.}~\bibnamefont {Wrachtrup}},\ and\ \bibinfo {author} {\bibfnamefont {X.}~\bibnamefont {Xu}},\ }\bibfield  {title} {\bibinfo {title} {Direct visualization of magnetic domains and moiré magnetism in twisted {2D} magnets},\ }\href {https://doi.org/10.1126/science.abj7478} {\bibfield
   {journal} {\bibinfo  {journal} {Science}\ }\textbf {\bibinfo {volume} {374}},\ \bibinfo {pages} {1140} (\bibinfo {year} {2021})}\BibitemShut {NoStop}%
\bibitem [{\citenamefont {Can}\ \emph {et~al.}(2021)\citenamefont {Can}, \citenamefont {Tummuru}, \citenamefont {Day}, \citenamefont {Elfimov}, \citenamefont {Damascelli},\ and\ \citenamefont {Franz}}]{can-natphys21}%
  \BibitemOpen
  \bibfield  {author} {\bibinfo {author} {\bibfnamefont {O.}~\bibnamefont {Can}}, \bibinfo {author} {\bibfnamefont {T.}~\bibnamefont {Tummuru}}, \bibinfo {author} {\bibfnamefont {R.~P.}\ \bibnamefont {Day}}, \bibinfo {author} {\bibfnamefont {I.}~\bibnamefont {Elfimov}}, \bibinfo {author} {\bibfnamefont {A.}~\bibnamefont {Damascelli}},\ and\ \bibinfo {author} {\bibfnamefont {M.}~\bibnamefont {Franz}},\ }\bibfield  {title} {\bibinfo {title} {High-temperature topological superconductivity in twisted double-layer copper oxides},\ }\href {https://doi.org/10.1038/s41567-020-01142-7} {\bibfield  {journal} {\bibinfo  {journal} {Nature Physics}\ }\textbf {\bibinfo {volume} {17}},\ \bibinfo {pages} {519–524} (\bibinfo {year} {2021})}\BibitemShut {NoStop}%
\bibitem [{\citenamefont {Zhao}\ \emph {et~al.}(2023)\citenamefont {Zhao}, \citenamefont {Cui}, \citenamefont {Volkov}, \citenamefont {Yoo}, \citenamefont {Lee}, \citenamefont {Gardener}, \citenamefont {Akey}, \citenamefont {Engelke}, \citenamefont {Ronen}, \citenamefont {Zhong}, \citenamefont {Gu}, \citenamefont {Plugge}, \citenamefont {Tummuru}, \citenamefont {Kim}, \citenamefont {Franz}, \citenamefont {Pixley}, \citenamefont {Poccia},\ and\ \citenamefont {Kim}}]{zhao-sci23}%
  \BibitemOpen
  \bibfield  {author} {\bibinfo {author} {\bibfnamefont {S.~Y.~F.}\ \bibnamefont {Zhao}}, \bibinfo {author} {\bibfnamefont {X.}~\bibnamefont {Cui}}, \bibinfo {author} {\bibfnamefont {P.~A.}\ \bibnamefont {Volkov}}, \bibinfo {author} {\bibfnamefont {H.}~\bibnamefont {Yoo}}, \bibinfo {author} {\bibfnamefont {S.}~\bibnamefont {Lee}}, \bibinfo {author} {\bibfnamefont {J.~A.}\ \bibnamefont {Gardener}}, \bibinfo {author} {\bibfnamefont {A.~J.}\ \bibnamefont {Akey}}, \bibinfo {author} {\bibfnamefont {R.}~\bibnamefont {Engelke}}, \bibinfo {author} {\bibfnamefont {Y.}~\bibnamefont {Ronen}}, \bibinfo {author} {\bibfnamefont {R.}~\bibnamefont {Zhong}}, \bibinfo {author} {\bibfnamefont {G.}~\bibnamefont {Gu}}, \bibinfo {author} {\bibfnamefont {S.}~\bibnamefont {Plugge}}, \bibinfo {author} {\bibfnamefont {T.}~\bibnamefont {Tummuru}}, \bibinfo {author} {\bibfnamefont {M.}~\bibnamefont {Kim}}, \bibinfo {author} {\bibfnamefont {M.}~\bibnamefont {Franz}}, \bibinfo {author} {\bibfnamefont {J.~H.}\ \bibnamefont {Pixley}},
  \bibinfo {author} {\bibfnamefont {N.}~\bibnamefont {Poccia}},\ and\ \bibinfo {author} {\bibfnamefont {P.}~\bibnamefont {Kim}},\ }\bibfield  {title} {\bibinfo {title} {Time-reversal symmetry breaking superconductivity between twisted cuprate superconductors},\ }\href {https://doi.org/10.1126/science.abl8371} {\bibfield  {journal} {\bibinfo  {journal} {Science}\ }\textbf {\bibinfo {volume} {382}},\ \bibinfo {pages} {1422} (\bibinfo {year} {2023})}\BibitemShut {NoStop}%
\bibitem [{\citenamefont {Li}\ \emph {et~al.}(2021{\natexlab{c}})\citenamefont {Li}, \citenamefont {Jiang}, \citenamefont {Shen}, \citenamefont {Zhang}, \citenamefont {Li}, \citenamefont {Tao}, \citenamefont {Devakul}, \citenamefont {Watanabe}, \citenamefont {Taniguchi}, \citenamefont {Fu}, \citenamefont {Shan},\ and\ \citenamefont {Mak}}]{li-nat21}%
  \BibitemOpen
  \bibfield  {author} {\bibinfo {author} {\bibfnamefont {T.}~\bibnamefont {Li}}, \bibinfo {author} {\bibfnamefont {S.}~\bibnamefont {Jiang}}, \bibinfo {author} {\bibfnamefont {B.}~\bibnamefont {Shen}}, \bibinfo {author} {\bibfnamefont {Y.}~\bibnamefont {Zhang}}, \bibinfo {author} {\bibfnamefont {L.}~\bibnamefont {Li}}, \bibinfo {author} {\bibfnamefont {Z.}~\bibnamefont {Tao}}, \bibinfo {author} {\bibfnamefont {T.}~\bibnamefont {Devakul}}, \bibinfo {author} {\bibfnamefont {K.}~\bibnamefont {Watanabe}}, \bibinfo {author} {\bibfnamefont {T.}~\bibnamefont {Taniguchi}}, \bibinfo {author} {\bibfnamefont {L.}~\bibnamefont {Fu}}, \bibinfo {author} {\bibfnamefont {J.}~\bibnamefont {Shan}},\ and\ \bibinfo {author} {\bibfnamefont {K.~F.}\ \bibnamefont {Mak}},\ }\bibfield  {title} {\bibinfo {title} {Quantum anomalous {Hall} effect from intertwined moiré bands},\ }\href {https://doi.org/10.1038/s41586-021-04171-1} {\bibfield  {journal} {\bibinfo  {journal} {Nature}\ }\textbf {\bibinfo {volume} {600}},\ \bibinfo {pages}
  {641–646} (\bibinfo {year} {2021}{\natexlab{c}})}\BibitemShut {NoStop}%
\bibitem [{\citenamefont {Guo}\ \emph {et~al.}(2024)\citenamefont {Guo}, \citenamefont {Pack}, \citenamefont {Swann}, \citenamefont {Holtzman}, \citenamefont {Cothrine}, \citenamefont {Watanabe}, \citenamefont {Taniguchi}, \citenamefont {Mandrus}, \citenamefont {Barmak}, \citenamefont {Hone}, \citenamefont {Millis}, \citenamefont {Pasupathy},\ and\ \citenamefont {Dean}}]{guo-arxiv24}%
  \BibitemOpen
  \bibfield  {author} {\bibinfo {author} {\bibfnamefont {Y.}~\bibnamefont {Guo}}, \bibinfo {author} {\bibfnamefont {J.}~\bibnamefont {Pack}}, \bibinfo {author} {\bibfnamefont {J.}~\bibnamefont {Swann}}, \bibinfo {author} {\bibfnamefont {L.}~\bibnamefont {Holtzman}}, \bibinfo {author} {\bibfnamefont {M.}~\bibnamefont {Cothrine}}, \bibinfo {author} {\bibfnamefont {K.}~\bibnamefont {Watanabe}}, \bibinfo {author} {\bibfnamefont {T.}~\bibnamefont {Taniguchi}}, \bibinfo {author} {\bibfnamefont {D.}~\bibnamefont {Mandrus}}, \bibinfo {author} {\bibfnamefont {K.}~\bibnamefont {Barmak}}, \bibinfo {author} {\bibfnamefont {J.}~\bibnamefont {Hone}}, \bibinfo {author} {\bibfnamefont {A.~J.}\ \bibnamefont {Millis}}, \bibinfo {author} {\bibfnamefont {A.~N.}\ \bibnamefont {Pasupathy}},\ and\ \bibinfo {author} {\bibfnamefont {C.~R.}\ \bibnamefont {Dean}},\ }\href {https://doi.org/10.48550/ARXIV.2406.03418} {\bibinfo {title} {Superconductivity in twisted bilayer {WSe$_2$}}} (\bibinfo {year} {2024})\BibitemShut {NoStop}%
\bibitem [{\citenamefont {Xia}\ \emph {et~al.}(2024)\citenamefont {Xia}, \citenamefont {Han}, \citenamefont {Watanabe}, \citenamefont {Taniguchi}, \citenamefont {Shan},\ and\ \citenamefont {Mak}}]{xia-arxiv24}%
  \BibitemOpen
  \bibfield  {author} {\bibinfo {author} {\bibfnamefont {Y.}~\bibnamefont {Xia}}, \bibinfo {author} {\bibfnamefont {Z.}~\bibnamefont {Han}}, \bibinfo {author} {\bibfnamefont {K.}~\bibnamefont {Watanabe}}, \bibinfo {author} {\bibfnamefont {T.}~\bibnamefont {Taniguchi}}, \bibinfo {author} {\bibfnamefont {J.}~\bibnamefont {Shan}},\ and\ \bibinfo {author} {\bibfnamefont {K.~F.}\ \bibnamefont {Mak}},\ }\href {https://doi.org/10.48550/ARXIV.2405.14784} {\bibinfo {title} {Unconventional superconductivity in twisted bilayer {WSe$_2$}}} (\bibinfo {year} {2024})\BibitemShut {NoStop}%
\bibitem [{\citenamefont {Ji}\ \emph {et~al.}(2024)\citenamefont {Ji}, \citenamefont {Park}, \citenamefont {Barber}, \citenamefont {Hu}, \citenamefont {Watanabe}, \citenamefont {Taniguchi}, \citenamefont {Chu}, \citenamefont {Xu},\ and\ \citenamefont {Shen}}]{ji-arxiv24}%
  \BibitemOpen
  \bibfield  {author} {\bibinfo {author} {\bibfnamefont {Z.}~\bibnamefont {Ji}}, \bibinfo {author} {\bibfnamefont {H.}~\bibnamefont {Park}}, \bibinfo {author} {\bibfnamefont {M.~E.}\ \bibnamefont {Barber}}, \bibinfo {author} {\bibfnamefont {C.}~\bibnamefont {Hu}}, \bibinfo {author} {\bibfnamefont {K.}~\bibnamefont {Watanabe}}, \bibinfo {author} {\bibfnamefont {T.}~\bibnamefont {Taniguchi}}, \bibinfo {author} {\bibfnamefont {J.-H.}\ \bibnamefont {Chu}}, \bibinfo {author} {\bibfnamefont {X.}~\bibnamefont {Xu}},\ and\ \bibinfo {author} {\bibfnamefont {Z.-x.}\ \bibnamefont {Shen}},\ }\href {https://doi.org/10.48550/ARXIV.2404.07157} {\bibinfo {title} {Local probe of bulk and edge states in a fractional {Chern} insulator}} (\bibinfo {year} {2024})\BibitemShut {NoStop}%
\bibitem [{\citenamefont {Redekop}\ \emph {et~al.}(2024)\citenamefont {Redekop}, \citenamefont {Zhang}, \citenamefont {Park}, \citenamefont {Cai}, \citenamefont {Anderson}, \citenamefont {Sheekey}, \citenamefont {Arp}, \citenamefont {Babikyan}, \citenamefont {Salters}, \citenamefont {Watanabe}, \citenamefont {Taniguchi}, \citenamefont {Xu},\ and\ \citenamefont {Young}}]{redekop-arxiv24}%
  \BibitemOpen
  \bibfield  {author} {\bibinfo {author} {\bibfnamefont {E.}~\bibnamefont {Redekop}}, \bibinfo {author} {\bibfnamefont {C.}~\bibnamefont {Zhang}}, \bibinfo {author} {\bibfnamefont {H.}~\bibnamefont {Park}}, \bibinfo {author} {\bibfnamefont {J.}~\bibnamefont {Cai}}, \bibinfo {author} {\bibfnamefont {E.}~\bibnamefont {Anderson}}, \bibinfo {author} {\bibfnamefont {O.}~\bibnamefont {Sheekey}}, \bibinfo {author} {\bibfnamefont {T.}~\bibnamefont {Arp}}, \bibinfo {author} {\bibfnamefont {G.}~\bibnamefont {Babikyan}}, \bibinfo {author} {\bibfnamefont {S.}~\bibnamefont {Salters}}, \bibinfo {author} {\bibfnamefont {K.}~\bibnamefont {Watanabe}}, \bibinfo {author} {\bibfnamefont {T.}~\bibnamefont {Taniguchi}}, \bibinfo {author} {\bibfnamefont {X.}~\bibnamefont {Xu}},\ and\ \bibinfo {author} {\bibfnamefont {A.~F.}\ \bibnamefont {Young}},\ }\href {https://doi.org/10.48550/ARXIV.2405.10269} {\bibinfo {title} {Direct magnetic imaging of fractional {Chern} insulators in twisted {MoTe$_2$} with a superconducting sensor}}
  (\bibinfo {year} {2024})\BibitemShut {NoStop}%
\bibitem [{\citenamefont {Bistritzer}\ and\ \citenamefont {MacDonald}(2011)}]{bistritzer-pnas11}%
  \BibitemOpen
  \bibfield  {author} {\bibinfo {author} {\bibfnamefont {R.}~\bibnamefont {Bistritzer}}\ and\ \bibinfo {author} {\bibfnamefont {A.~H.}\ \bibnamefont {MacDonald}},\ }\bibfield  {title} {\bibinfo {title} {Moiré bands in twisted double-layer graphene},\ }\href {https://doi.org/10.1073/pnas.1108174108} {\bibfield  {journal} {\bibinfo  {journal} {Proceedings of the National Academy of Sciences}\ }\textbf {\bibinfo {volume} {108}},\ \bibinfo {pages} {12233–12237} (\bibinfo {year} {2011})}\BibitemShut {NoStop}%
\bibitem [{\citenamefont {Lau}\ \emph {et~al.}(2022)\citenamefont {Lau}, \citenamefont {Bockrath}, \citenamefont {Mak},\ and\ \citenamefont {Zhang}}]{lau-nat22}%
  \BibitemOpen
  \bibfield  {author} {\bibinfo {author} {\bibfnamefont {C.~N.}\ \bibnamefont {Lau}}, \bibinfo {author} {\bibfnamefont {M.~W.}\ \bibnamefont {Bockrath}}, \bibinfo {author} {\bibfnamefont {K.~F.}\ \bibnamefont {Mak}},\ and\ \bibinfo {author} {\bibfnamefont {F.}~\bibnamefont {Zhang}},\ }\bibfield  {title} {\bibinfo {title} {Reproducibility in the fabrication and physics of moiré materials},\ }\href {https://doi.org/10.1038/s41586-021-04173-z} {\bibfield  {journal} {\bibinfo  {journal} {Nature}\ }\textbf {\bibinfo {volume} {602}},\ \bibinfo {pages} {41–50} (\bibinfo {year} {2022})}\BibitemShut {NoStop}%
\bibitem [{\citenamefont {Ghorashi}\ \emph {et~al.}(2023)\citenamefont {Ghorashi}, \citenamefont {Dunbrack}, \citenamefont {Abouelkomsan}, \citenamefont {Sun}, \citenamefont {Du},\ and\ \citenamefont {Cano}}]{ghorashi-prl23}%
  \BibitemOpen
  \bibfield  {author} {\bibinfo {author} {\bibfnamefont {S.~A.~A.}\ \bibnamefont {Ghorashi}}, \bibinfo {author} {\bibfnamefont {A.}~\bibnamefont {Dunbrack}}, \bibinfo {author} {\bibfnamefont {A.}~\bibnamefont {Abouelkomsan}}, \bibinfo {author} {\bibfnamefont {J.}~\bibnamefont {Sun}}, \bibinfo {author} {\bibfnamefont {X.}~\bibnamefont {Du}},\ and\ \bibinfo {author} {\bibfnamefont {J.}~\bibnamefont {Cano}},\ }\bibfield  {title} {\bibinfo {title} {Topological and stacked flat bands in bilayer graphene with a superlattice potential},\ }\href {https://doi.org/10.1103/PhysRevLett.130.196201} {\bibfield  {journal} {\bibinfo  {journal} {Phys. Rev. Lett.}\ }\textbf {\bibinfo {volume} {130}},\ \bibinfo {pages} {196201} (\bibinfo {year} {2023})}\BibitemShut {NoStop}%
\bibitem [{\citenamefont {Ghorashi}\ and\ \citenamefont {Cano}(2023)}]{ghorashi-prb23}%
  \BibitemOpen
  \bibfield  {author} {\bibinfo {author} {\bibfnamefont {S.~A.~A.}\ \bibnamefont {Ghorashi}}\ and\ \bibinfo {author} {\bibfnamefont {J.}~\bibnamefont {Cano}},\ }\bibfield  {title} {\bibinfo {title} {Multilayer graphene with a superlattice potential},\ }\href {https://doi.org/10.1103/PhysRevB.107.195423} {\bibfield  {journal} {\bibinfo  {journal} {Phys. Rev. B}\ }\textbf {\bibinfo {volume} {107}},\ \bibinfo {pages} {195423} (\bibinfo {year} {2023})}\BibitemShut {NoStop}%
\bibitem [{\citenamefont {Zeng}\ \emph {et~al.}(2024)\citenamefont {Zeng}, \citenamefont {Wolf}, \citenamefont {Huang}, \citenamefont {Wei}, \citenamefont {Ghorashi}, \citenamefont {MacDonald},\ and\ \citenamefont {Cano}}]{zeng-prb24}%
  \BibitemOpen
  \bibfield  {author} {\bibinfo {author} {\bibfnamefont {Y.}~\bibnamefont {Zeng}}, \bibinfo {author} {\bibfnamefont {T.~M.~R.}\ \bibnamefont {Wolf}}, \bibinfo {author} {\bibfnamefont {C.}~\bibnamefont {Huang}}, \bibinfo {author} {\bibfnamefont {N.}~\bibnamefont {Wei}}, \bibinfo {author} {\bibfnamefont {S.~A.~A.}\ \bibnamefont {Ghorashi}}, \bibinfo {author} {\bibfnamefont {A.~H.}\ \bibnamefont {MacDonald}},\ and\ \bibinfo {author} {\bibfnamefont {J.}~\bibnamefont {Cano}},\ }\bibfield  {title} {\bibinfo {title} {Gate-tunable topological phases in superlattice modulated bilayer graphene},\ }\href {https://doi.org/10.1103/PhysRevB.109.195406} {\bibfield  {journal} {\bibinfo  {journal} {Phys. Rev. B}\ }\textbf {\bibinfo {volume} {109}},\ \bibinfo {pages} {195406} (\bibinfo {year} {2024})}\BibitemShut {NoStop}%
\bibitem [{\citenamefont {Wang}\ \emph {et~al.}(2024)\citenamefont {Wang}, \citenamefont {Zhang}, \citenamefont {Liu}, \citenamefont {He}, \citenamefont {Xu}, \citenamefont {Ran}, \citenamefont {Cao},\ and\ \citenamefont {Xiao}}]{wang-prl24}%
  \BibitemOpen
  \bibfield  {author} {\bibinfo {author} {\bibfnamefont {C.}~\bibnamefont {Wang}}, \bibinfo {author} {\bibfnamefont {X.-W.}\ \bibnamefont {Zhang}}, \bibinfo {author} {\bibfnamefont {X.}~\bibnamefont {Liu}}, \bibinfo {author} {\bibfnamefont {Y.}~\bibnamefont {He}}, \bibinfo {author} {\bibfnamefont {X.}~\bibnamefont {Xu}}, \bibinfo {author} {\bibfnamefont {Y.}~\bibnamefont {Ran}}, \bibinfo {author} {\bibfnamefont {T.}~\bibnamefont {Cao}},\ and\ \bibinfo {author} {\bibfnamefont {D.}~\bibnamefont {Xiao}},\ }\bibfield  {title} {\bibinfo {title} {Fractional {Chern} insulator in twisted bilayer {MoTe$_2$}},\ }\href {https://doi.org/10.1103/PhysRevLett.132.036501} {\bibfield  {journal} {\bibinfo  {journal} {Phys. Rev. Lett.}\ }\textbf {\bibinfo {volume} {132}},\ \bibinfo {pages} {036501} (\bibinfo {year} {2024})}\BibitemShut {NoStop}%
\bibitem [{\citenamefont {Liu}\ \emph {et~al.}(2012)\citenamefont {Liu}, \citenamefont {Bergholtz}, \citenamefont {Fan},\ and\ \citenamefont {L\"auchli}}]{liu-prl12}%
  \BibitemOpen
  \bibfield  {author} {\bibinfo {author} {\bibfnamefont {Z.}~\bibnamefont {Liu}}, \bibinfo {author} {\bibfnamefont {E.~J.}\ \bibnamefont {Bergholtz}}, \bibinfo {author} {\bibfnamefont {H.}~\bibnamefont {Fan}},\ and\ \bibinfo {author} {\bibfnamefont {A.~M.}\ \bibnamefont {L\"auchli}},\ }\bibfield  {title} {\bibinfo {title} {Fractional {Chern} insulators in topological flat bands with higher {Chern} number},\ }\href {https://doi.org/10.1103/PhysRevLett.109.186805} {\bibfield  {journal} {\bibinfo  {journal} {Phys. Rev. Lett.}\ }\textbf {\bibinfo {volume} {109}},\ \bibinfo {pages} {186805} (\bibinfo {year} {2012})}\BibitemShut {NoStop}%
\bibitem [{\citenamefont {Yang}\ \emph {et~al.}(2012)\citenamefont {Yang}, \citenamefont {Gu}, \citenamefont {Sun},\ and\ \citenamefont {Das~Sarma}}]{yang-prb12}%
  \BibitemOpen
  \bibfield  {author} {\bibinfo {author} {\bibfnamefont {S.}~\bibnamefont {Yang}}, \bibinfo {author} {\bibfnamefont {Z.-C.}\ \bibnamefont {Gu}}, \bibinfo {author} {\bibfnamefont {K.}~\bibnamefont {Sun}},\ and\ \bibinfo {author} {\bibfnamefont {S.}~\bibnamefont {Das~Sarma}},\ }\bibfield  {title} {\bibinfo {title} {Topological flat band models with arbitrary {Chern} numbers},\ }\href {https://doi.org/10.1103/PhysRevB.86.241112} {\bibfield  {journal} {\bibinfo  {journal} {Phys. Rev. B}\ }\textbf {\bibinfo {volume} {86}},\ \bibinfo {pages} {241112} (\bibinfo {year} {2012})}\BibitemShut {NoStop}%
\bibitem [{\citenamefont {Wang}\ \emph {et~al.}(2012)\citenamefont {Wang}, \citenamefont {Yao}, \citenamefont {Gong},\ and\ \citenamefont {Sheng}}]{wang-prb12}%
  \BibitemOpen
  \bibfield  {author} {\bibinfo {author} {\bibfnamefont {Y.-F.}\ \bibnamefont {Wang}}, \bibinfo {author} {\bibfnamefont {H.}~\bibnamefont {Yao}}, \bibinfo {author} {\bibfnamefont {C.-D.}\ \bibnamefont {Gong}},\ and\ \bibinfo {author} {\bibfnamefont {D.~N.}\ \bibnamefont {Sheng}},\ }\bibfield  {title} {\bibinfo {title} {Fractional quantum {Hall} effect in topological flat bands with {Chern} number two},\ }\href {https://doi.org/10.1103/PhysRevB.86.201101} {\bibfield  {journal} {\bibinfo  {journal} {Phys. Rev. B}\ }\textbf {\bibinfo {volume} {86}},\ \bibinfo {pages} {201101} (\bibinfo {year} {2012})}\BibitemShut {NoStop}%
\bibitem [{\citenamefont {Sterdyniak}\ \emph {et~al.}(2013)\citenamefont {Sterdyniak}, \citenamefont {Repellin}, \citenamefont {Bernevig},\ and\ \citenamefont {Regnault}}]{sterdyniak-prb13}%
  \BibitemOpen
  \bibfield  {author} {\bibinfo {author} {\bibfnamefont {A.}~\bibnamefont {Sterdyniak}}, \bibinfo {author} {\bibfnamefont {C.}~\bibnamefont {Repellin}}, \bibinfo {author} {\bibfnamefont {B.~A.}\ \bibnamefont {Bernevig}},\ and\ \bibinfo {author} {\bibfnamefont {N.}~\bibnamefont {Regnault}},\ }\bibfield  {title} {\bibinfo {title} {Series of {Abelian} and non-{Abelian} states in {$C>1$} fractional {Chern} insulators},\ }\href {https://doi.org/10.1103/PhysRevB.87.205137} {\bibfield  {journal} {\bibinfo  {journal} {Phys. Rev. B}\ }\textbf {\bibinfo {volume} {87}},\ \bibinfo {pages} {205137} (\bibinfo {year} {2013})}\BibitemShut {NoStop}%
\bibitem [{\citenamefont {Wu}\ \emph {et~al.}(2013)\citenamefont {Wu}, \citenamefont {Regnault},\ and\ \citenamefont {Bernevig}}]{wu-prl13}%
  \BibitemOpen
  \bibfield  {author} {\bibinfo {author} {\bibfnamefont {Y.-L.}\ \bibnamefont {Wu}}, \bibinfo {author} {\bibfnamefont {N.}~\bibnamefont {Regnault}},\ and\ \bibinfo {author} {\bibfnamefont {B.~A.}\ \bibnamefont {Bernevig}},\ }\bibfield  {title} {\bibinfo {title} {Bloch model wave functions and pseudopotentials for all fractional {Chern} insulators},\ }\href {https://doi.org/10.1103/PhysRevLett.110.106802} {\bibfield  {journal} {\bibinfo  {journal} {Phys. Rev. Lett.}\ }\textbf {\bibinfo {volume} {110}},\ \bibinfo {pages} {106802} (\bibinfo {year} {2013})}\BibitemShut {NoStop}%
\bibitem [{\citenamefont {Barkeshli}\ and\ \citenamefont {Qi}(2012)}]{barkeshli-prx12}%
  \BibitemOpen
  \bibfield  {author} {\bibinfo {author} {\bibfnamefont {M.}~\bibnamefont {Barkeshli}}\ and\ \bibinfo {author} {\bibfnamefont {X.-L.}\ \bibnamefont {Qi}},\ }\bibfield  {title} {\bibinfo {title} {Topological nematic states and non-{Abelian} lattice dislocations},\ }\href {https://doi.org/10.1103/PhysRevX.2.031013} {\bibfield  {journal} {\bibinfo  {journal} {Phys. Rev. X}\ }\textbf {\bibinfo {volume} {2}},\ \bibinfo {pages} {031013} (\bibinfo {year} {2012})}\BibitemShut {NoStop}%
\bibitem [{\citenamefont {Barkeshli}\ \emph {et~al.}(2013)\citenamefont {Barkeshli}, \citenamefont {Jian},\ and\ \citenamefont {Qi}}]{barkeshli-prb13}%
  \BibitemOpen
  \bibfield  {author} {\bibinfo {author} {\bibfnamefont {M.}~\bibnamefont {Barkeshli}}, \bibinfo {author} {\bibfnamefont {C.-M.}\ \bibnamefont {Jian}},\ and\ \bibinfo {author} {\bibfnamefont {X.-L.}\ \bibnamefont {Qi}},\ }\bibfield  {title} {\bibinfo {title} {Twist defects and projective non-{Abelian} braiding statistics},\ }\href {https://doi.org/10.1103/PhysRevB.87.045130} {\bibfield  {journal} {\bibinfo  {journal} {Phys. Rev. B}\ }\textbf {\bibinfo {volume} {87}},\ \bibinfo {pages} {045130} (\bibinfo {year} {2013})}\BibitemShut {NoStop}%
\bibitem [{\citenamefont {M\"oller}\ and\ \citenamefont {Cooper}(2015)}]{moller-prl15}%
  \BibitemOpen
  \bibfield  {author} {\bibinfo {author} {\bibfnamefont {G.}~\bibnamefont {M\"oller}}\ and\ \bibinfo {author} {\bibfnamefont {N.~R.}\ \bibnamefont {Cooper}},\ }\bibfield  {title} {\bibinfo {title} {Fractional {Chern} insulators in {Harper-Hofstadter} bands with higher {Chern} number},\ }\href {https://doi.org/10.1103/PhysRevLett.115.126401} {\bibfield  {journal} {\bibinfo  {journal} {Phys. Rev. Lett.}\ }\textbf {\bibinfo {volume} {115}},\ \bibinfo {pages} {126401} (\bibinfo {year} {2015})}\BibitemShut {NoStop}%
\bibitem [{\citenamefont {Wu}\ \emph {et~al.}(2015)\citenamefont {Wu}, \citenamefont {Jain},\ and\ \citenamefont {Sun}}]{wu-prb15}%
  \BibitemOpen
  \bibfield  {author} {\bibinfo {author} {\bibfnamefont {Y.-H.}\ \bibnamefont {Wu}}, \bibinfo {author} {\bibfnamefont {J.~K.}\ \bibnamefont {Jain}},\ and\ \bibinfo {author} {\bibfnamefont {K.}~\bibnamefont {Sun}},\ }\bibfield  {title} {\bibinfo {title} {Fractional topological phases in generalized {Hofstadter} bands with arbitrary {Chern} numbers},\ }\href {https://doi.org/10.1103/PhysRevB.91.041119} {\bibfield  {journal} {\bibinfo  {journal} {Phys. Rev. B}\ }\textbf {\bibinfo {volume} {91}},\ \bibinfo {pages} {041119} (\bibinfo {year} {2015})}\BibitemShut {NoStop}%
\bibitem [{\citenamefont {Behrmann}\ \emph {et~al.}(2016)\citenamefont {Behrmann}, \citenamefont {Liu},\ and\ \citenamefont {Bergholtz}}]{behrmann-prl16}%
  \BibitemOpen
  \bibfield  {author} {\bibinfo {author} {\bibfnamefont {J.}~\bibnamefont {Behrmann}}, \bibinfo {author} {\bibfnamefont {Z.}~\bibnamefont {Liu}},\ and\ \bibinfo {author} {\bibfnamefont {E.~J.}\ \bibnamefont {Bergholtz}},\ }\bibfield  {title} {\bibinfo {title} {Model fractional {Chern} insulators},\ }\href {https://doi.org/10.1103/PhysRevLett.116.216802} {\bibfield  {journal} {\bibinfo  {journal} {Phys. Rev. Lett.}\ }\textbf {\bibinfo {volume} {116}},\ \bibinfo {pages} {216802} (\bibinfo {year} {2016})}\BibitemShut {NoStop}%
\bibitem [{\citenamefont {Andrews}\ and\ \citenamefont {M\"oller}(2018)}]{andrews-prb18}%
  \BibitemOpen
  \bibfield  {author} {\bibinfo {author} {\bibfnamefont {B.}~\bibnamefont {Andrews}}\ and\ \bibinfo {author} {\bibfnamefont {G.}~\bibnamefont {M\"oller}},\ }\bibfield  {title} {\bibinfo {title} {Stability of fractional {Chern} insulators in the effective continuum limit of {Harper-Hofstadter} bands with {Chern} number {$|C|>1$}},\ }\href {https://doi.org/10.1103/PhysRevB.97.035159} {\bibfield  {journal} {\bibinfo  {journal} {Phys. Rev. B}\ }\textbf {\bibinfo {volume} {97}},\ \bibinfo {pages} {035159} (\bibinfo {year} {2018})}\BibitemShut {NoStop}%
\bibitem [{\citenamefont {Wang}\ and\ \citenamefont {Liu}(2022)}]{wang-prl22}%
  \BibitemOpen
  \bibfield  {author} {\bibinfo {author} {\bibfnamefont {J.}~\bibnamefont {Wang}}\ and\ \bibinfo {author} {\bibfnamefont {Z.}~\bibnamefont {Liu}},\ }\bibfield  {title} {\bibinfo {title} {Hierarchy of ideal flatbands in chiral twisted multilayer graphene models},\ }\href {https://doi.org/10.1103/PhysRevLett.128.176403} {\bibfield  {journal} {\bibinfo  {journal} {Phys. Rev. Lett.}\ }\textbf {\bibinfo {volume} {128}},\ \bibinfo {pages} {176403} (\bibinfo {year} {2022})}\BibitemShut {NoStop}%
\bibitem [{\citenamefont {Liu}\ and\ \citenamefont {Bergholtz}(2024)}]{liu-book24}%
  \BibitemOpen
  \bibfield  {author} {\bibinfo {author} {\bibfnamefont {Z.}~\bibnamefont {Liu}}\ and\ \bibinfo {author} {\bibfnamefont {E.~J.}\ \bibnamefont {Bergholtz}},\ }\bibfield  {title} {\bibinfo {title} {Recent developments in fractional {Chern} insulators},\ }in\ \href {https://doi.org/https://doi.org/10.1016/B978-0-323-90800-9.00136-0} {\emph {\bibinfo {booktitle} {Encyclopedia of Condensed Matter Physics (Second Edition)}}},\ \bibinfo {editor} {edited by\ \bibinfo {editor} {\bibfnamefont {T.}~\bibnamefont {Chakraborty}}}\ (\bibinfo  {publisher} {Academic Press},\ \bibinfo {address} {Oxford},\ \bibinfo {year} {2024})\ \bibinfo {edition} {second edition}\ ed.,\ pp.\ \bibinfo {pages} {515--538}\BibitemShut {NoStop}%
\bibitem [{\citenamefont {Khalaf}\ \emph {et~al.}(2021)\citenamefont {Khalaf}, \citenamefont {Chatterjee}, \citenamefont {Bultinck}, \citenamefont {Zaletel},\ and\ \citenamefont {Vishwanath}}]{khalaf-sciadv21}%
  \BibitemOpen
  \bibfield  {author} {\bibinfo {author} {\bibfnamefont {E.}~\bibnamefont {Khalaf}}, \bibinfo {author} {\bibfnamefont {S.}~\bibnamefont {Chatterjee}}, \bibinfo {author} {\bibfnamefont {N.}~\bibnamefont {Bultinck}}, \bibinfo {author} {\bibfnamefont {M.~P.}\ \bibnamefont {Zaletel}},\ and\ \bibinfo {author} {\bibfnamefont {A.}~\bibnamefont {Vishwanath}},\ }\bibfield  {title} {\bibinfo {title} {Charged skyrmions and topological origin of superconductivity in magic-angle graphene},\ }\href {https://doi.org/10.1126/sciadv.abf5299} {\bibfield  {journal} {\bibinfo  {journal} {Science Advances}\ }\textbf {\bibinfo {volume} {7}},\ \bibinfo {pages} {eabf5299} (\bibinfo {year} {2021})}\BibitemShut {NoStop}%
\bibitem [{\citenamefont {Chatterjee}\ \emph {et~al.}(2022)\citenamefont {Chatterjee}, \citenamefont {Ippoliti},\ and\ \citenamefont {Zaletel}}]{chatterjee-prb22}%
  \BibitemOpen
  \bibfield  {author} {\bibinfo {author} {\bibfnamefont {S.}~\bibnamefont {Chatterjee}}, \bibinfo {author} {\bibfnamefont {M.}~\bibnamefont {Ippoliti}},\ and\ \bibinfo {author} {\bibfnamefont {M.~P.}\ \bibnamefont {Zaletel}},\ }\bibfield  {title} {\bibinfo {title} {Skyrmion superconductivity: {DMRG} evidence for a topological route to superconductivity},\ }\href {https://doi.org/10.1103/PhysRevB.106.035421} {\bibfield  {journal} {\bibinfo  {journal} {Phys. Rev. B}\ }\textbf {\bibinfo {volume} {106}},\ \bibinfo {pages} {035421} (\bibinfo {year} {2022})}\BibitemShut {NoStop}%
\bibitem [{\citenamefont {T\"{o}rm\"{a}}\ \emph {et~al.}(2022)\citenamefont {T\"{o}rm\"{a}}, \citenamefont {Peotta},\ and\ \citenamefont {Bernevig}}]{torma-natrevphys22}%
  \BibitemOpen
  \bibfield  {author} {\bibinfo {author} {\bibfnamefont {P.}~\bibnamefont {T\"{o}rm\"{a}}}, \bibinfo {author} {\bibfnamefont {S.}~\bibnamefont {Peotta}},\ and\ \bibinfo {author} {\bibfnamefont {B.~A.}\ \bibnamefont {Bernevig}},\ }\bibfield  {title} {\bibinfo {title} {Superconductivity, superfluidity and quantum geometry in twisted multilayer systems},\ }\href {https://doi.org/10.1038/s42254-022-00466-y} {\bibfield  {journal} {\bibinfo  {journal} {Nature Reviews Physics}\ }\textbf {\bibinfo {volume} {4}},\ \bibinfo {pages} {528–542} (\bibinfo {year} {2022})}\BibitemShut {NoStop}%
\bibitem [{\citenamefont {Sun}\ \emph {et~al.}(2023)\citenamefont {Sun}, \citenamefont {Ghorashi}, \citenamefont {Watanabe}, \citenamefont {Taniguchi}, \citenamefont {Camino}, \citenamefont {Cano},\ and\ \citenamefont {Du}}]{sun-arxiv23}%
  \BibitemOpen
  \bibfield  {author} {\bibinfo {author} {\bibfnamefont {J.}~\bibnamefont {Sun}}, \bibinfo {author} {\bibfnamefont {S.~A.~A.}\ \bibnamefont {Ghorashi}}, \bibinfo {author} {\bibfnamefont {K.}~\bibnamefont {Watanabe}}, \bibinfo {author} {\bibfnamefont {T.}~\bibnamefont {Taniguchi}}, \bibinfo {author} {\bibfnamefont {F.}~\bibnamefont {Camino}}, \bibinfo {author} {\bibfnamefont {J.}~\bibnamefont {Cano}},\ and\ \bibinfo {author} {\bibfnamefont {X.}~\bibnamefont {Du}},\ }\href {https://doi.org/10.48550/ARXIV.2306.06848} {\bibinfo {title} {Signature of correlated insulator in electric field controlled superlattice}} (\bibinfo {year} {2023})\BibitemShut {NoStop}%
\bibitem [{\citenamefont {Dean}\ \emph {et~al.}(2013)\citenamefont {Dean}, \citenamefont {Wang}, \citenamefont {Maher}, \citenamefont {Forsythe}, \citenamefont {Ghahari}, \citenamefont {Gao}, \citenamefont {Katoch}, \citenamefont {Ishigami}, \citenamefont {Moon}, \citenamefont {Koshino}, \citenamefont {Taniguchi}, \citenamefont {Watanabe}, \citenamefont {Shepard}, \citenamefont {Hone},\ and\ \citenamefont {Kim}}]{dean-nat13}%
  \BibitemOpen
  \bibfield  {author} {\bibinfo {author} {\bibfnamefont {C.~R.}\ \bibnamefont {Dean}}, \bibinfo {author} {\bibfnamefont {L.}~\bibnamefont {Wang}}, \bibinfo {author} {\bibfnamefont {P.}~\bibnamefont {Maher}}, \bibinfo {author} {\bibfnamefont {C.}~\bibnamefont {Forsythe}}, \bibinfo {author} {\bibfnamefont {F.}~\bibnamefont {Ghahari}}, \bibinfo {author} {\bibfnamefont {Y.}~\bibnamefont {Gao}}, \bibinfo {author} {\bibfnamefont {J.}~\bibnamefont {Katoch}}, \bibinfo {author} {\bibfnamefont {M.}~\bibnamefont {Ishigami}}, \bibinfo {author} {\bibfnamefont {P.}~\bibnamefont {Moon}}, \bibinfo {author} {\bibfnamefont {M.}~\bibnamefont {Koshino}}, \bibinfo {author} {\bibfnamefont {T.}~\bibnamefont {Taniguchi}}, \bibinfo {author} {\bibfnamefont {K.}~\bibnamefont {Watanabe}}, \bibinfo {author} {\bibfnamefont {K.~L.}\ \bibnamefont {Shepard}}, \bibinfo {author} {\bibfnamefont {J.}~\bibnamefont {Hone}},\ and\ \bibinfo {author} {\bibfnamefont {P.}~\bibnamefont {Kim}},\ }\bibfield  {title} {\bibinfo {title} {Hofstadter’s
  butterfly and the fractal quantum {Hall} effect in moiré superlattices},\ }\href {https://doi.org/10.1038/nature12186} {\bibfield  {journal} {\bibinfo  {journal} {Nature}\ }\textbf {\bibinfo {volume} {497}},\ \bibinfo {pages} {598–602} (\bibinfo {year} {2013})}\BibitemShut {NoStop}%
\bibitem [{\citenamefont {Pierce}\ \emph {et~al.}(2021)\citenamefont {Pierce}, \citenamefont {Xie}, \citenamefont {Park}, \citenamefont {Khalaf}, \citenamefont {Lee}, \citenamefont {Cao}, \citenamefont {Parker}, \citenamefont {Forrester}, \citenamefont {Chen}, \citenamefont {Watanabe}, \citenamefont {Taniguchi}, \citenamefont {Vishwanath}, \citenamefont {Jarillo-Herrero},\ and\ \citenamefont {Yacoby}}]{pierce-natphys21}%
  \BibitemOpen
  \bibfield  {author} {\bibinfo {author} {\bibfnamefont {A.~T.}\ \bibnamefont {Pierce}}, \bibinfo {author} {\bibfnamefont {Y.}~\bibnamefont {Xie}}, \bibinfo {author} {\bibfnamefont {J.~M.}\ \bibnamefont {Park}}, \bibinfo {author} {\bibfnamefont {E.}~\bibnamefont {Khalaf}}, \bibinfo {author} {\bibfnamefont {S.~H.}\ \bibnamefont {Lee}}, \bibinfo {author} {\bibfnamefont {Y.}~\bibnamefont {Cao}}, \bibinfo {author} {\bibfnamefont {D.~E.}\ \bibnamefont {Parker}}, \bibinfo {author} {\bibfnamefont {P.~R.}\ \bibnamefont {Forrester}}, \bibinfo {author} {\bibfnamefont {S.}~\bibnamefont {Chen}}, \bibinfo {author} {\bibfnamefont {K.}~\bibnamefont {Watanabe}}, \bibinfo {author} {\bibfnamefont {T.}~\bibnamefont {Taniguchi}}, \bibinfo {author} {\bibfnamefont {A.}~\bibnamefont {Vishwanath}}, \bibinfo {author} {\bibfnamefont {P.}~\bibnamefont {Jarillo-Herrero}},\ and\ \bibinfo {author} {\bibfnamefont {A.}~\bibnamefont {Yacoby}},\ }\bibfield  {title} {\bibinfo {title} {Unconventional sequence of correlated {Chern} insulators in
  magic-angle twisted bilayer graphene},\ }\href {https://doi.org/10.1038/s41567-021-01347-4} {\bibfield  {journal} {\bibinfo  {journal} {Nature Physics}\ }\textbf {\bibinfo {volume} {17}},\ \bibinfo {pages} {1210–1215} (\bibinfo {year} {2021})}\BibitemShut {NoStop}%
\bibitem [{\citenamefont {Saito}\ \emph {et~al.}(2021)\citenamefont {Saito}, \citenamefont {Ge}, \citenamefont {Rademaker}, \citenamefont {Watanabe}, \citenamefont {Taniguchi}, \citenamefont {Abanin},\ and\ \citenamefont {Young}}]{saito-natphys21}%
  \BibitemOpen
  \bibfield  {author} {\bibinfo {author} {\bibfnamefont {Y.}~\bibnamefont {Saito}}, \bibinfo {author} {\bibfnamefont {J.}~\bibnamefont {Ge}}, \bibinfo {author} {\bibfnamefont {L.}~\bibnamefont {Rademaker}}, \bibinfo {author} {\bibfnamefont {K.}~\bibnamefont {Watanabe}}, \bibinfo {author} {\bibfnamefont {T.}~\bibnamefont {Taniguchi}}, \bibinfo {author} {\bibfnamefont {D.~A.}\ \bibnamefont {Abanin}},\ and\ \bibinfo {author} {\bibfnamefont {A.~F.}\ \bibnamefont {Young}},\ }\bibfield  {title} {\bibinfo {title} {Hofstadter subband ferromagnetism and symmetry-broken {Chern} insulators in twisted bilayer graphene},\ }\href {https://doi.org/10.1038/s41567-020-01129-4} {\bibfield  {journal} {\bibinfo  {journal} {Nature Physics}\ }\textbf {\bibinfo {volume} {17}},\ \bibinfo {pages} {478–481} (\bibinfo {year} {2021})}\BibitemShut {NoStop}%
\bibitem [{\citenamefont {Das}\ \emph {et~al.}(2022)\citenamefont {Das}, \citenamefont {Shen}, \citenamefont {Jaoui}, \citenamefont {Herzog-Arbeitman}, \citenamefont {Chew}, \citenamefont {Cho}, \citenamefont {Watanabe}, \citenamefont {Taniguchi}, \citenamefont {Piot}, \citenamefont {Bernevig},\ and\ \citenamefont {Efetov}}]{das-prl22}%
  \BibitemOpen
  \bibfield  {author} {\bibinfo {author} {\bibfnamefont {I.}~\bibnamefont {Das}}, \bibinfo {author} {\bibfnamefont {C.}~\bibnamefont {Shen}}, \bibinfo {author} {\bibfnamefont {A.}~\bibnamefont {Jaoui}}, \bibinfo {author} {\bibfnamefont {J.}~\bibnamefont {Herzog-Arbeitman}}, \bibinfo {author} {\bibfnamefont {A.}~\bibnamefont {Chew}}, \bibinfo {author} {\bibfnamefont {C.-W.}\ \bibnamefont {Cho}}, \bibinfo {author} {\bibfnamefont {K.}~\bibnamefont {Watanabe}}, \bibinfo {author} {\bibfnamefont {T.}~\bibnamefont {Taniguchi}}, \bibinfo {author} {\bibfnamefont {B.~A.}\ \bibnamefont {Piot}}, \bibinfo {author} {\bibfnamefont {B.~A.}\ \bibnamefont {Bernevig}},\ and\ \bibinfo {author} {\bibfnamefont {D.~K.}\ \bibnamefont {Efetov}},\ }\bibfield  {title} {\bibinfo {title} {Observation of reentrant correlated insulators and interaction-driven {Fermi}-surface reconstructions at one magnetic flux quantum per moir\'e unit cell in magic-angle twisted bilayer graphene},\ }\href {https://doi.org/10.1103/PhysRevLett.128.217701}
  {\bibfield  {journal} {\bibinfo  {journal} {Phys. Rev. Lett.}\ }\textbf {\bibinfo {volume} {128}},\ \bibinfo {pages} {217701} (\bibinfo {year} {2022})}\BibitemShut {NoStop}%
\bibitem [{\citenamefont {Herzog-Arbeitman}\ \emph {et~al.}(2022{\natexlab{a}})\citenamefont {Herzog-Arbeitman}, \citenamefont {Chew}, \citenamefont {Efetov},\ and\ \citenamefont {Bernevig}}]{herzog-prl22}%
  \BibitemOpen
  \bibfield  {author} {\bibinfo {author} {\bibfnamefont {J.}~\bibnamefont {Herzog-Arbeitman}}, \bibinfo {author} {\bibfnamefont {A.}~\bibnamefont {Chew}}, \bibinfo {author} {\bibfnamefont {D.~K.}\ \bibnamefont {Efetov}},\ and\ \bibinfo {author} {\bibfnamefont {B.~A.}\ \bibnamefont {Bernevig}},\ }\bibfield  {title} {\bibinfo {title} {Reentrant correlated insulators in twisted bilayer graphene at 25 {T} ($2\ensuremath{\pi}$ flux)},\ }\href {https://doi.org/10.1103/PhysRevLett.129.076401} {\bibfield  {journal} {\bibinfo  {journal} {Phys. Rev. Lett.}\ }\textbf {\bibinfo {volume} {129}},\ \bibinfo {pages} {076401} (\bibinfo {year} {2022}{\natexlab{a}})}\BibitemShut {NoStop}%
\bibitem [{\citenamefont {Finney}\ \emph {et~al.}(2022)\citenamefont {Finney}, \citenamefont {Sharpe}, \citenamefont {Fox}, \citenamefont {Hsueh}, \citenamefont {Parker}, \citenamefont {Yankowitz}, \citenamefont {Chen}, \citenamefont {Watanabe}, \citenamefont {Taniguchi}, \citenamefont {Dean}, \citenamefont {Vishwanath}, \citenamefont {Kastner},\ and\ \citenamefont {Goldhaber-Gordon}}]{finney-pnas22}%
  \BibitemOpen
  \bibfield  {author} {\bibinfo {author} {\bibfnamefont {J.}~\bibnamefont {Finney}}, \bibinfo {author} {\bibfnamefont {A.~L.}\ \bibnamefont {Sharpe}}, \bibinfo {author} {\bibfnamefont {E.~J.}\ \bibnamefont {Fox}}, \bibinfo {author} {\bibfnamefont {C.~L.}\ \bibnamefont {Hsueh}}, \bibinfo {author} {\bibfnamefont {D.~E.}\ \bibnamefont {Parker}}, \bibinfo {author} {\bibfnamefont {M.}~\bibnamefont {Yankowitz}}, \bibinfo {author} {\bibfnamefont {S.}~\bibnamefont {Chen}}, \bibinfo {author} {\bibfnamefont {K.}~\bibnamefont {Watanabe}}, \bibinfo {author} {\bibfnamefont {T.}~\bibnamefont {Taniguchi}}, \bibinfo {author} {\bibfnamefont {C.~R.}\ \bibnamefont {Dean}}, \bibinfo {author} {\bibfnamefont {A.}~\bibnamefont {Vishwanath}}, \bibinfo {author} {\bibfnamefont {M.~A.}\ \bibnamefont {Kastner}},\ and\ \bibinfo {author} {\bibfnamefont {D.}~\bibnamefont {Goldhaber-Gordon}},\ }\bibfield  {title} {\bibinfo {title} {Unusual magnetotransport in twisted bilayer graphene},\ }\bibfield  {journal} {\bibinfo  {journal}
  {Proceedings of the National Academy of Sciences}\ }\textbf {\bibinfo {volume} {119}},\ \href {https://doi.org/10.1073/pnas.2118482119} {10.1073/pnas.2118482119} (\bibinfo {year} {2022})\BibitemShut {NoStop}%
\bibitem [{\citenamefont {Forsythe}\ \emph {et~al.}(2018)\citenamefont {Forsythe}, \citenamefont {Zhou}, \citenamefont {Watanabe}, \citenamefont {Taniguchi}, \citenamefont {Pasupathy}, \citenamefont {Moon}, \citenamefont {Koshino}, \citenamefont {Kim},\ and\ \citenamefont {Dean}}]{forsythe-natnano18}%
  \BibitemOpen
  \bibfield  {author} {\bibinfo {author} {\bibfnamefont {C.}~\bibnamefont {Forsythe}}, \bibinfo {author} {\bibfnamefont {X.}~\bibnamefont {Zhou}}, \bibinfo {author} {\bibfnamefont {K.}~\bibnamefont {Watanabe}}, \bibinfo {author} {\bibfnamefont {T.}~\bibnamefont {Taniguchi}}, \bibinfo {author} {\bibfnamefont {A.}~\bibnamefont {Pasupathy}}, \bibinfo {author} {\bibfnamefont {P.}~\bibnamefont {Moon}}, \bibinfo {author} {\bibfnamefont {M.}~\bibnamefont {Koshino}}, \bibinfo {author} {\bibfnamefont {P.}~\bibnamefont {Kim}},\ and\ \bibinfo {author} {\bibfnamefont {C.~R.}\ \bibnamefont {Dean}},\ }\bibfield  {title} {\bibinfo {title} {Band structure engineering of {2D} materials using patterned dielectric superlattices},\ }\href {https://doi.org/10.1038/s41565-018-0138-7} {\bibfield  {journal} {\bibinfo  {journal} {Nature Nanotechnology}\ }\textbf {\bibinfo {volume} {13}},\ \bibinfo {pages} {566–571} (\bibinfo {year} {2018})}\BibitemShut {NoStop}%
\bibitem [{\citenamefont {Li}\ \emph {et~al.}(2021{\natexlab{d}})\citenamefont {Li}, \citenamefont {Dietrich}, \citenamefont {Forsythe}, \citenamefont {Taniguchi}, \citenamefont {Watanabe}, \citenamefont {Moon},\ and\ \citenamefont {Dean}}]{li-natnano21}%
  \BibitemOpen
  \bibfield  {author} {\bibinfo {author} {\bibfnamefont {Y.}~\bibnamefont {Li}}, \bibinfo {author} {\bibfnamefont {S.}~\bibnamefont {Dietrich}}, \bibinfo {author} {\bibfnamefont {C.}~\bibnamefont {Forsythe}}, \bibinfo {author} {\bibfnamefont {T.}~\bibnamefont {Taniguchi}}, \bibinfo {author} {\bibfnamefont {K.}~\bibnamefont {Watanabe}}, \bibinfo {author} {\bibfnamefont {P.}~\bibnamefont {Moon}},\ and\ \bibinfo {author} {\bibfnamefont {C.~R.}\ \bibnamefont {Dean}},\ }\bibfield  {title} {\bibinfo {title} {Anisotropic band flattening in graphene with one-dimensional superlattices},\ }\href {https://doi.org/10.1038/s41565-021-00849-9} {\bibfield  {journal} {\bibinfo  {journal} {Nature Nanotechnology}\ }\textbf {\bibinfo {volume} {16}},\ \bibinfo {pages} {525–530} (\bibinfo {year} {2021}{\natexlab{d}})}\BibitemShut {NoStop}%
\bibitem [{\citenamefont {Barcons~Ruiz}\ \emph {et~al.}(2022)\citenamefont {Barcons~Ruiz}, \citenamefont {Herzig~Sheinfux}, \citenamefont {Hoffmann}, \citenamefont {Torre}, \citenamefont {Agarwal}, \citenamefont {Kumar}, \citenamefont {Vistoli}, \citenamefont {Taniguchi}, \citenamefont {Watanabe}, \citenamefont {Bachtold},\ and\ \citenamefont {Koppens}}]{barcons-natcomm22}%
  \BibitemOpen
  \bibfield  {author} {\bibinfo {author} {\bibfnamefont {D.}~\bibnamefont {Barcons~Ruiz}}, \bibinfo {author} {\bibfnamefont {H.}~\bibnamefont {Herzig~Sheinfux}}, \bibinfo {author} {\bibfnamefont {R.}~\bibnamefont {Hoffmann}}, \bibinfo {author} {\bibfnamefont {I.}~\bibnamefont {Torre}}, \bibinfo {author} {\bibfnamefont {H.}~\bibnamefont {Agarwal}}, \bibinfo {author} {\bibfnamefont {R.~K.}\ \bibnamefont {Kumar}}, \bibinfo {author} {\bibfnamefont {L.}~\bibnamefont {Vistoli}}, \bibinfo {author} {\bibfnamefont {T.}~\bibnamefont {Taniguchi}}, \bibinfo {author} {\bibfnamefont {K.}~\bibnamefont {Watanabe}}, \bibinfo {author} {\bibfnamefont {A.}~\bibnamefont {Bachtold}},\ and\ \bibinfo {author} {\bibfnamefont {F.~H.~L.}\ \bibnamefont {Koppens}},\ }\bibfield  {title} {\bibinfo {title} {Engineering high quality graphene superlattices via ion milled ultra-thin etching masks},\ }\bibfield  {journal} {\bibinfo  {journal} {Nature Communications}\ }\textbf {\bibinfo {volume} {13}},\ \href
  {https://doi.org/10.1038/s41467-022-34734-3} {10.1038/s41467-022-34734-3} (\bibinfo {year} {2022})\BibitemShut {NoStop}%
\bibitem [{\citenamefont {McCann}(2006)}]{mccann-prb06}%
  \BibitemOpen
  \bibfield  {author} {\bibinfo {author} {\bibfnamefont {E.}~\bibnamefont {McCann}},\ }\bibfield  {title} {\bibinfo {title} {Asymmetry gap in the electronic band structure of bilayer graphene},\ }\href {https://doi.org/10.1103/PhysRevB.74.161403} {\bibfield  {journal} {\bibinfo  {journal} {Phys. Rev. B}\ }\textbf {\bibinfo {volume} {74}},\ \bibinfo {pages} {161403} (\bibinfo {year} {2006})}\BibitemShut {NoStop}%
\bibitem [{\citenamefont {McCann}\ and\ \citenamefont {Fal'ko}(2006)}]{mccann-prl06}%
  \BibitemOpen
  \bibfield  {author} {\bibinfo {author} {\bibfnamefont {E.}~\bibnamefont {McCann}}\ and\ \bibinfo {author} {\bibfnamefont {V.~I.}\ \bibnamefont {Fal'ko}},\ }\bibfield  {title} {\bibinfo {title} {Landau-level degeneracy and quantum {Hall} effect in a graphite bilayer},\ }\href {https://doi.org/10.1103/PhysRevLett.96.086805} {\bibfield  {journal} {\bibinfo  {journal} {Phys. Rev. Lett.}\ }\textbf {\bibinfo {volume} {96}},\ \bibinfo {pages} {086805} (\bibinfo {year} {2006})}\BibitemShut {NoStop}%
\bibitem [{\citenamefont {Castro~Neto}\ \emph {et~al.}(2009)\citenamefont {Castro~Neto}, \citenamefont {Guinea}, \citenamefont {Peres}, \citenamefont {Novoselov},\ and\ \citenamefont {Geim}}]{neto-rmp09}%
  \BibitemOpen
  \bibfield  {author} {\bibinfo {author} {\bibfnamefont {A.~H.}\ \bibnamefont {Castro~Neto}}, \bibinfo {author} {\bibfnamefont {F.}~\bibnamefont {Guinea}}, \bibinfo {author} {\bibfnamefont {N.~M.~R.}\ \bibnamefont {Peres}}, \bibinfo {author} {\bibfnamefont {K.~S.}\ \bibnamefont {Novoselov}},\ and\ \bibinfo {author} {\bibfnamefont {A.~K.}\ \bibnamefont {Geim}},\ }\bibfield  {title} {\bibinfo {title} {The electronic properties of graphene},\ }\href {https://doi.org/10.1103/RevModPhys.81.109} {\bibfield  {journal} {\bibinfo  {journal} {Rev. Mod. Phys.}\ }\textbf {\bibinfo {volume} {81}},\ \bibinfo {pages} {109} (\bibinfo {year} {2009})}\BibitemShut {NoStop}%
\bibitem [{\citenamefont {McCann}\ and\ \citenamefont {Koshino}(2013)}]{mccann-ropp13}%
  \BibitemOpen
  \bibfield  {author} {\bibinfo {author} {\bibfnamefont {E.}~\bibnamefont {McCann}}\ and\ \bibinfo {author} {\bibfnamefont {M.}~\bibnamefont {Koshino}},\ }\bibfield  {title} {\bibinfo {title} {The electronic properties of bilayer graphene},\ }\href {https://doi.org/10.1088/0034-4885/76/5/056503} {\bibfield  {journal} {\bibinfo  {journal} {Reports on Progress in Physics}\ }\textbf {\bibinfo {volume} {76}},\ \bibinfo {pages} {056503} (\bibinfo {year} {2013})}\BibitemShut {NoStop}%
\bibitem [{\citenamefont {Ohta}\ \emph {et~al.}(2006)\citenamefont {Ohta}, \citenamefont {Bostwick}, \citenamefont {Seyller}, \citenamefont {Horn},\ and\ \citenamefont {Rotenberg}}]{ohta-sci06}%
  \BibitemOpen
  \bibfield  {author} {\bibinfo {author} {\bibfnamefont {T.}~\bibnamefont {Ohta}}, \bibinfo {author} {\bibfnamefont {A.}~\bibnamefont {Bostwick}}, \bibinfo {author} {\bibfnamefont {T.}~\bibnamefont {Seyller}}, \bibinfo {author} {\bibfnamefont {K.}~\bibnamefont {Horn}},\ and\ \bibinfo {author} {\bibfnamefont {E.}~\bibnamefont {Rotenberg}},\ }\bibfield  {title} {\bibinfo {title} {Controlling the electronic structure of bilayer graphene},\ }\href {https://doi.org/10.1126/science.1130681} {\bibfield  {journal} {\bibinfo  {journal} {Science}\ }\textbf {\bibinfo {volume} {313}},\ \bibinfo {pages} {951–954} (\bibinfo {year} {2006})}\BibitemShut {NoStop}%
\bibitem [{\citenamefont {Min}\ \emph {et~al.}(2007)\citenamefont {Min}, \citenamefont {Sahu}, \citenamefont {Banerjee},\ and\ \citenamefont {MacDonald}}]{min-prb07}%
  \BibitemOpen
  \bibfield  {author} {\bibinfo {author} {\bibfnamefont {H.}~\bibnamefont {Min}}, \bibinfo {author} {\bibfnamefont {B.}~\bibnamefont {Sahu}}, \bibinfo {author} {\bibfnamefont {S.~K.}\ \bibnamefont {Banerjee}},\ and\ \bibinfo {author} {\bibfnamefont {A.~H.}\ \bibnamefont {MacDonald}},\ }\bibfield  {title} {\bibinfo {title} {\textit{Ab initio} theory of gate induced gaps in graphene bilayers},\ }\href {https://doi.org/10.1103/PhysRevB.75.155115} {\bibfield  {journal} {\bibinfo  {journal} {Phys. Rev. B}\ }\textbf {\bibinfo {volume} {75}},\ \bibinfo {pages} {155115} (\bibinfo {year} {2007})}\BibitemShut {NoStop}%
\bibitem [{\citenamefont {Castro}\ \emph {et~al.}(2007)\citenamefont {Castro}, \citenamefont {Novoselov}, \citenamefont {Morozov}, \citenamefont {Peres}, \citenamefont {dos Santos}, \citenamefont {Nilsson}, \citenamefont {Guinea}, \citenamefont {Geim},\ and\ \citenamefont {Neto}}]{castro-prl07}%
  \BibitemOpen
  \bibfield  {author} {\bibinfo {author} {\bibfnamefont {E.~V.}\ \bibnamefont {Castro}}, \bibinfo {author} {\bibfnamefont {K.~S.}\ \bibnamefont {Novoselov}}, \bibinfo {author} {\bibfnamefont {S.~V.}\ \bibnamefont {Morozov}}, \bibinfo {author} {\bibfnamefont {N.~M.~R.}\ \bibnamefont {Peres}}, \bibinfo {author} {\bibfnamefont {J.~M. B.~L.}\ \bibnamefont {dos Santos}}, \bibinfo {author} {\bibfnamefont {J.}~\bibnamefont {Nilsson}}, \bibinfo {author} {\bibfnamefont {F.}~\bibnamefont {Guinea}}, \bibinfo {author} {\bibfnamefont {A.~K.}\ \bibnamefont {Geim}},\ and\ \bibinfo {author} {\bibfnamefont {A.~H.~C.}\ \bibnamefont {Neto}},\ }\bibfield  {title} {\bibinfo {title} {Biased bilayer graphene: Semiconductor with a gap tunable by the electric field effect},\ }\href {https://doi.org/10.1103/PhysRevLett.99.216802} {\bibfield  {journal} {\bibinfo  {journal} {Phys. Rev. Lett.}\ }\textbf {\bibinfo {volume} {99}},\ \bibinfo {pages} {216802} (\bibinfo {year} {2007})}\BibitemShut {NoStop}%
\bibitem [{\citenamefont {Martin}\ \emph {et~al.}(2008)\citenamefont {Martin}, \citenamefont {Blanter},\ and\ \citenamefont {Morpurgo}}]{martin-prl08}%
  \BibitemOpen
  \bibfield  {author} {\bibinfo {author} {\bibfnamefont {I.}~\bibnamefont {Martin}}, \bibinfo {author} {\bibfnamefont {Y.~M.}\ \bibnamefont {Blanter}},\ and\ \bibinfo {author} {\bibfnamefont {A.~F.}\ \bibnamefont {Morpurgo}},\ }\bibfield  {title} {\bibinfo {title} {Topological confinement in bilayer graphene},\ }\href {https://doi.org/10.1103/PhysRevLett.100.036804} {\bibfield  {journal} {\bibinfo  {journal} {Phys. Rev. Lett.}\ }\textbf {\bibinfo {volume} {100}},\ \bibinfo {pages} {036804} (\bibinfo {year} {2008})}\BibitemShut {NoStop}%
\bibitem [{\citenamefont {Zhang}\ \emph {et~al.}(2011)\citenamefont {Zhang}, \citenamefont {Jung}, \citenamefont {Fiete}, \citenamefont {Niu},\ and\ \citenamefont {MacDonald}}]{zhang-prl11}%
  \BibitemOpen
  \bibfield  {author} {\bibinfo {author} {\bibfnamefont {F.}~\bibnamefont {Zhang}}, \bibinfo {author} {\bibfnamefont {J.}~\bibnamefont {Jung}}, \bibinfo {author} {\bibfnamefont {G.~A.}\ \bibnamefont {Fiete}}, \bibinfo {author} {\bibfnamefont {Q.}~\bibnamefont {Niu}},\ and\ \bibinfo {author} {\bibfnamefont {A.~H.}\ \bibnamefont {MacDonald}},\ }\bibfield  {title} {\bibinfo {title} {Spontaneous quantum {Hall} states in chirally stacked few-layer graphene systems},\ }\href {https://doi.org/10.1103/PhysRevLett.106.156801} {\bibfield  {journal} {\bibinfo  {journal} {Phys. Rev. Lett.}\ }\textbf {\bibinfo {volume} {106}},\ \bibinfo {pages} {156801} (\bibinfo {year} {2011})}\BibitemShut {NoStop}%
\bibitem [{\citenamefont {Rokni}\ and\ \citenamefont {Lu}(2017)}]{rokni-scirep17}%
  \BibitemOpen
  \bibfield  {author} {\bibinfo {author} {\bibfnamefont {H.}~\bibnamefont {Rokni}}\ and\ \bibinfo {author} {\bibfnamefont {W.}~\bibnamefont {Lu}},\ }\bibfield  {title} {\bibinfo {title} {Layer-by-layer insight into electrostatic charge distribution of few-layer graphene},\ }\bibfield  {journal} {\bibinfo  {journal} {Scientific Reports}\ }\textbf {\bibinfo {volume} {7}},\ \href {https://doi.org/10.1038/srep42821} {10.1038/srep42821} (\bibinfo {year} {2017})\BibitemShut {NoStop}%
\bibitem [{\citenamefont {Eich}\ \emph {et~al.}(2018)\citenamefont {Eich}, \citenamefont {Herman}, \citenamefont {Pisoni}, \citenamefont {Overweg}, \citenamefont {Kurzmann}, \citenamefont {Lee}, \citenamefont {Rickhaus}, \citenamefont {Watanabe}, \citenamefont {Taniguchi}, \citenamefont {Sigrist}, \citenamefont {Ihn},\ and\ \citenamefont {Ensslin}}]{eich-prx18}%
  \BibitemOpen
  \bibfield  {author} {\bibinfo {author} {\bibfnamefont {M.}~\bibnamefont {Eich}}, \bibinfo {author} {\bibfnamefont {F.~c.~v.}\ \bibnamefont {Herman}}, \bibinfo {author} {\bibfnamefont {R.}~\bibnamefont {Pisoni}}, \bibinfo {author} {\bibfnamefont {H.}~\bibnamefont {Overweg}}, \bibinfo {author} {\bibfnamefont {A.}~\bibnamefont {Kurzmann}}, \bibinfo {author} {\bibfnamefont {Y.}~\bibnamefont {Lee}}, \bibinfo {author} {\bibfnamefont {P.}~\bibnamefont {Rickhaus}}, \bibinfo {author} {\bibfnamefont {K.}~\bibnamefont {Watanabe}}, \bibinfo {author} {\bibfnamefont {T.}~\bibnamefont {Taniguchi}}, \bibinfo {author} {\bibfnamefont {M.}~\bibnamefont {Sigrist}}, \bibinfo {author} {\bibfnamefont {T.}~\bibnamefont {Ihn}},\ and\ \bibinfo {author} {\bibfnamefont {K.}~\bibnamefont {Ensslin}},\ }\bibfield  {title} {\bibinfo {title} {Spin and valley states in gate-defined bilayer graphene quantum dots},\ }\href {https://doi.org/10.1103/PhysRevX.8.031023} {\bibfield  {journal} {\bibinfo  {journal} {Phys. Rev. X}\ }\textbf {\bibinfo
  {volume} {8}},\ \bibinfo {pages} {031023} (\bibinfo {year} {2018})}\BibitemShut {NoStop}%
\bibitem [{\citenamefont {Kurzmann}\ \emph {et~al.}(2019)\citenamefont {Kurzmann}, \citenamefont {Eich}, \citenamefont {Overweg}, \citenamefont {Mangold}, \citenamefont {Herman}, \citenamefont {Rickhaus}, \citenamefont {Pisoni}, \citenamefont {Lee}, \citenamefont {Garreis}, \citenamefont {Tong}, \citenamefont {Watanabe}, \citenamefont {Taniguchi}, \citenamefont {Ensslin},\ and\ \citenamefont {Ihn}}]{kurzmann-prl19}%
  \BibitemOpen
  \bibfield  {author} {\bibinfo {author} {\bibfnamefont {A.}~\bibnamefont {Kurzmann}}, \bibinfo {author} {\bibfnamefont {M.}~\bibnamefont {Eich}}, \bibinfo {author} {\bibfnamefont {H.}~\bibnamefont {Overweg}}, \bibinfo {author} {\bibfnamefont {M.}~\bibnamefont {Mangold}}, \bibinfo {author} {\bibfnamefont {F.}~\bibnamefont {Herman}}, \bibinfo {author} {\bibfnamefont {P.}~\bibnamefont {Rickhaus}}, \bibinfo {author} {\bibfnamefont {R.}~\bibnamefont {Pisoni}}, \bibinfo {author} {\bibfnamefont {Y.}~\bibnamefont {Lee}}, \bibinfo {author} {\bibfnamefont {R.}~\bibnamefont {Garreis}}, \bibinfo {author} {\bibfnamefont {C.}~\bibnamefont {Tong}}, \bibinfo {author} {\bibfnamefont {K.}~\bibnamefont {Watanabe}}, \bibinfo {author} {\bibfnamefont {T.}~\bibnamefont {Taniguchi}}, \bibinfo {author} {\bibfnamefont {K.}~\bibnamefont {Ensslin}},\ and\ \bibinfo {author} {\bibfnamefont {T.}~\bibnamefont {Ihn}},\ }\bibfield  {title} {\bibinfo {title} {Excited states in bilayer graphene quantum dots},\ }\href
  {https://doi.org/10.1103/PhysRevLett.123.026803} {\bibfield  {journal} {\bibinfo  {journal} {Phys. Rev. Lett.}\ }\textbf {\bibinfo {volume} {123}},\ \bibinfo {pages} {026803} (\bibinfo {year} {2019})}\BibitemShut {NoStop}%
\bibitem [{\citenamefont {Lee}\ \emph {et~al.}(2020)\citenamefont {Lee}, \citenamefont {Knothe}, \citenamefont {Overweg}, \citenamefont {Eich}, \citenamefont {Gold}, \citenamefont {Kurzmann}, \citenamefont {Klasovika}, \citenamefont {Taniguchi}, \citenamefont {Wantanabe}, \citenamefont {Fal'ko}, \citenamefont {Ihn}, \citenamefont {Ensslin},\ and\ \citenamefont {Rickhaus}}]{lee-prl20}%
  \BibitemOpen
  \bibfield  {author} {\bibinfo {author} {\bibfnamefont {Y.}~\bibnamefont {Lee}}, \bibinfo {author} {\bibfnamefont {A.}~\bibnamefont {Knothe}}, \bibinfo {author} {\bibfnamefont {H.}~\bibnamefont {Overweg}}, \bibinfo {author} {\bibfnamefont {M.}~\bibnamefont {Eich}}, \bibinfo {author} {\bibfnamefont {C.}~\bibnamefont {Gold}}, \bibinfo {author} {\bibfnamefont {A.}~\bibnamefont {Kurzmann}}, \bibinfo {author} {\bibfnamefont {V.}~\bibnamefont {Klasovika}}, \bibinfo {author} {\bibfnamefont {T.}~\bibnamefont {Taniguchi}}, \bibinfo {author} {\bibfnamefont {K.}~\bibnamefont {Wantanabe}}, \bibinfo {author} {\bibfnamefont {V.}~\bibnamefont {Fal'ko}}, \bibinfo {author} {\bibfnamefont {T.}~\bibnamefont {Ihn}}, \bibinfo {author} {\bibfnamefont {K.}~\bibnamefont {Ensslin}},\ and\ \bibinfo {author} {\bibfnamefont {P.}~\bibnamefont {Rickhaus}},\ }\bibfield  {title} {\bibinfo {title} {Tunable valley splitting due to topological orbital magnetic moment in bilayer graphene quantum point contacts},\ }\href
  {https://doi.org/10.1103/PhysRevLett.124.126802} {\bibfield  {journal} {\bibinfo  {journal} {Phys. Rev. Lett.}\ }\textbf {\bibinfo {volume} {124}},\ \bibinfo {pages} {126802} (\bibinfo {year} {2020})}\BibitemShut {NoStop}%
\bibitem [{\citenamefont {Kurzmann}\ \emph {et~al.}(2021)\citenamefont {Kurzmann}, \citenamefont {Kleeorin}, \citenamefont {Tong}, \citenamefont {Garreis}, \citenamefont {Knothe}, \citenamefont {Eich}, \citenamefont {Mittag}, \citenamefont {Gold}, \citenamefont {de~Vries}, \citenamefont {Watanabe}, \citenamefont {Taniguchi}, \citenamefont {Fal’ko}, \citenamefont {Meir}, \citenamefont {Ihn},\ and\ \citenamefont {Ensslin}}]{kurzmann-natcomm2021}%
  \BibitemOpen
  \bibfield  {author} {\bibinfo {author} {\bibfnamefont {A.}~\bibnamefont {Kurzmann}}, \bibinfo {author} {\bibfnamefont {Y.}~\bibnamefont {Kleeorin}}, \bibinfo {author} {\bibfnamefont {C.}~\bibnamefont {Tong}}, \bibinfo {author} {\bibfnamefont {R.}~\bibnamefont {Garreis}}, \bibinfo {author} {\bibfnamefont {A.}~\bibnamefont {Knothe}}, \bibinfo {author} {\bibfnamefont {M.}~\bibnamefont {Eich}}, \bibinfo {author} {\bibfnamefont {C.}~\bibnamefont {Mittag}}, \bibinfo {author} {\bibfnamefont {C.}~\bibnamefont {Gold}}, \bibinfo {author} {\bibfnamefont {F.~K.}\ \bibnamefont {de~Vries}}, \bibinfo {author} {\bibfnamefont {K.}~\bibnamefont {Watanabe}}, \bibinfo {author} {\bibfnamefont {T.}~\bibnamefont {Taniguchi}}, \bibinfo {author} {\bibfnamefont {V.}~\bibnamefont {Fal’ko}}, \bibinfo {author} {\bibfnamefont {Y.}~\bibnamefont {Meir}}, \bibinfo {author} {\bibfnamefont {T.}~\bibnamefont {Ihn}},\ and\ \bibinfo {author} {\bibfnamefont {K.}~\bibnamefont {Ensslin}},\ }\bibfield  {title} {\bibinfo {title} {Kondo effect and
  spin–orbit coupling in graphene quantum dots},\ }\bibfield  {journal} {\bibinfo  {journal} {Nature Communications}\ }\textbf {\bibinfo {volume} {12}},\ \href {https://doi.org/10.1038/s41467-021-26149-3} {10.1038/s41467-021-26149-3} (\bibinfo {year} {2021})\BibitemShut {NoStop}%
\bibitem [{\citenamefont {Vanderbilt}(2018)}]{vanderbilt-book2018}%
  \BibitemOpen
  \bibfield  {author} {\bibinfo {author} {\bibfnamefont {D.}~\bibnamefont {Vanderbilt}},\ }\href {https://doi.org/10.1017/9781316662205} {\emph {\bibinfo {title} {Berry Phases in Electronic Structure Theory: Electric Polarization, Orbital Magnetization and Topological Insulators}}}\ (\bibinfo  {publisher} {Cambridge University Press},\ \bibinfo {year} {2018})\BibitemShut {NoStop}%
\bibitem [{\citenamefont {Brown}(1964)}]{brown-pr64}%
  \BibitemOpen
  \bibfield  {author} {\bibinfo {author} {\bibfnamefont {E.}~\bibnamefont {Brown}},\ }\bibfield  {title} {\bibinfo {title} {Bloch electrons in a uniform magnetic field},\ }\href {https://doi.org/10.1103/PhysRev.133.A1038} {\bibfield  {journal} {\bibinfo  {journal} {Phys. Rev.}\ }\textbf {\bibinfo {volume} {133}},\ \bibinfo {pages} {A1038} (\bibinfo {year} {1964})}\BibitemShut {NoStop}%
\bibitem [{\citenamefont {Zak}(1964{\natexlab{a}})}]{zak-pr64a}%
  \BibitemOpen
  \bibfield  {author} {\bibinfo {author} {\bibfnamefont {J.}~\bibnamefont {Zak}},\ }\bibfield  {title} {\bibinfo {title} {Magnetic translation group},\ }\href {https://doi.org/10.1103/PhysRev.134.A1602} {\bibfield  {journal} {\bibinfo  {journal} {Phys. Rev.}\ }\textbf {\bibinfo {volume} {134}},\ \bibinfo {pages} {A1602} (\bibinfo {year} {1964}{\natexlab{a}})}\BibitemShut {NoStop}%
\bibitem [{\citenamefont {Zak}(1964{\natexlab{b}})}]{zak-pr64b}%
  \BibitemOpen
  \bibfield  {author} {\bibinfo {author} {\bibfnamefont {J.}~\bibnamefont {Zak}},\ }\bibfield  {title} {\bibinfo {title} {Magnetic translation group. {II. Irreducible} representations},\ }\href {https://doi.org/10.1103/PhysRev.134.A1607} {\bibfield  {journal} {\bibinfo  {journal} {Phys. Rev.}\ }\textbf {\bibinfo {volume} {134}},\ \bibinfo {pages} {A1607} (\bibinfo {year} {1964}{\natexlab{b}})}\BibitemShut {NoStop}%
\bibitem [{\citenamefont {Zak}(1964{\natexlab{c}})}]{zak-pr64c}%
  \BibitemOpen
  \bibfield  {author} {\bibinfo {author} {\bibfnamefont {J.}~\bibnamefont {Zak}},\ }\bibfield  {title} {\bibinfo {title} {Group-theoretical consideration of {Landau} level broadening in crystals},\ }\href {https://doi.org/10.1103/PhysRev.136.A776} {\bibfield  {journal} {\bibinfo  {journal} {Phys. Rev.}\ }\textbf {\bibinfo {volume} {136}},\ \bibinfo {pages} {A776} (\bibinfo {year} {1964}{\natexlab{c}})}\BibitemShut {NoStop}%
\bibitem [{\citenamefont {Zak}(1965)}]{zak-pr65}%
  \BibitemOpen
  \bibfield  {author} {\bibinfo {author} {\bibfnamefont {J.}~\bibnamefont {Zak}},\ }\bibfield  {title} {\bibinfo {title} {Proper functions for a {Bloch} electron in a magnetic field},\ }\href {https://doi.org/10.1103/PhysRev.139.A1159} {\bibfield  {journal} {\bibinfo  {journal} {Phys. Rev.}\ }\textbf {\bibinfo {volume} {139}},\ \bibinfo {pages} {A1159} (\bibinfo {year} {1965})}\BibitemShut {NoStop}%
\bibitem [{\citenamefont {Hofstadter}(1976)}]{hofstadter-prb76}%
  \BibitemOpen
  \bibfield  {author} {\bibinfo {author} {\bibfnamefont {D.~R.}\ \bibnamefont {Hofstadter}},\ }\bibfield  {title} {\bibinfo {title} {Energy levels and wave functions of {Bloch} electrons in rational and irrational magnetic fields},\ }\href {https://doi.org/10.1103/PhysRevB.14.2239} {\bibfield  {journal} {\bibinfo  {journal} {Phys. Rev. B}\ }\textbf {\bibinfo {volume} {14}},\ \bibinfo {pages} {2239} (\bibinfo {year} {1976})}\BibitemShut {NoStop}%
\bibitem [{\citenamefont {Thouless}\ \emph {et~al.}(1982)\citenamefont {Thouless}, \citenamefont {Kohmoto}, \citenamefont {Nightingale},\ and\ \citenamefont {den Nijs}}]{thouless-prl82}%
  \BibitemOpen
  \bibfield  {author} {\bibinfo {author} {\bibfnamefont {D.~J.}\ \bibnamefont {Thouless}}, \bibinfo {author} {\bibfnamefont {M.}~\bibnamefont {Kohmoto}}, \bibinfo {author} {\bibfnamefont {M.~P.}\ \bibnamefont {Nightingale}},\ and\ \bibinfo {author} {\bibfnamefont {M.}~\bibnamefont {den Nijs}},\ }\bibfield  {title} {\bibinfo {title} {Quantized {Hall} conductance in a two-dimensional periodic potential},\ }\href {https://doi.org/10.1103/PhysRevLett.49.405} {\bibfield  {journal} {\bibinfo  {journal} {Phys. Rev. Lett.}\ }\textbf {\bibinfo {volume} {49}},\ \bibinfo {pages} {405} (\bibinfo {year} {1982})}\BibitemShut {NoStop}%
\bibitem [{\citenamefont {Xiao}\ \emph {et~al.}(2010)\citenamefont {Xiao}, \citenamefont {Chang},\ and\ \citenamefont {Niu}}]{xiao-rmp10}%
  \BibitemOpen
  \bibfield  {author} {\bibinfo {author} {\bibfnamefont {D.}~\bibnamefont {Xiao}}, \bibinfo {author} {\bibfnamefont {M.-C.}\ \bibnamefont {Chang}},\ and\ \bibinfo {author} {\bibfnamefont {Q.}~\bibnamefont {Niu}},\ }\bibfield  {title} {\bibinfo {title} {Berry phase effects on electronic properties},\ }\href {https://doi.org/10.1103/RevModPhys.82.1959} {\bibfield  {journal} {\bibinfo  {journal} {Rev. Mod. Phys.}\ }\textbf {\bibinfo {volume} {82}},\ \bibinfo {pages} {1959} (\bibinfo {year} {2010})}\BibitemShut {NoStop}%
\bibitem [{\citenamefont {Herzog-Arbeitman}\ \emph {et~al.}(2020)\citenamefont {Herzog-Arbeitman}, \citenamefont {Song}, \citenamefont {Regnault},\ and\ \citenamefont {Bernevig}}]{herzog-prl20}%
  \BibitemOpen
  \bibfield  {author} {\bibinfo {author} {\bibfnamefont {J.}~\bibnamefont {Herzog-Arbeitman}}, \bibinfo {author} {\bibfnamefont {Z.-D.}\ \bibnamefont {Song}}, \bibinfo {author} {\bibfnamefont {N.}~\bibnamefont {Regnault}},\ and\ \bibinfo {author} {\bibfnamefont {B.~A.}\ \bibnamefont {Bernevig}},\ }\bibfield  {title} {\bibinfo {title} {Hofstadter topology: Noncrystalline topological materials at high flux},\ }\href {https://doi.org/10.1103/PhysRevLett.125.236804} {\bibfield  {journal} {\bibinfo  {journal} {Phys. Rev. Lett.}\ }\textbf {\bibinfo {volume} {125}},\ \bibinfo {pages} {236804} (\bibinfo {year} {2020})}\BibitemShut {NoStop}%
\bibitem [{\citenamefont {Ozawa}\ and\ \citenamefont {Mera}(2021)}]{ozawa-prb21}%
  \BibitemOpen
  \bibfield  {author} {\bibinfo {author} {\bibfnamefont {T.}~\bibnamefont {Ozawa}}\ and\ \bibinfo {author} {\bibfnamefont {B.}~\bibnamefont {Mera}},\ }\bibfield  {title} {\bibinfo {title} {Relations between topology and the quantum metric for {Chern} insulators},\ }\href {https://doi.org/10.1103/PhysRevB.104.045103} {\bibfield  {journal} {\bibinfo  {journal} {Phys. Rev. B}\ }\textbf {\bibinfo {volume} {104}},\ \bibinfo {pages} {045103} (\bibinfo {year} {2021})}\BibitemShut {NoStop}%
\bibitem [{\citenamefont {Wang}\ \emph {et~al.}(2021)\citenamefont {Wang}, \citenamefont {Cano}, \citenamefont {Millis}, \citenamefont {Liu},\ and\ \citenamefont {Yang}}]{wang-prl21}%
  \BibitemOpen
  \bibfield  {author} {\bibinfo {author} {\bibfnamefont {J.}~\bibnamefont {Wang}}, \bibinfo {author} {\bibfnamefont {J.}~\bibnamefont {Cano}}, \bibinfo {author} {\bibfnamefont {A.~J.}\ \bibnamefont {Millis}}, \bibinfo {author} {\bibfnamefont {Z.}~\bibnamefont {Liu}},\ and\ \bibinfo {author} {\bibfnamefont {B.}~\bibnamefont {Yang}},\ }\bibfield  {title} {\bibinfo {title} {Exact {Landau} level description of geometry and interaction in a flatband},\ }\href {https://doi.org/10.1103/PhysRevLett.127.246403} {\bibfield  {journal} {\bibinfo  {journal} {Phys. Rev. Lett.}\ }\textbf {\bibinfo {volume} {127}},\ \bibinfo {pages} {246403} (\bibinfo {year} {2021})}\BibitemShut {NoStop}%
\bibitem [{\citenamefont {Herzog-Arbeitman}\ \emph {et~al.}(2022{\natexlab{b}})\citenamefont {Herzog-Arbeitman}, \citenamefont {Chew},\ and\ \citenamefont {Bernevig}}]{herzog-prb22}%
  \BibitemOpen
  \bibfield  {author} {\bibinfo {author} {\bibfnamefont {J.}~\bibnamefont {Herzog-Arbeitman}}, \bibinfo {author} {\bibfnamefont {A.}~\bibnamefont {Chew}},\ and\ \bibinfo {author} {\bibfnamefont {B.~A.}\ \bibnamefont {Bernevig}},\ }\bibfield  {title} {\bibinfo {title} {Magnetic {Bloch} theorem and reentrant flat bands in twisted bilayer graphene at $2\ensuremath{\pi}$ flux},\ }\href {https://doi.org/10.1103/PhysRevB.106.085140} {\bibfield  {journal} {\bibinfo  {journal} {Phys. Rev. B}\ }\textbf {\bibinfo {volume} {106}},\ \bibinfo {pages} {085140} (\bibinfo {year} {2022}{\natexlab{b}})}\BibitemShut {NoStop}%
\bibitem [{\citenamefont {Wang}\ and\ \citenamefont {Vafek}(2022)}]{wang-prb22}%
  \BibitemOpen
  \bibfield  {author} {\bibinfo {author} {\bibfnamefont {X.}~\bibnamefont {Wang}}\ and\ \bibinfo {author} {\bibfnamefont {O.}~\bibnamefont {Vafek}},\ }\bibfield  {title} {\bibinfo {title} {Narrow bands in magnetic field and strong-coupling {Hofstadter} spectra},\ }\href {https://doi.org/10.1103/PhysRevB.106.L121111} {\bibfield  {journal} {\bibinfo  {journal} {Phys. Rev. B}\ }\textbf {\bibinfo {volume} {106}},\ \bibinfo {pages} {L121111} (\bibinfo {year} {2022})}\BibitemShut {NoStop}%
\bibitem [{\citenamefont {Wang}\ and\ \citenamefont {Vafek}(2023)}]{wang-prb23}%
  \BibitemOpen
  \bibfield  {author} {\bibinfo {author} {\bibfnamefont {X.}~\bibnamefont {Wang}}\ and\ \bibinfo {author} {\bibfnamefont {O.}~\bibnamefont {Vafek}},\ }\bibfield  {title} {\bibinfo {title} {Revisiting {Bloch} electrons in a magnetic field: {Hofstadter} physics via hybrid {Wannier} states},\ }\href {https://doi.org/10.1103/PhysRevB.108.245109} {\bibfield  {journal} {\bibinfo  {journal} {Phys. Rev. B}\ }\textbf {\bibinfo {volume} {108}},\ \bibinfo {pages} {245109} (\bibinfo {year} {2023})}\BibitemShut {NoStop}%
\bibitem [{\citenamefont {Fang}\ and\ \citenamefont {Cano}(2023)}]{fang-prb23}%
  \BibitemOpen
  \bibfield  {author} {\bibinfo {author} {\bibfnamefont {Y.}~\bibnamefont {Fang}}\ and\ \bibinfo {author} {\bibfnamefont {J.}~\bibnamefont {Cano}},\ }\bibfield  {title} {\bibinfo {title} {Symmetry indicators in commensurate magnetic flux},\ }\href {https://doi.org/10.1103/PhysRevB.107.245108} {\bibfield  {journal} {\bibinfo  {journal} {Phys. Rev. B}\ }\textbf {\bibinfo {volume} {107}},\ \bibinfo {pages} {245108} (\bibinfo {year} {2023})}\BibitemShut {NoStop}%
\bibitem [{\citenamefont {Singh}\ \emph {et~al.}(2024)\citenamefont {Singh}, \citenamefont {Chew}, \citenamefont {Herzog-Arbeitman}, \citenamefont {Bernevig},\ and\ \citenamefont {Vafek}}]{singh-natcomm24}%
  \BibitemOpen
  \bibfield  {author} {\bibinfo {author} {\bibfnamefont {K.}~\bibnamefont {Singh}}, \bibinfo {author} {\bibfnamefont {A.}~\bibnamefont {Chew}}, \bibinfo {author} {\bibfnamefont {J.}~\bibnamefont {Herzog-Arbeitman}}, \bibinfo {author} {\bibfnamefont {B.~A.}\ \bibnamefont {Bernevig}},\ and\ \bibinfo {author} {\bibfnamefont {O.}~\bibnamefont {Vafek}},\ }\bibfield  {title} {\bibinfo {title} {Topological heavy fermions in magnetic field},\ }\bibfield  {journal} {\bibinfo  {journal} {Nature Communications}\ }\textbf {\bibinfo {volume} {15}},\ \href {https://doi.org/10.1038/s41467-024-49531-3} {10.1038/s41467-024-49531-3} (\bibinfo {year} {2024})\BibitemShut {NoStop}%
\bibitem [{\citenamefont {Herzog-Arbeitman}(2023)}]{herzog-privcomm23}%
  \BibitemOpen
  \bibfield  {author} {\bibinfo {author} {\bibfnamefont {J.}~\bibnamefont {Herzog-Arbeitman}},\ }\href@noop {} {}\bibinfo {howpublished} {{Private Communication}} (\bibinfo {year} {2023})\BibitemShut {NoStop}%
\bibitem [{\citenamefont {Wang}\ \emph {et~al.}(2006)\citenamefont {Wang}, \citenamefont {Yates}, \citenamefont {Souza},\ and\ \citenamefont {Vanderbilt}}]{wang-prb06}%
  \BibitemOpen
  \bibfield  {author} {\bibinfo {author} {\bibfnamefont {X.}~\bibnamefont {Wang}}, \bibinfo {author} {\bibfnamefont {J.~R.}\ \bibnamefont {Yates}}, \bibinfo {author} {\bibfnamefont {I.}~\bibnamefont {Souza}},\ and\ \bibinfo {author} {\bibfnamefont {D.}~\bibnamefont {Vanderbilt}},\ }\bibfield  {title} {\bibinfo {title} {\textit{Ab initio} calculation of the anomalous {Hall} conductivity by {Wannier} interpolation},\ }\href {https://doi.org/10.1103/PhysRevB.74.195118} {\bibfield  {journal} {\bibinfo  {journal} {Phys. Rev. B}\ }\textbf {\bibinfo {volume} {74}},\ \bibinfo {pages} {195118} (\bibinfo {year} {2006})}\BibitemShut {NoStop}%
\bibitem [{\citenamefont {Wang}\ \emph {et~al.}(2007)\citenamefont {Wang}, \citenamefont {Vanderbilt}, \citenamefont {Yates},\ and\ \citenamefont {Souza}}]{wang-prb07}%
  \BibitemOpen
  \bibfield  {author} {\bibinfo {author} {\bibfnamefont {X.}~\bibnamefont {Wang}}, \bibinfo {author} {\bibfnamefont {D.}~\bibnamefont {Vanderbilt}}, \bibinfo {author} {\bibfnamefont {J.~R.}\ \bibnamefont {Yates}},\ and\ \bibinfo {author} {\bibfnamefont {I.}~\bibnamefont {Souza}},\ }\bibfield  {title} {\bibinfo {title} {Fermi-surface calculation of the anomalous {Hall} conductivity},\ }\href {https://doi.org/10.1103/PhysRevB.76.195109} {\bibfield  {journal} {\bibinfo  {journal} {Phys. Rev. B}\ }\textbf {\bibinfo {volume} {76}},\ \bibinfo {pages} {195109} (\bibinfo {year} {2007})}\BibitemShut {NoStop}%
\bibitem [{\citenamefont {Nielsch}\ \emph {et~al.}(2001)\citenamefont {Nielsch}, \citenamefont {Wehrspohn}, \citenamefont {Barthel}, \citenamefont {Kirschner}, \citenamefont {G\"{o}sele}, \citenamefont {Fischer},\ and\ \citenamefont {Kronm\"{u}ller}}]{nielsch-apl01}%
  \BibitemOpen
  \bibfield  {author} {\bibinfo {author} {\bibfnamefont {K.}~\bibnamefont {Nielsch}}, \bibinfo {author} {\bibfnamefont {R.~B.}\ \bibnamefont {Wehrspohn}}, \bibinfo {author} {\bibfnamefont {J.}~\bibnamefont {Barthel}}, \bibinfo {author} {\bibfnamefont {J.}~\bibnamefont {Kirschner}}, \bibinfo {author} {\bibfnamefont {U.}~\bibnamefont {G\"{o}sele}}, \bibinfo {author} {\bibfnamefont {S.~F.}\ \bibnamefont {Fischer}},\ and\ \bibinfo {author} {\bibfnamefont {H.}~\bibnamefont {Kronm\"{u}ller}},\ }\bibfield  {title} {\bibinfo {title} {Hexagonally ordered 100 nm period nickel nanowire arrays},\ }\href {https://doi.org/10.1063/1.1399006} {\bibfield  {journal} {\bibinfo  {journal} {Applied Physics Letters}\ }\textbf {\bibinfo {volume} {79}},\ \bibinfo {pages} {1360–1362} (\bibinfo {year} {2001})}\BibitemShut {NoStop}%
\bibitem [{\citenamefont {Nielsch}\ \emph {et~al.}(2002)\citenamefont {Nielsch}, \citenamefont {Wehrspohn}, \citenamefont {Barthel}, \citenamefont {Kirschner}, \citenamefont {Fischer}, \citenamefont {Kronmüller}, \citenamefont {Schweinböck}, \citenamefont {Weiss},\ and\ \citenamefont {Gösele}}]{nielsch-jmmm02}%
  \BibitemOpen
  \bibfield  {author} {\bibinfo {author} {\bibfnamefont {K.}~\bibnamefont {Nielsch}}, \bibinfo {author} {\bibfnamefont {R.~B.}\ \bibnamefont {Wehrspohn}}, \bibinfo {author} {\bibfnamefont {J.}~\bibnamefont {Barthel}}, \bibinfo {author} {\bibfnamefont {J.}~\bibnamefont {Kirschner}}, \bibinfo {author} {\bibfnamefont {S.~F.}\ \bibnamefont {Fischer}}, \bibinfo {author} {\bibfnamefont {H.}~\bibnamefont {Kronmüller}}, \bibinfo {author} {\bibfnamefont {T.}~\bibnamefont {Schweinböck}}, \bibinfo {author} {\bibfnamefont {D.}~\bibnamefont {Weiss}},\ and\ \bibinfo {author} {\bibfnamefont {U.}~\bibnamefont {Gösele}},\ }\bibfield  {title} {\bibinfo {title} {High density hexagonal nickel nanowire array},\ }\href {https://doi.org/https://doi.org/10.1016/S0304-8853(02)00536-X} {\bibfield  {journal} {\bibinfo  {journal} {Journal of Magnetism and Magnetic Materials}\ }\textbf {\bibinfo {volume} {249}},\ \bibinfo {pages} {234} (\bibinfo {year} {2002})}\BibitemShut {NoStop}%
\bibitem [{\citenamefont {Bruno}\ \emph {et~al.}(2004)\citenamefont {Bruno}, \citenamefont {Dugaev},\ and\ \citenamefont {Taillefumier}}]{bruno-prl04}%
  \BibitemOpen
  \bibfield  {author} {\bibinfo {author} {\bibfnamefont {P.}~\bibnamefont {Bruno}}, \bibinfo {author} {\bibfnamefont {V.~K.}\ \bibnamefont {Dugaev}},\ and\ \bibinfo {author} {\bibfnamefont {M.}~\bibnamefont {Taillefumier}},\ }\bibfield  {title} {\bibinfo {title} {Topological {Hall} effect and {Berry} phase in magnetic nanostructures},\ }\href {https://doi.org/10.1103/PhysRevLett.93.096806} {\bibfield  {journal} {\bibinfo  {journal} {Phys. Rev. Lett.}\ }\textbf {\bibinfo {volume} {93}},\ \bibinfo {pages} {096806} (\bibinfo {year} {2004})}\BibitemShut {NoStop}%
\bibitem [{\citenamefont {Taillefumier}\ \emph {et~al.}(2008)\citenamefont {Taillefumier}, \citenamefont {Dugaev}, \citenamefont {Canals}, \citenamefont {Lacroix},\ and\ \citenamefont {Bruno}}]{taillefumier-prb08}%
  \BibitemOpen
  \bibfield  {author} {\bibinfo {author} {\bibfnamefont {M.}~\bibnamefont {Taillefumier}}, \bibinfo {author} {\bibfnamefont {V.~K.}\ \bibnamefont {Dugaev}}, \bibinfo {author} {\bibfnamefont {B.}~\bibnamefont {Canals}}, \bibinfo {author} {\bibfnamefont {C.}~\bibnamefont {Lacroix}},\ and\ \bibinfo {author} {\bibfnamefont {P.}~\bibnamefont {Bruno}},\ }\bibfield  {title} {\bibinfo {title} {Chiral two-dimensional electron gas in a periodic magnetic field: Persistent current and quantized anomalous {Hall} effect},\ }\href {https://doi.org/10.1103/PhysRevB.78.155330} {\bibfield  {journal} {\bibinfo  {journal} {Phys. Rev. B}\ }\textbf {\bibinfo {volume} {78}},\ \bibinfo {pages} {155330} (\bibinfo {year} {2008})}\BibitemShut {NoStop}%
\bibitem [{\citenamefont {Yaron}\ \emph {et~al.}(1996)\citenamefont {Yaron}, \citenamefont {Gammel}, \citenamefont {Ramirez}, \citenamefont {Huse}, \citenamefont {Bishop}, \citenamefont {Goldman}, \citenamefont {Stassis}, \citenamefont {Canfield}, \citenamefont {Mortensen},\ and\ \citenamefont {Eskildsen}}]{yaron-nat96}%
  \BibitemOpen
  \bibfield  {author} {\bibinfo {author} {\bibfnamefont {U.}~\bibnamefont {Yaron}}, \bibinfo {author} {\bibfnamefont {P.~L.}\ \bibnamefont {Gammel}}, \bibinfo {author} {\bibfnamefont {A.~P.}\ \bibnamefont {Ramirez}}, \bibinfo {author} {\bibfnamefont {D.~A.}\ \bibnamefont {Huse}}, \bibinfo {author} {\bibfnamefont {D.~J.}\ \bibnamefont {Bishop}}, \bibinfo {author} {\bibfnamefont {A.~I.}\ \bibnamefont {Goldman}}, \bibinfo {author} {\bibfnamefont {C.}~\bibnamefont {Stassis}}, \bibinfo {author} {\bibfnamefont {P.~C.}\ \bibnamefont {Canfield}}, \bibinfo {author} {\bibfnamefont {K.}~\bibnamefont {Mortensen}},\ and\ \bibinfo {author} {\bibfnamefont {M.~R.}\ \bibnamefont {Eskildsen}},\ }\bibfield  {title} {\bibinfo {title} {Microscopic coexistence of magnetism and superconductivity in {ErNi$_2$B$_2$C}},\ }\href {https://doi.org/10.1038/382236a0} {\bibfield  {journal} {\bibinfo  {journal} {Nature}\ }\textbf {\bibinfo {volume} {382}},\ \bibinfo {pages} {236–238} (\bibinfo {year} {1996})}\BibitemShut {NoStop}%
\bibitem [{\citenamefont {De~Wilde}\ \emph {et~al.}(1997)\citenamefont {De~Wilde}, \citenamefont {Iavarone}, \citenamefont {Welp}, \citenamefont {Metlushko}, \citenamefont {Koshelev}, \citenamefont {Aranson}, \citenamefont {Crabtree},\ and\ \citenamefont {Canfield}}]{dewilde-prl97}%
  \BibitemOpen
  \bibfield  {author} {\bibinfo {author} {\bibfnamefont {Y.}~\bibnamefont {De~Wilde}}, \bibinfo {author} {\bibfnamefont {M.}~\bibnamefont {Iavarone}}, \bibinfo {author} {\bibfnamefont {U.}~\bibnamefont {Welp}}, \bibinfo {author} {\bibfnamefont {V.}~\bibnamefont {Metlushko}}, \bibinfo {author} {\bibfnamefont {A.~E.}\ \bibnamefont {Koshelev}}, \bibinfo {author} {\bibfnamefont {I.}~\bibnamefont {Aranson}}, \bibinfo {author} {\bibfnamefont {G.~W.}\ \bibnamefont {Crabtree}},\ and\ \bibinfo {author} {\bibfnamefont {P.~C.}\ \bibnamefont {Canfield}},\ }\bibfield  {title} {\bibinfo {title} {Scanning tunneling microscopy observation of a square {Abrikosov} lattice in {LuNi$_2$B$_2$C}},\ }\href {https://doi.org/10.1103/PhysRevLett.78.4273} {\bibfield  {journal} {\bibinfo  {journal} {Phys. Rev. Lett.}\ }\textbf {\bibinfo {volume} {78}},\ \bibinfo {pages} {4273} (\bibinfo {year} {1997})}\BibitemShut {NoStop}%
\bibitem [{\citenamefont {Eskildsen}\ \emph {et~al.}(1997)\citenamefont {Eskildsen}, \citenamefont {Gammel}, \citenamefont {Barber}, \citenamefont {Yaron}, \citenamefont {Ramirez}, \citenamefont {Huse}, \citenamefont {Bishop}, \citenamefont {Bolle}, \citenamefont {Lieber}, \citenamefont {Oxx}, \citenamefont {Sridhar}, \citenamefont {Andersen}, \citenamefont {Mortensen},\ and\ \citenamefont {Canfield}}]{eskildsen-prl97}%
  \BibitemOpen
  \bibfield  {author} {\bibinfo {author} {\bibfnamefont {M.~R.}\ \bibnamefont {Eskildsen}}, \bibinfo {author} {\bibfnamefont {P.~L.}\ \bibnamefont {Gammel}}, \bibinfo {author} {\bibfnamefont {B.~P.}\ \bibnamefont {Barber}}, \bibinfo {author} {\bibfnamefont {U.}~\bibnamefont {Yaron}}, \bibinfo {author} {\bibfnamefont {A.~P.}\ \bibnamefont {Ramirez}}, \bibinfo {author} {\bibfnamefont {D.~A.}\ \bibnamefont {Huse}}, \bibinfo {author} {\bibfnamefont {D.~J.}\ \bibnamefont {Bishop}}, \bibinfo {author} {\bibfnamefont {C.}~\bibnamefont {Bolle}}, \bibinfo {author} {\bibfnamefont {C.~M.}\ \bibnamefont {Lieber}}, \bibinfo {author} {\bibfnamefont {S.}~\bibnamefont {Oxx}}, \bibinfo {author} {\bibfnamefont {S.}~\bibnamefont {Sridhar}}, \bibinfo {author} {\bibfnamefont {N.~H.}\ \bibnamefont {Andersen}}, \bibinfo {author} {\bibfnamefont {K.}~\bibnamefont {Mortensen}},\ and\ \bibinfo {author} {\bibfnamefont {P.~C.}\ \bibnamefont {Canfield}},\ }\bibfield  {title} {\bibinfo {title} {Observation of a field-driven structural phase
  transition in the flux line lattice in {ErNi$_2$B$_2$C}},\ }\href {https://doi.org/10.1103/PhysRevLett.78.1968} {\bibfield  {journal} {\bibinfo  {journal} {Phys. Rev. Lett.}\ }\textbf {\bibinfo {volume} {78}},\ \bibinfo {pages} {1968} (\bibinfo {year} {1997})}\BibitemShut {NoStop}%
\bibitem [{\citenamefont {Yethiraj}\ \emph {et~al.}(1997)\citenamefont {Yethiraj}, \citenamefont {Paul}, \citenamefont {Tomy},\ and\ \citenamefont {Forgan}}]{yethiraj-prl97}%
  \BibitemOpen
  \bibfield  {author} {\bibinfo {author} {\bibfnamefont {M.}~\bibnamefont {Yethiraj}}, \bibinfo {author} {\bibfnamefont {D.~M.}\ \bibnamefont {Paul}}, \bibinfo {author} {\bibfnamefont {C.~V.}\ \bibnamefont {Tomy}},\ and\ \bibinfo {author} {\bibfnamefont {E.~M.}\ \bibnamefont {Forgan}},\ }\bibfield  {title} {\bibinfo {title} {Neutron scattering study of the flux lattice in {YNi$_2$B$_2$C}},\ }\href {https://doi.org/10.1103/PhysRevLett.78.4849} {\bibfield  {journal} {\bibinfo  {journal} {Phys. Rev. Lett.}\ }\textbf {\bibinfo {volume} {78}},\ \bibinfo {pages} {4849} (\bibinfo {year} {1997})}\BibitemShut {NoStop}%
\bibitem [{\citenamefont {Brown}\ \emph {et~al.}(2004)\citenamefont {Brown}, \citenamefont {Charalambous}, \citenamefont {Jones}, \citenamefont {Forgan}, \citenamefont {Kealey}, \citenamefont {Erb},\ and\ \citenamefont {Kohlbrecher}}]{brown-prl04}%
  \BibitemOpen
  \bibfield  {author} {\bibinfo {author} {\bibfnamefont {S.~P.}\ \bibnamefont {Brown}}, \bibinfo {author} {\bibfnamefont {D.}~\bibnamefont {Charalambous}}, \bibinfo {author} {\bibfnamefont {E.~C.}\ \bibnamefont {Jones}}, \bibinfo {author} {\bibfnamefont {E.~M.}\ \bibnamefont {Forgan}}, \bibinfo {author} {\bibfnamefont {P.~G.}\ \bibnamefont {Kealey}}, \bibinfo {author} {\bibfnamefont {A.}~\bibnamefont {Erb}},\ and\ \bibinfo {author} {\bibfnamefont {J.}~\bibnamefont {Kohlbrecher}},\ }\bibfield  {title} {\bibinfo {title} {Triangular to square flux lattice phase transition in {YBa$_2$Cu$_3$O$_7$}},\ }\href {https://doi.org/10.1103/PhysRevLett.92.067004} {\bibfield  {journal} {\bibinfo  {journal} {Phys. Rev. Lett.}\ }\textbf {\bibinfo {volume} {92}},\ \bibinfo {pages} {067004} (\bibinfo {year} {2004})}\BibitemShut {NoStop}%
\bibitem [{\citenamefont {Reichhardt}\ and\ \citenamefont {Gr\o{}nbech-Jensen}(2001)}]{reichhardt-prb01}%
  \BibitemOpen
  \bibfield  {author} {\bibinfo {author} {\bibfnamefont {C.}~\bibnamefont {Reichhardt}}\ and\ \bibinfo {author} {\bibfnamefont {N.}~\bibnamefont {Gr\o{}nbech-Jensen}},\ }\bibfield  {title} {\bibinfo {title} {Critical currents and vortex states at fractional matching fields in superconductors with periodic pinning},\ }\href {https://doi.org/10.1103/PhysRevB.63.054510} {\bibfield  {journal} {\bibinfo  {journal} {Phys. Rev. B}\ }\textbf {\bibinfo {volume} {63}},\ \bibinfo {pages} {054510} (\bibinfo {year} {2001})}\BibitemShut {NoStop}%
\bibitem [{\citenamefont {Han}\ \emph {et~al.}(2004)\citenamefont {Han}, \citenamefont {Wang}, \citenamefont {Wang},\ and\ \citenamefont {Xia}}]{han-prl04}%
  \BibitemOpen
  \bibfield  {author} {\bibinfo {author} {\bibfnamefont {Q.}~\bibnamefont {Han}}, \bibinfo {author} {\bibfnamefont {Z.~D.}\ \bibnamefont {Wang}}, \bibinfo {author} {\bibfnamefont {Q.-H.}\ \bibnamefont {Wang}},\ and\ \bibinfo {author} {\bibfnamefont {T.}~\bibnamefont {Xia}},\ }\bibfield  {title} {\bibinfo {title} {Vortex state in {Na$_x$CoO$_2\cdot$yH$_2$O}: ${p}_{x}\ifmmode\pm\else\textpm\fi{}i{p}_{y}$--wave versus ${d}_{{x}^{2}\ensuremath{-}{y}^{2}}\ifmmode\pm\else\textpm\fi{}i{d}_{xy}$--wave pairing},\ }\href {https://doi.org/10.1103/PhysRevLett.92.027004} {\bibfield  {journal} {\bibinfo  {journal} {Phys. Rev. Lett.}\ }\textbf {\bibinfo {volume} {92}},\ \bibinfo {pages} {027004} (\bibinfo {year} {2004})}\BibitemShut {NoStop}%
\bibitem [{\citenamefont {Meng}\ \emph {et~al.}(2014)\citenamefont {Meng}, \citenamefont {Varney}, \citenamefont {Fangohr},\ and\ \citenamefont {Babaev}}]{meng-prb14}%
  \BibitemOpen
  \bibfield  {author} {\bibinfo {author} {\bibfnamefont {Q.}~\bibnamefont {Meng}}, \bibinfo {author} {\bibfnamefont {C.~N.}\ \bibnamefont {Varney}}, \bibinfo {author} {\bibfnamefont {H.}~\bibnamefont {Fangohr}},\ and\ \bibinfo {author} {\bibfnamefont {E.}~\bibnamefont {Babaev}},\ }\bibfield  {title} {\bibinfo {title} {Honeycomb, square, and kagome vortex lattices in superconducting systems with multiscale intervortex interactions},\ }\href {https://doi.org/10.1103/PhysRevB.90.020509} {\bibfield  {journal} {\bibinfo  {journal} {Phys. Rev. B}\ }\textbf {\bibinfo {volume} {90}},\ \bibinfo {pages} {020509} (\bibinfo {year} {2014})}\BibitemShut {NoStop}%
\end{thebibliography}%

\end{document}